\documentclass[longauth,traditabstract]{aa}

\usepackage[nonamebreak]{natbib}
\usepackage[stable]{footmisc}
\usepackage{amsmath}
\usepackage{txfonts}
\usepackage{placeins}
\usepackage{natbib}
\bibpunct{(}{)}{;}{a}{}{,} %
\usepackage{graphicx}
\usepackage{epstopdf}
\usepackage{ifthen}
\usepackage[breaklinks, colorlinks, citecolor=blue]{hyperref}
\usepackage{overpic}
\usepackage{fixltx2e}
\usepackage{rotating}

\usepackage[table,usenames,dvipsnames]{xcolor}

\def\setsymbol#1#2{\expandafter\def\csname #1\endcsname{#2}}
\def\getsymbol#1{\csname #1\endcsname}

\def\Planck{\textit{Planck}}

\newbox\tablebox    \newdimen\tablewidth
\def\leaderfil{\leaders\hbox to 5pt{\hss.\hss}\hfil}
\def\endPlancktable{\tablewidth=\columnwidth 
    $$\hss\copy\tablebox\hss$$
    \vskip-\lastskip\vskip -2pt}
\def\endPlancktablewide{\tablewidth=\textwidth 
    $$\hss\copy\tablebox\hss$$
    \vskip-\lastskip\vskip -2pt}
\def\tablenote#1 #2\par{\begingroup \parindent=0.8em
    \abovedisplayshortskip=0pt\belowdisplayshortskip=0pt
    \noindent
    $$\hss\vbox{\hsize\tablewidth \hangindent=\parindent \hangafter=1 \noindent
    \hbox to \parindent{$^#1$\hss}\strut#2\strut\par}\hss$$
    \endgroup}
\def\doubleline{\vskip 3pt\hrule \vskip 1.5pt \hrule \vskip 5pt}

\def\L2{\ifmmode L_2\else $L_2$\fi}

\def\DeltaT{\ifmmode \Delta T\else $\Delta T$\fi}
\def\deltat{\ifmmode \Delta t\else $\Delta t$\fi}
\def\fknee{\ifmmode f_{\rm knee}\else $f_{\rm knee}$\fi}
\def\Fmax{\ifmmode F_{\rm max}\else $F_{\rm max}$\fi}
\def\solar{\ifmmode{\rm M}_{\mathord\odot}\else${\rm M}_{\mathord\odot}$\fi}
\def\Msolar{\ifmmode{\rm M}_{\mathord\odot}\else${\rm M}_{\mathord\odot}$\fi}
\def\Lsolar{\ifmmode{\rm L}_{\mathord\odot}\else${\rm L}_{\mathord\odot}$\fi}
\def\inv{\ifmmode^{-1}\else$^{-1}$\fi}
\def\mo{\ifmmode^{-1}\else$^{-1}$\fi}
\def\sup#1{\ifmmode ^{\rm #1}\else $^{\rm #1}$\fi}
\def\expo#1{\ifmmode \times 10^{#1}\else $\times 10^{#1}$\fi}
\def\,{\thinspace}
\def\lsim{\mathrel{\raise .4ex\hbox{\rlap{$<$}\lower 1.2ex\hbox{$\sim$}}}}
\def\gsim{\mathrel{\raise .4ex\hbox{\rlap{$>$}\lower 1.2ex\hbox{$\sim$}}}}

\def\simprop{\mathrel{\raise .4ex\hbox{\rlap{$\propto$}\lower 1.2ex\hbox{$\sim$}}}}
\def\deg{\ifmmode^\circ\else$^\circ$\fi}
\def\pdeg{\ifmmode $\setbox0=\hbox{$^{\circ}$}\rlap{\hskip.11\wd0 .}$^{\circ}
          \else \setbox0=\hbox{$^{\circ}$}\rlap{\hskip.11\wd0 .}$^{\circ}$\fi}
\def\arcs{\ifmmode {^{\scriptstyle\prime\prime}}
          \else $^{\scriptstyle\prime\prime}$\fi}
\def\arcm{\ifmmode {^{\scriptstyle\prime}}
          \else $^{\scriptstyle\prime}$\fi}
\newdimen\sa  \newdimen\sb
\def\parcs{\sa=.07em \sb=.03em
     \ifmmode \hbox{\rlap{.}}^{\scriptstyle\prime\kern -\sb\prime}\hbox{\kern -\sa}
     \else \rlap{.}$^{\scriptstyle\prime\kern -\sb\prime}$\kern -\sa\fi}
\def\parcm{\sa=.08em \sb=.03em
     \ifmmode \hbox{\rlap{.}\kern\sa}^{\scriptstyle\prime}\hbox{\kern-\sb}
     \else \rlap{.}\kern\sa$^{\scriptstyle\prime}$\kern-\sb\fi}
\def\ra[#1 #2 #3.#4]{#1\sup{h}#2\sup{m}#3\sup{s}\llap.#4}
\def\dec[#1 #2 #3.#4]{#1\deg#2\arcm#3\arcs\llap.#4}
\def\deco[#1 #2 #3]{#1\deg#2\arcm#3\arcs}
\def\rra[#1 #2]{#1\sup{h}#2\sup{m}}

\def\dots{\relax\ifmmode \ldots\else $\ldots$\fi}
\def\WHzsr{\ifmmode $W\,Hz\mo\,sr\mo$\else W\,Hz\mo\,sr\mo\fi}
\def\mHz{\ifmmode $\,mHz$\else \,mHz\fi}
\def\GHz{\ifmmode $\,GHz$\else \,GHz\fi}
\def\mKs{\ifmmode $\,mK\,s$^{1/2}\else \,mK\,s$^{1/2}$\fi}
\def\muKs{\ifmmode \,\mu$K\,s$^{1/2}\else \,$\mu$K\,s$^{1/2}$\fi}
\def\muKRJs{\ifmmode \,\mu$K$_{\rm RJ}$\,s$^{1/2}\else \,$\mu$K$_{\rm RJ}$\,s$^{1/2}$\fi}
\def\muKHz{\ifmmode \,\mu$K\,Hz$^{-1/2}\else \,$\mu$K\,Hz$^{-1/2}$\fi}
\def\MJysr{\ifmmode \,$MJy\,sr\mo$\else \,MJy\,sr\mo\fi}
\def\MJysrmK{\ifmmode \,$MJy\,sr\mo$\,mK$_{\rm CMB}\mo\else \,MJy\,sr\mo\,mK$_{\rm CMB}\mo$\fi}
\def\microns{\ifmmode \,\mu$m$\else \,$\mu$m\fi}

\def\muK{\ifmmode \,\mu$K$\else \,$\mu$\hbox{K}\fi}
\def\microK{\ifmmode \,\mu$K$\else \,$\mu$\hbox{K}\fi}
\def\muW{\ifmmode \,\mu$W$\else \,$\mu$\hbox{W}\fi}
\def\kms{\ifmmode $\,km\,s$^{-1}\else \,km\,s$^{-1}$\fi}
\def\kmsMpc{\ifmmode $\,\kms\,Mpc\mo$\else \,\kms\,Mpc\mo\fi}
\providecommand{\sorthelp}[1]{}

\def\LCDM{$\Lambda$CDM}
\def\NHUNIT{\ifmmode {\rm \,cm^{-2}} \else $\rm \,cm^{-2}$ \fi} % NH units

\def\wmap{\WMAP}
\def\aap{{A\&A}}

\def\apjs{{ApJ Supp.}}
\def\muKcmb{\ifmmode \,\mu$K$_{\rm CMB}$ \else \,$\mu$K$_{\rm CMB}$ \fi}
\newcommand{\hi}{\ensuremath{\mathsc {Hi}}}
\newcommand{\planck}{\Planck}
\newcommand{\WMAP}{WMAP}

\newcommand{\OmegaM}{\ifmmode \Omega_{\rm M} \else $\Omega_{\rm M}$ \fi}

\newcommand{\feff}{f_{\rm sky}^{\rm eff}}
\newcommand{\fsky}{f_{\rm sky}}

\newcommand{\ellbin}{\ell_{\rm bin}}
\newcommand{\XPol}{\texttt{XPol}}
\newcommand{\HFI}{HFI}
\newcommand{\LFI}{LFI}
\newcommand{\PR}{PR3}

\newcommand{\ndofspect}{23} % degrees of freedom in Sect 4 spectral model fit. 28 - 4 (or 5) parameters.
\newcommand{\ainten}{\langle I_{353}\rangle} % for use in math mode

\newcommand{\valpee}{-2.42} % alpha_EE in largest region
\newcommand{\valpeeu}{0.02} % alpha_EE uncertainty
\newcommand{\valpbb}{-2.54} % alpha_BB
\newcommand{\valpbbu}{0.02} % alpha_BB uncertainty

\newcommand{\mvalpee}{-2.39} % alpha_EE weighted mean
\newcommand{\mvalpeed}{0.05} % alpha_EE weighted mean individual dispersions
\newcommand{\mvalpbb}{-2.51} % alpha_BB weighted mean
\newcommand{\mvalpbbd}{0.06} % alpha_BB weighted mean individual dispersions

\newcommand{\vbetap}{1.53} % betaP in Section 5.1
\newcommand{\vbetapu}{0.02} % betaP uncertainty
\newcommand{\vdbetapi}{0.05} % betaP - betaI in Section 5.3
\newcommand{\vdbetapiu}{0.03} % betaP - betaI uncertainty

\begin{document}
%

%This author list corresponds to \title{Author list for L11\_Dust}
%Prepared by M. Lopez-Caniego (Marcos.Lopez.Caniego@sciops.esa.int), ESAC/ESA
%This version is from Wed Jul 11 15:42:44 2018 CET
%\subtitle{There are 133 co-authors in this list}
\author{\small
Planck Collaboration: Y.~Akrami\inst{46, 48}
\and
M.~Ashdown\inst{55, 4}
\and
J.~Aumont\inst{82}
\and
C.~Baccigalupi\inst{69}
\and
M.~Ballardini\inst{17, 32}
\and
A.~J.~Banday\inst{82, 7}
\and
R.~B.~Barreiro\inst{50}
\and
N.~Bartolo\inst{22, 51}
\and
S.~Basak\inst{75}
\and
K.~Benabed\inst{44, 81}
% ...
}
\institute{\small
AIM, CEA, CNRS,ÊUniversit\'{e} Paris-Saclay, F-91191 Gif sur Yvette, France. AIM, Universit\'{e} Paris Diderot, Sorbonne Paris Cit\'{e}, F-91191 Gif sur Yvette, France.\goodbreak
\and
APC, AstroParticule et Cosmologie, Universit\'{e} Paris Diderot, CNRS/IN2P3, CEA/lrfu, Observatoire de Paris, Sorbonne Paris Cit\'{e}, 10, rue Alice Domon et L\'{e}onie Duquet, 75205 Paris Cedex 13, France\goodbreak
\and
African Institute for Mathematical Sciences, 6-8 Melrose Road, Muizenberg, Cape Town, South Africa\goodbreak
\and
Astrophysics Group, Cavendish Laboratory, University of Cambridge, J J Thomson Avenue, Cambridge CB3 0HE, U.K.\goodbreak
\and
Astrophysics \& Cosmology Research Unit, School of Mathematics, Statistics \& Computer Science, University of KwaZulu-Natal, Westville Campus, Private Bag X54001, Durban 4000, South Africa\goodbreak
\and
CITA, University of Toronto, 60 St. George St., Toronto, ON M5S 3H8, Canada\goodbreak
\and
CNRS, IRAP, 9 Av. colonel Roche, BP 44346, F-31028 Toulouse cedex 4, France\goodbreak
\and
Cahill Center for Astronomy and Astrophysics, California Institute of Technology, Pasadena CA,  91125, USA\goodbreak
\and
California Institute of Technology, Pasadena, California, U.S.A.\goodbreak
\and
Computational Cosmology Center, Lawrence Berkeley National Laboratory, Berkeley, California, U.S.A.\goodbreak
\and
D\'{e}partement de Physique Th\'{e}orique, Universit\'{e} de Gen\`{e}ve, 24, Quai E. Ansermet,1211 Gen\`{e}ve 4, Switzerland\goodbreak
\and
Departamento de Astrof\'{i}sica, Universidad de La Laguna (ULL), E-38206 La Laguna, Tenerife, Spain\goodbreak
\and
Departamento de F\'{\i}sica, Universidad de Oviedo, C/ Federico Garc\'{\i}a Lorca, 18 , Oviedo, Spain\goodbreak
\and
Department of Astrophysics/IMAPP, Radboud University, P.O. Box 9010, 6500 GL Nijmegen, The Netherlands\goodbreak
\and
Department of Mathematics, University of Stellenbosch, Stellenbosch 7602, South Africa\goodbreak
\and
Department of Physics \& Astronomy, University of British Columbia, 6224 Agricultural Road, Vancouver, British Columbia, Canada\goodbreak
\and
Department of Physics \& Astronomy, University of the Western Cape, Cape Town 7535, South Africa\goodbreak
\and
Department of Physics and Astronomy, University of Sussex, Brighton BN1 9QH, U.K.\goodbreak
\and
Department of Physics, Gustaf H\"{a}llstr\"{o}min katu 2a, University of Helsinki, Helsinki, Finland\goodbreak
\and
Department of Physics, Princeton University, Princeton, New Jersey, U.S.A.\goodbreak
\and
Department of Physics, University of California, Santa Barbara, California, U.S.A.\goodbreak
\and
Dipartimento di Fisica e Astronomia G. Galilei, Universit\`{a} degli Studi di Padova, via Marzolo 8, 35131 Padova, Italy\goodbreak
\and
Dipartimento di Fisica e Scienze della Terra, Universit\`{a} di Ferrara, Via Saragat 1, 44122 Ferrara, Italy\goodbreak
\and
Dipartimento di Fisica, Universit\`{a} La Sapienza, P. le A. Moro 2, Roma, Italy\goodbreak
\and
Dipartimento di Fisica, Universit\`{a} degli Studi di Milano, Via Celoria, 16, Milano, Italy\goodbreak
\and
Dipartimento di Fisica, Universit\`{a} degli Studi di Trieste, via A. Valerio 2, Trieste, Italy\goodbreak
\and
Dipartimento di Fisica, Universit\`{a} di Roma Tor Vergata, Via della Ricerca Scientifica, 1, Roma, Italy\goodbreak
\and
European Space Agency, ESAC, Planck Science Office, Camino bajo del Castillo, s/n, Urbanizaci\'{o}n Villafranca del Castillo, Villanueva de la Ca\~{n}ada, Madrid, Spain\goodbreak
\and
European Space Agency, ESTEC, Keplerlaan 1, 2201 AZ Noordwijk, The Netherlands\goodbreak
\and
Gran Sasso Science Institute, INFN, viale F. Crispi 7, 67100 L'Aquila, Italy\goodbreak
\and
Helsinki Institute of Physics, Gustaf H\"{a}llstr\"{o}min katu 2, University of Helsinki, Helsinki, Finland\goodbreak
\and
INAF - OAS Bologna, Istituto Nazionale di Astrofisica - Osservatorio di Astrofisica e Scienza dello Spazio di Bologna, Area della Ricerca del CNR, Via Gobetti 101, 40129, Bologna, Italy\goodbreak
\and
INAF - Osservatorio Astronomico di Padova, Vicolo dell'Osservatorio 5, Padova, Italy\goodbreak
\and
INAF - Osservatorio Astronomico di Trieste, Via G.B. Tiepolo 11, Trieste, Italy\goodbreak
\and
INAF, Istituto di Radioastronomia, Via Piero Gobetti 101, I-40129 Bologna, Italy\goodbreak
\and
INAF/IASF Milano, Via E. Bassini 15, Milano, Italy\goodbreak
\and
INFN, Sezione di Bologna, viale Berti Pichat 6/2, 40127 Bologna, Italy\goodbreak
\and
INFN, Sezione di Ferrara, Via Saragat 1, 44122 Ferrara, Italy\goodbreak
\and
INFN, Sezione di Milano, Via Celoria 16, Milano, Italy\goodbreak
\and
INFN, Sezione di Roma 1, Universit\`{a} di Roma Sapienza, Piazzale Aldo Moro 2, 00185, Roma, Italy\goodbreak
\and
INFN, Sezione di Roma 2, Universit\`{a} di Roma Tor Vergata, Via della Ricerca Scientifica, 1, Roma, Italy\goodbreak
\and
Imperial College London, Astrophysics group, Blackett Laboratory, Prince Consort Road, London, SW7 2AZ, U.K.\goodbreak
\and
Institut d'Astrophysique Spatiale, CNRS, Univ. Paris-Sud, Universit\'{e} Paris-Saclay, B\^{a}t. 121, 91405 Orsay cedex, France\goodbreak
\and
Institut d'Astrophysique de Paris, CNRS (UMR7095), 98 bis Boulevard Arago, F-75014, Paris, France\goodbreak
\and
Institut dÕAstrophysique Spatiale, CNRS, Univ. Paris-Sud, Universite Paris-Saclay, Bat. 121, 91405 Orsay cedex, France\goodbreak
\and
Institute Lorentz, Leiden University, PO Box 9506, Leiden 2300 RA, The Netherlands\goodbreak
\and
Institute of Astronomy, University of Cambridge, Madingley Road, Cambridge CB3 0HA, U.K.\goodbreak
\and
Institute of Theoretical Astrophysics, University of Oslo, Blindern, Oslo, Norway\goodbreak
\and
Instituto de Astrof\'{\i}sica de Canarias, C/V\'{\i}a L\'{a}ctea s/n, La Laguna, Tenerife, Spain\goodbreak
\and
Instituto de F\'{\i}sica de Cantabria (CSIC-Universidad de Cantabria), Avda. de los Castros s/n, Santander, Spain\goodbreak
\and
Istituto Nazionale di Fisica Nucleare, Sezione di Padova, via Marzolo 8, I-35131 Padova, Italy\goodbreak
\and
Jet Propulsion Laboratory, California Institute of Technology, 4800 Oak Grove Drive, Pasadena, California, U.S.A.\goodbreak
\and
Jodrell Bank Centre for Astrophysics, Alan Turing Building, School of Physics and Astronomy, The University of Manchester, Oxford Road, Manchester, M13 9PL, U.K.\goodbreak
\and
Kavli Institute for Cosmological Physics, University of Chicago, Chicago, IL 60637, USA\goodbreak
\and
Kavli Institute for Cosmology Cambridge, Madingley Road, Cambridge, CB3 0HA, U.K.\goodbreak
\and
LERMA, CNRS, Observatoire de Paris, 61 Avenue de l'Observatoire, Paris, France\goodbreak
\and
LERMA/LRA, Observatoire de Paris, PSL Research University, CNRS, Ecole Normale Sup\'erieure, 75005 Paris, France\goodbreak
\and
Laboratoire AIM, CEA - Universit\'{e} Paris-Saclay, 91191 Gif-sur-Yvette, France\goodbreak
\and
Laboratoire Univers et Particules de Montpellier, UniversitŽ de Montpellier, CNRS/IN2P3, CC 72, Place Eugne Bataillon, 34095 Montpellier Cedex 5, France\goodbreak
\and
Laboratoire de Physique Subatomique et Cosmologie, Universit\'{e} Grenoble-Alpes, CNRS/IN2P3, 53, rue des Martyrs, 38026 Grenoble Cedex, France\goodbreak
\and
Laboratoire de Physique Th\'{e}orique, Universit\'{e} Paris-Sud 11 \& CNRS, B\^{a}timent 210, 91405 Orsay, France\goodbreak
\and
Low Temperature Laboratory, Department of Applied Physics, Aalto University, Espoo, FI-00076 AALTO, Finland\goodbreak
\and
Max Planck Institute for Astronomy, Kšnigstuhl 17, 69117 Heidelberg, Germany\goodbreak
\and
Max-Planck-Institut f\"{u}r Astrophysik, Karl-Schwarzschild-Str. 1, 85741 Garching, Germany\goodbreak
\and
Mullard Space Science Laboratory, University College London, Surrey RH5 6NT, U.K.\goodbreak
\and
NAOC-UKZN Computational Astrophysics Centre (NUCAC), University of KwaZulu-Natal, Durban 4000, South Africa\goodbreak
\and
Nicolaus Copernicus Astronomical Center, Polish Academy of Sciences, Bartycka 18, 00-716 Warsaw, Poland\goodbreak
\and
Nordita, KTH Royal Institute of Technology and Stockholm University, Roslagstullsbacken 23, 10691 Stockholm, Sweden\goodbreak
\and
SISSA, Astrophysics Sector, via Bonomea 265, 34136, Trieste, Italy\goodbreak
\and
San Diego Supercomputer Center, University of California, San Diego, 9500 Gilman Drive, La Jolla, CA 92093, USA\goodbreak
\and
School of Chemistry and Physics, University of KwaZulu-Natal, Westville Campus, Private Bag X54001, Durban, 4000, South Africa\goodbreak
\and
School of Physical Sciences, National Institute of Science Education and Research, HBNI, Jatni-752050, Odissa, India\goodbreak
\and
School of Physics and Astronomy, Cardiff University, Queens Buildings, The Parade, Cardiff, CF24 3AA, U.K.\goodbreak
\and
School of Physics and Astronomy, Sun Yat-sen University, 2 Daxue Rd, Tangjia, Zhuhai, China\goodbreak
\and
School of Physics, Indian Institute of Science Education and Research Thiruvananthapuram, Maruthamala PO, Vithura, Thiruvananthapuram 695551, Kerala, India\goodbreak
\and
Simon Fraser University, Department of Physics, 8888 University Drive, Burnaby BC, Canada\goodbreak
\and
Sorbonne Universit\'{e}-UPMC, UMR7095, Institut d'Astrophysique de Paris, 98 bis Boulevard Arago, F-75014, Paris, France\goodbreak
\and
Space Science Data Center - Agenzia Spaziale Italiana, Via del Politecnico snc, 00133, Roma, Italy\goodbreak
\and
Space Sciences Laboratory, University of California, Berkeley, California, U.S.A.\goodbreak
\and
The Oskar Klein Centre for Cosmoparticle Physics, Department of Physics, Stockholm University, AlbaNova, SE-106 91 Stockholm, Sweden\goodbreak
\and
UPMC Univ Paris 06, UMR7095, 98 bis Boulevard Arago, F-75014, Paris, France\goodbreak
\and
Universit\'{e} de Toulouse, UPS-OMP, IRAP, F-31028 Toulouse cedex 4, France\goodbreak
% ...
}

\title{\vglue -10mm{\Planck} 2018 results. XI.  Polarized dust foregrounds}

\abstract{\vglue -3mm 
The study of polarized dust emission has become entwined with the analysis of the cosmic microwave background (CMB) polarization in 
the quest for the curl-like $B$-mode polarization from primordial gravitational waves and the low-multipole $E$-mode polarization associated with the reionization of the Universe. 
% 2
We used the new \Planck\ \PR\ maps to characterize Galactic dust emission at high latitudes as a foreground to the CMB polarization and use 
end-to-end simulations to compute uncertainties and assess the statistical significance of our measurements.
% 3.1
We present \Planck\ $EE$, $BB$, and $TE$ power spectra of dust polarization at 353\,GHz for a set of six nested high-Galactic-latitude sky regions covering from 24 to 71\,\% of the sky.
% 3.2
We present power-law fits to the angular power spectra, yielding evidence for 
statistically significant variations of the exponents over sky regions and a difference between the 
%mean 
values for the $EE$ and $BB$ spectra,
which for the largest sky region are
$\alpha_{EE} = \valpee\pm\valpeeu$ and $\alpha_{BB} = \valpbb\pm\valpbbu$, respectively.
% 3.5, 3.4
The spectra show that the $TE$ correlation and $E/B$ power asymmetry discovered by \Planck\ extend to low multipoles that were not included in earlier \Planck\ polarization papers due to residual data systematics. We also report evidence for a positive $TB$ dust signal.
% 4
Combining data from \Planck\ and \WMAP, we have determined the amplitudes and spectral energy distributions (SEDs) of polarized foregrounds, including the correlation between dust and 
synchrotron polarized emission, for the six sky regions as a function of multipole.  This quantifies the challenge of the component-separation procedure that is required for measuring the low-$\ell$ reionization CMB $E$-mode signal and detecting 
the reionization and recombination peaks of primordial CMB $B$ modes. 
% 5
The SED of polarized dust emission is fit well by a single-temperature modified black-body emission law from 353\,GHz to below 70\,GHz.
For a dust temperature of $19.6$\,K, the mean dust spectral index for dust polarization is $\beta_{\rm d}^{P} = 
\vbetap\pm\vbetapu$. 
The difference between indices for polarization and total intensity is
$\beta_{\rm d}^P - \beta_{\rm d}^I = 
\vdbetapi\pm\vdbetapiu$.
%6
By fitting multi-frequency cross-spectra between \Planck\ data at 100, 143, 217, and 353\,GHz, we
examine the correlation of the dust polarization maps across frequency.  We find no evidence for a loss of correlation and provide lower limits to the correlation ratio that are tighter than values we 
derive from the correlation of the 217- and 353-GHz maps alone. 
If the \Planck\ limit on decorrelation for the largest sky region applies to the smaller sky regions observed by sub-orbital experiments, 
then frequency decorrelation of dust polarization might not be a problem for CMB experiments aiming at a primordial $B$-mode detection limit on the tensor-to-scalar ratio $r\simeq0.01$ at the recombination peak.
However, the \Planck\ sensitivity precludes identifying how difficult the component-separation problem will be
for more ambitious experiments targeting lower limits on $r$.
}
 
\keywords{CMB -- Foregrounds -- Galaxy -- Polarization -- Magnetic fields -- Interstellar dust }
\date{Received 11 January 2018 / Accepted 14 September 2018}

\authorrunning{Planck Collaboration}
\titlerunning{Dust polarized foregrounds}
\maketitle

\section{Introduction}

The polarization of the cosmic microwave background (CMB)
offers an opportunity for detecting primordial
gravitational waves, a key experimental manifestation of quantum gravity \citep{Starobinsky79}. 
Inflation generates tensor (gravitational waves) together with scalar (energy density) inhomogeneities. 
The polarization curl-like signal, referred to as primordial $B$ modes,
is a generic signature of gravitational waves produced during the inflation era in the very early Universe \citep{Guth81,Linde82}.
However, the ratio of tensor-to-scalar power, denoted $r$, varies considerably among models \citep{Baumann09}. 
Improvement of
the present limit, $r\,{<}\,0.07$ \citep[95\,\% confidence,][]{PhysRevLett.116.031302},
might be achieved by combining data from new sub-orbital experiments with data from \Planck,\footnote{\Planck\ (\url{http://www.esa.int/Planck}) is a project of
the European Space Agency (ESA) with instruments provided by two scientific
consortia funded by ESA member states and led by Principal Investigators from
France and Italy, telescope reflectors provided through a collaboration between
ESA and a scientific consortium led and funded by Denmark, and additional
contributions from NASA (USA).} as pioneered by the BICEP/Keck and \Planck\ joint analysis \citep{pb2015}.

Until the next CMB space mission, the \Planck\ data will remain unique, both for the all-sky coverage, required to measure 
CMB polarization at very low multipoles, and for its sensitive 353-GHz dust polarization maps.
At microwave frequencies, the sensitivity of \Planck\ is limited by the small number of detectors (12 per channel for the High Frequency Instrument \HFI), while 
today the most sensitive sub-orbital experiments have array sizes up to of
order $10^3$ detectors.
Further in the future, the CMB stage III and IV development plans in the United States
include array sizes increasing to more than $10^5$ detectors,
with a goal of detecting primordial $B$ modes down to $r\,{\simeq}\,10^{-3}$. 
On-going sub-orbital projects, including Advanced ACTPol \citep{Naess14}, BICEP2/3 and the Keck Array \citep{Grayson16}, CLASS \citep{Essinger-Hileman14}, 
PIPER \citep{Kogut11}, POLARBEAR and the Simons Array \citep{Arnold14}, the Simons Observatory,\footnote{\url{https://simonsobservatory.org/}} SPIDER \citep{Fraisse13}, and SPTPol \citep{Austermann12},
are paving this ambitious path. 

Indeed, the primordial $B$ modes might have high enough amplitude to be discovered by these experiments, but this exciting prospect does not depend solely on the data sensitivity. Discovery
depends on component separation, because
the cosmological signal is contaminated by polarized foreground emission from the Galaxy that has a higher amplitude \citep{Dunkley09,pb2015,Errard16,Hensley17,Remazeilles17}.
Component separation is also a key issue in the definition of future CMB space experiments, for example LiteBIRD \citep{Ishino16}.
This component-separation challenge binds the search for primordial $B$ modes to the statistical characterization, and the astrophysics, of polarized emission
from the magnetized interstellar medium (ISM). 

The spin axis of a non-spherical dust grain is both perpendicular to its long axis and aligned, statistically, with the orientation of the ambient Galactic magnetic field.
This alignment makes dust emission polarized perpendicular to the magnetic field projection on the plane of the sky \citep{Stein66,Hildebrand88,Martin07}, and also perpendicular to the optical interstellar polarization from the same grains, as confirmed by \citet{planck2014-XXI}.
Dust emission is the dominant polarized foreground at frequencies larger than around $70\,$GHz \citep{Dunkley09,planck2014-a12}. 
The \Planck\ maps greatly supersede, in sensitivity and statistical power, the data available from earlier ground-based and balloon-borne observations. 

Several studies have already used the \Planck\ data to investigate the link between the dust polarization maps and the structure of the ISM and of the Galactic magnetic field (GMF). \citet{planck2014-XIX} presented the first analysis of the polarized sky as seen at 353\,GHz (the most sensitive \Planck\ channel for polarized thermal dust emission), focusing on the statistics of the polarization fraction and angle, $p$ and $\psi$. Comparison with synthetic polarized emission maps, computed from simulations of magneto-hydrodynamical (MHD) turbulence, shows that the turbulent structure of the GMF is able to reproduce the main statistical properties in interstellar clouds \citep{planck2014-XX}. 

\citet{planck2014-XXX} (hereafter PXXX) present the polarized dust angular power spectra computed with the \Planck\ data over the high-Galactic-latitude sky that is best 
suited for the analysis of CMB anisotropies.  An $E/B$ asymmetry (usually quantified as the power ratio $C_\ell^{BB}/C_\ell^{EE}$) was discovered,
as well as significant $TE$ power.
A correlation between the filamentary structure of cold gas identified in the \Planck\ dust total intensity maps, 
and the local orientation of the GMF, derived from the dust polarization angle, 
has shown the two fields to be aligned statistically 
\citep{planck2014-XXXII,planck2015-XXXV,Kalberla16}. 
This alignment has also been reported for filamentary structures identified in spectroscopic \hi\ data cubes \citep{McClure06,Clark13,Clark15}.
The structures identified in \hi\ channel maps could, at least partly, correspond to gas velocity caustics. In that case,
the correlation between gas velocity and magnetic field orientation \citep{Lazarian18} would contribute to the observed alignment. 
However, the Planck dust total intensity maps trace the dust column density, and for these data the observed correlation with the GMF is 
unambiguously an alignment of density structures with the magnetic field. 
\citet{planck2015-XXXVIII} showed 
that this correlation could account for the $E/B$ asymmetry and also
the $TE$ correlation.

These observational results have been discussed in the context of interstellar  turbulence.
The alignment between density structures and magnetic field 
is observed in MHD simulations of the diffuse ISM and discussed by \citet{Hennebelle13}, \citet{Inoue16} and \citet{Soler17}.
The $E/B$ asymmetry and the $TE$ correlation have been considered as statistical signatures of turbulence in the magnetized ISM 
from different theoretical perspectives by \citet{Caldwell17,Kandel17,Kritsuk17,Kandel18}. This hypothesis is still debated.  There is no consensus
on whether it holds, and what we may be learning about interstellar turbulence. 

\citet{planck2016-XLIV} introduced a phenomenological framework that relates the dust polarization to the GMF structure, its mean orientation and a 
statistical description of its random (turbulent) component. This framework has been used to model dust polarization power spectra 
and to produce simulated maps that can be used to 
assess component-separation methods and residuals in the analysis of CMB polarization \citep{Ghosh16,Vansyngel16} and also underlies the dust sky model in the end-to-end (E2E) simulations used in this paper (see Appendix~\ref{appendix:methodology}). 

The \Planck\ data on polarized thermal dust emission allowed \citet{planck2014-XXII} to determine
the spectral energy distributions (SEDs) of dust polarized emission and dust total intensity at microwave frequencies ($\nu \le 353\,$GHz).  The combination of BLASTPol submillimetre data with \Planck\ \citep{Gandilo2016,Ashton2017} also shows that the frequency dependence of the polarization fraction $p$ is not strong.
New constraints like this, along with the ratio of dust polarized emission to the polarization fraction of optical interstellar polarization \citep{planck2014-XXI}, can be used to refine dust models \citep{Guillet17}.
The modelling of the dust SED is also essential to component separation for CMB studies \citep{Chluba17,Hensley17}.

\citet{planck2016-L} (hereafter PL) studied the correlation between dust polarization maps from the \HFI\ channels at 217 and 353\,GHz.  In developing the analysis for this paper, we found that systematic errors and noise question the evidence for spectral decorrelation proposed in that earlier paper. This conclusion is in agreement with the results of \citet{Sheehy17}, who discovered this independently. 

In this paper, one of a set associated with the 2018 release of data from the \Planck\ mission \citep{planck2016-l01}, we make use of \Planck\ maps from this third public release
hereafter referred to as \PR, to extend the characterization of
the polarized Galactic dust emission that is foreground to CMB polarization. Our data analysis procedure has three main new directions.

\noindent
(1) We expand the power-spectrum analysis of dust polarization into the low-multipole regime relevant for $E$- and $B$-mode CMB polarization associated with the reionization of the Universe.
This part of our analysis includes a validation of the dust polarization maps through running the mapmaking pipeline on simulated time-line data built from simulations of the sky, including a model of polarized dust emission.

\noindent
(2) We characterize the mean SED of polarized Galactic foregrounds away from the Galactic plane in harmonic space as a function of multipole. 

\noindent
(3) We analyse the correlation of dust polarization maps over all four polarized \HFI\ channels from 100 to 353\,GHz.
 
We focus on presenting results of direct relevance to component separation, leaving the astrophysical modelling of the results to follow-up studies. 
A second paper \citep{planck2016-l11B} presents a complementary perspective on dust polarization from an astrophysics perspective, focusing on the statistics of the polarization fraction and angle
derived from the 353-GHz \Planck\ maps.

The paper is organized as follows. 
In Sect.~\ref{sec:data}, we present the \planck\ sky maps and their validation.
Results from the power-spectrum analysis of the dust polarization maps at 353\,GHz are described in Sect.~\ref{sec:power_spectra}.
In Sect.~\ref{sec:dust_sed}, the \Planck\ \HFI\ maps in the frequency range 100 to 353\,GHz are combined with lower frequency maps from the \Planck\ Low Frequency Instrument (\LFI; \citealp{planck2016-l02}) and \WMAP\ \citep{bennett2012} to characterize polarized foregrounds across microwave frequencies and multipoles, including the correlation between dust and synchrotron polarization.
We compare the microwave SEDs of dust polarized emission and total intensity in Sect.~\ref{sec:betaT}.
We quantify the correlation between \Planck-HFI\ polarized dust maps in Sect.~\ref{sec:dust_deco}. 
Section~\ref{sec:summary} summarizes the main results of the paper.
The paper also has three appendices. Data simulations used to estimate uncertainties in our data analysis are presented in Appendix~\ref{appendix:methodology}. 
In Appendix~\ref{appendix:pdf_Rell}, we revisit the correlation analysis of the 217- and 353-GHz \Planck\ polarization maps investigated previously in PL, using the \PR\ data and E2E simulations.
Large tables, to be published electronically, are gathered in Appendix~\ref{appendix:tables}.

%============================================================
\section{The \Planck\ \PR\ polarization maps}
\label{sec:data}
%============================================================

%\subsection{\Planck\ polarization maps}
%============================================================

\Planck\ observed the sky in seven frequency
bands from 30 to $353\,$GHz for polarization, and in two additional bands at 545 and $857\,$GHz for intensity, with an angular resolution
from 31\arcmin\ to 5\arcmin\ \citep{planck2013-p01}. 
The in-flight performance of the two focal-plane instruments, \HFI\ and \LFI, are described in \citet{planck2011-1.5} and \citet{planck2011-1.4}, respectively. 
For this study, we use the new \Planck\ \PR\ maps. 
The processing of the \HFI\ data is described in
\citet{planck2016-l03} and that of \LFI\ data in \citet{planck2016-l02}.

The 100-, 143-, and 217-GHz \HFI\ maps are made using data from all bolometers, while the 353-GHz maps are constructed using only data from the polarization-sensitive bolometers (PSBs), 
as recommended in \citet{planck2016-l03}. 
To characterize the data noise and to compute power spectra at one given frequency that are unbiased by noise, we use maps built from data subsets, specifically the two half-mission and the 
two odd-even
survey maps \citep{planck2016-l03}.\footnote{The 'odd-even' split means the odd-numbered surveys
versus the even-numbered surveys, where a 'survey' is roughly six months of
data.}  In this paper, we focus on results obtained using half-mission maps, but have checked that 
conclusions would not be changed if we had used odd-even surveys instead.
The \Planck-HFI\ data noise and systematics are quantified and discussed in \citet{planck2016-l03} 
using the E2E simulations of \Planck\ observations introduced there.
The related methodology that we follow to estimate uncertainties from detector noise and residual systematic effects, and to propagate them to the results of our data analysis, 
is presented in Appendix~\ref{appendix:methodology}.

A posteriori characterization of polarization efficiencies \citep{planck2016-l03} suggests small modifications relative to the values used to produce the delivered frequency maps available on the \Planck\ legacy archive.
%CMB data analysis \citep{planck2016-l04,planck2016-l05} yields small modifications to the \HFI\ polarization efficiencies adopted in \citet{planck2016-l03} to produce the \HFI\ polarization maps.
Accordingly, we multiply the \PR\ \HFI\ polarization maps at 100, 143, and $217\,$GHz by 1.005, 0.98, and 1.015, respectively; uncertainties in these factors are of order 0.005.
For 353\,GHz, no such factor has been determined but we expect it to have the same magnitude as at the other HFI frequencies. Thus, we consider that there is a 1.5\,\% photometric uncertainty on the 353\,GHz polarized emission.

In addition, in Sect.~\ref{sec:dust_sed} we use polarization maps from \LFI\ at 30\,GHz, and the $K$ and $K\,a$ \WMAP\ channels \citep{bennett2012}
to separate dust and synchrotron polarized emission and quantify the
correlation between the two sources of emission. 
Because E2E simulations are not available for these data, we compute maps of uncertainties from Gaussian realizations of the data noise.
Power spectra of the data noise are derived from the half-difference of half-mission \Planck-LFI\ maps 
and the difference of year maps
for \WMAP. We note that it is easy to produce a large number (1000 or more) of data realizations with Gaussian data noise,
while only 300 E2E realizations are available for \HFI.

%============================================================
\section{Angular power spectra of dust polarization}
\label{sec:power_spectra}
%============================================================

In this section, we derive angular power spectra of dust polarization from the \PR\
maps at $353\,$GHz. Key improvements in the correction of data systematics 
allow us to extend earlier work on dust polarization (power spectra and SED) to the lowest multipoles.

\begin{figure}[!htbp]
\includegraphics[width=0.48\textwidth]{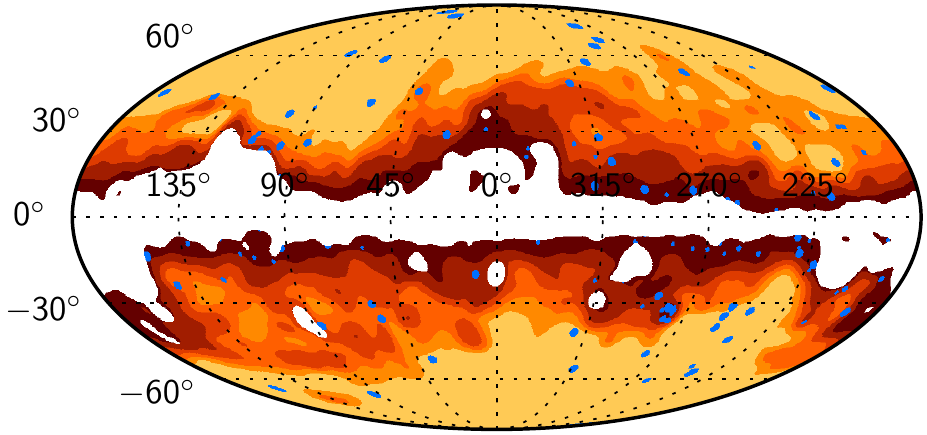}
\caption[]{All-sky map showing the sky regions used to measure power spectra, indicated with colours varying from yellow to orange and dark-red. 
The white region represents the area where the 
CO line brightness is larger than $0.4\,$K\,km\,s$^{-1}$, which is excluded from all the sky regions in our analysis. 
The blue dots represent the areas masked around point sources.}
\label{fig:sky_regions}
\end{figure}

\begin{figure*}[!htbp]
\includegraphics[width=18cm]{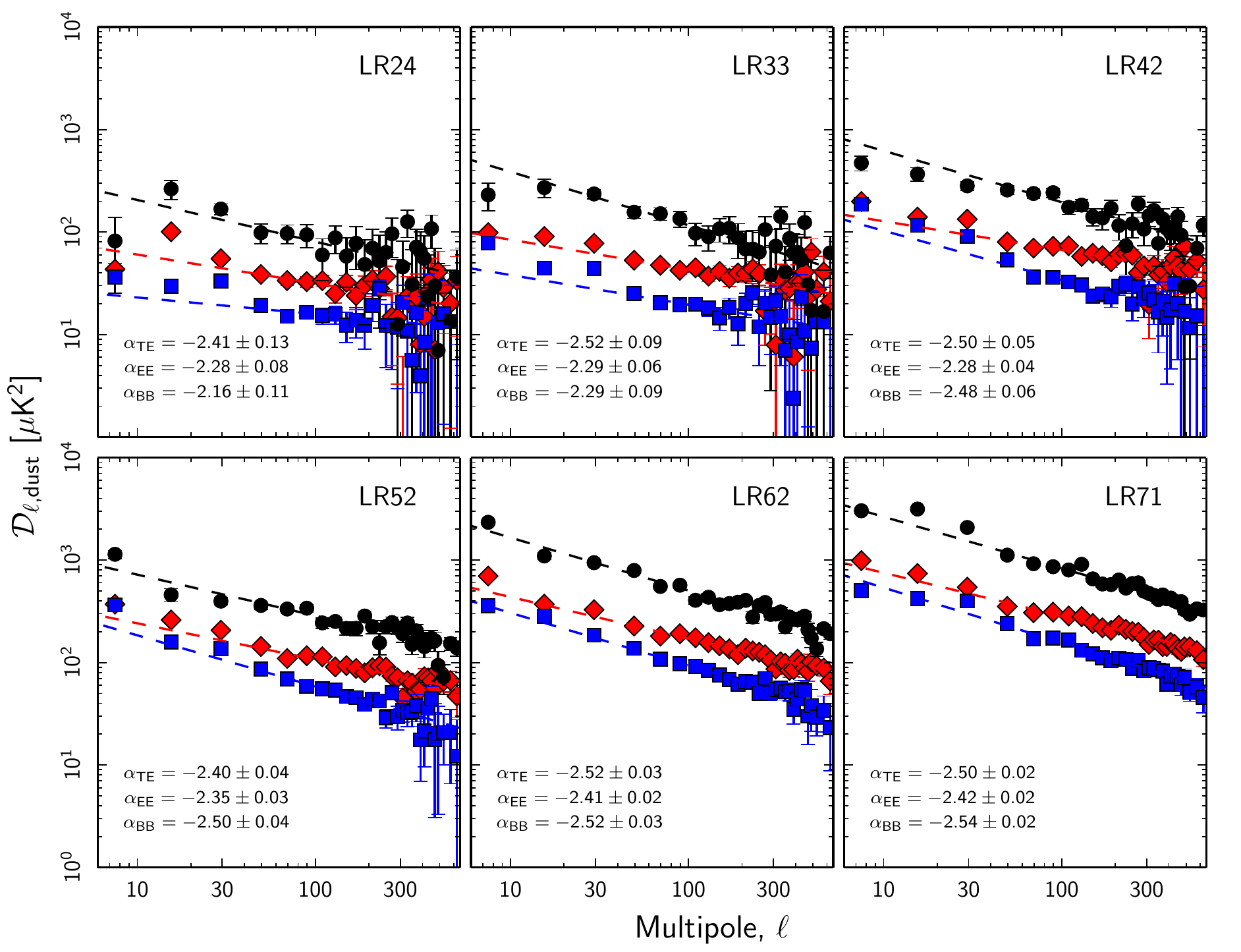}
\caption[]{CMB-corrected $EE$ (red diamonds), $BB$ (blue squares), and $TE$ (black circles) power spectra at 353\,GHz,
for each of the six sky regions that we analyse.  The dashed lines represent power-law fits to the data points 
from $\ell = 40$ to $600$.  The exponents of these fits, $\alpha_{\rm TE}$, $\alpha_{\rm EE}$, and $\alpha_{\rm BB}$, appear on each panel. }
\label{fig:spectra}
\end{figure*}

\begin{figure}[!htbp]
\includegraphics[width=9.0cm]{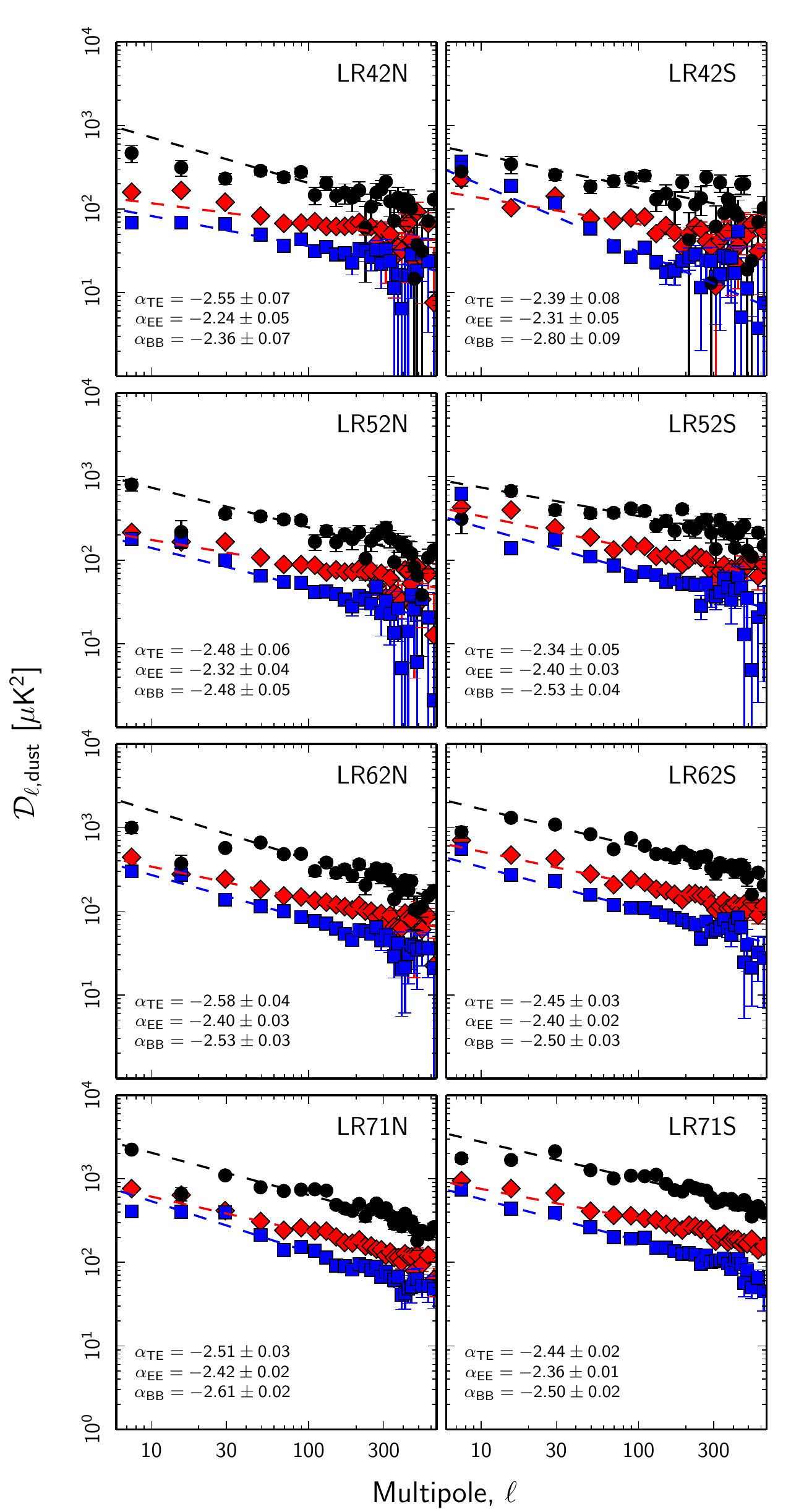}
\caption[]{Power spectra, as in Fig.~\ref{fig:spectra}, but for the northern and southern parts of the LR42, LR52, LR62, and LR71 regions. }
\label{fig:spectra2}
\end{figure}

\subsection{$\Planck$ angular power spectra at 353\,GHz}
\label{subsec:dust_power_spectra}

\begin{table*}[tbp!]
\newdimen\tblskip \tblskip=5pt
\caption{Parameters and $\chi^2$ of the power-law fits (Eq.~(\ref{eq:power_law})) to $EE$ and $BB$ dust power spectra over the multipole range $40 \le \ell \le 600$.}
\label{tab:ptes}
\vskip -5mm
\footnotesize
\setbox\tablebox=\vbox{
 \newdimen\digitwidth
 \setbox0=\hbox{\rm 0}
 \digitwidth=\wd0
 \catcode`*=\active
 \def*{\kern\digitwidth}
 \newdimen\signwidth
 \setbox0=\hbox{+}
 \signwidth=\wd0
 \catcode`!=\active
 \def!{\kern\signwidth}
  \newdimen\dpwidth
  \setbox0=\hbox{.}
  \dpwidth=\wd0
  \catcode`?=\active
  \def?{\kern\dpwidth}
\halign{\tabskip 0pt\hbox to 4.3cm{#\leaderfil}\tabskip 1em&
\hfil#\hfil\tabskip 0.75em& \hfil#\hfil& \hfil#\hfil&
\hfil#\hfil& \hfil#\hfil& \hfil#\hfil\tabskip 0em\cr
\noalign{\doubleline}
\omit& LR24& LR33& LR42& LR52& LR62& LR71\cr
\noalign{\vskip 3pt\hrule\vskip 5pt}
$\feff$ [\%]& 24& 33& 42& 52& 62& 71\cr
$\ainten$ [MJy\,sr$^{-1}$]& 0.066& 0.083& 0.104& 0.130& 0.164& 0.217\cr
$N_{\rm H} $ [$10^{20}$\,cm$^{-2}$]& 1.73& 2.18& 2.74& 3.48& 4.40& 5.85\cr
 \noalign{\vskip 3pt\hrule\vskip 5pt}
$\alpha_{TE}$& $-2.41\pm0.13$& $-2.52\pm0.09$& $-2.50\pm0.05$& $-2.40\pm0.04$& $-2.52\pm0.03$& $-2.50\pm0.02$\cr
$\alpha_{EE}$& $-2.28\pm0.08$& $-2.29\pm0.06$& $-2.28\pm0.04$& $-2.35\pm0.03$& $-2.41\pm0.02$& $-2.42\pm0.02$\cr
$\alpha_{BB}$& $-2.16\pm0.11$& $-2.29\pm0.09$& $-2.48\pm0.06$& $-2.50\pm0.04$& $-2.52\pm0.03$& $-2.54\pm0.02$\cr
\noalign{\vskip 3pt\hrule\vskip 5pt}
$\chi_{TE}^2 (\alpha_{TE}=-2.44, N_{\rm dof}=24)$& 16.0& $21.8$& $29.0$& $35.0$& $57.7$& $61.8$\cr
\noalign{\vskip 2pt}
$\chi_{EE}^2 (\alpha_{EE}=-2.44, N_{\rm dof}=24)$& 18.8& $25.2$& $37.5$& $37.1$& $30.4$& $53.8$\cr
\noalign{\vskip 2pt}
$\chi_{BB}^2 (\alpha_{BB}=-2.44, N_{\rm dof}=24)$& 19.6& $14.5$& $15.9$& $17.8$& $23.7$& $67.4$\cr
 \noalign{\vskip 3pt\hrule\vskip 5pt}
$A^{EE} (\ell = 80$)& $34.3\pm1.9$& $47.3\pm2.2$& $74.7\pm2.9$& $120.1\pm4.2$& $190.7\pm6.2$& $315.4\pm9.9$\cr 
 \noalign{\vskip 3pt\hrule\vskip 5pt}
$\langle A^{BB}/A^{EE}\rangle$& $0.48\pm0.04$& $0.45\pm0.03$& $0.50\pm0.02$& $0.53\pm0.01$& $0.53\pm0.01$& $0.53\pm0.01$\cr
$\langle A^{TE}/A^{EE}\rangle$& $2.60\pm0.27$& $2.68\pm0.20$& $2.83\pm0.13$& $2.68\pm0.09$& $2.78\pm0.07$& $2.77\pm0.05$\cr
\noalign{\vskip 3pt\hrule\vskip 5pt}}}
\endPlancktablewide
\end{table*} 

The power-spectrum analysis of \Planck\ dust polarization in PXXX was limited to multipoles $\ell > 40$, 
due to residual systematics in the available maps. 
The improvements made in correcting \Planck\ systematics for the new data release
allow us to extend the range of scales over which we can characterize dust polarization. 
 
The $EE$, $BB$, $TE$, $TB$, and $EB$ power spectra are computed with the \XPol\ code \citep{Tristram05}.  Following the approach in PXXX and PL, to avoid a bias arising from the noise, we compute all of the \Planck\ power spectra using cross-correlations of maps
with independent noise, specifically the half-mission maps.
To present a characterization of foregrounds that is independent of component-separation methods, 
we chose not to use the CMB polarization maps described in \citet{planck2016-l04}.
Instead, the CMB contribution is subtracted from the power spectra using 
the \Planck\ 2015 \LCDM\ model \citep{planck2014-a15}. 
The power spectra shown in the figures and tables below 
are in terms of ${\cal D_\ell}^{XY}\equiv\ell(\ell+1)C_\ell^{XY}/(2\pi)$, where $X \in \{T,E,B\}$, $Y \in \{E,B\}$, and $C_\ell^{XY}$ is the $XY$ angular power spectrum. 
The error bars are derived from the simulations described in Appendix~\ref{appendix:methodology};
they include the cosmic variance of the CMB computed for each sky region, because the CMB is subtracted using the \Planck\ 2015 \LCDM\ model. % \citep{planck2016-l06}. \DS{Again -- 2015?}

We examine six nested regions at high Galactic latitude, with an effective sky fraction $f_{\rm sky}^{\rm eff}$ ranging from 24 to 71\,\%.
These regions are defined using the same set of criteria as in PXXX,
meant to minimize dust polarization power for a given sky fraction,
and with the same apodization (see Fig.~\ref{fig:sky_regions}). 
The regions differ only in the masking of point sources; we mask a smaller number of sources that are polarized. 
We keep the same ``LRnm'' nomenclature, where ``nm'' is $f_{\rm sky}^{\rm eff}$ as a percentage. 
Table~\ref{tab:dust_spectra} lists other properties of the regions, including
the mean specific intensity at 353\,GHz, $\ainten$ in MJy\,sr$^{-1}$, and the mean \hi\ column density, $N_{\rm H}$ in units of $10^{20}$\,cm$^{-2}$,
inferred as in PL from the \Planck\ dust opacity map
in \citet{planck2016-XLVIII}.

The $EE$ and $BB$ spectra are tabulated in Table~\ref{tab:dust_spectra} and presented in Fig.~\ref{fig:spectra} for each of our six sky regions.
For the lowest multipole bin ($\ell=2$--3), we report a value for only the largest sky region LR71, over which it is best measured. 
In Fig.~\ref{fig:spectra2}, we present spectra computed on the northern and southern parts of the LR42, LR52, LR62, and LR71 regions.

\subsection{Power-law fits}
\label{subsec:power_law_fits}

We performed a $\chi^2$ fit to the power spectra
over the multipole range $40 \le \ell \le 600$, as in PXXX, using the equation: 
\begin{equation}
\label{eq:power_law}
{\cal D_\ell}^{XY} \equiv A^{XY} \, (\ell/80)^{\alpha_{XY}+2},
\end{equation}
where $XY \in \{EE, BB, TE\}$.
The power-law fits are displayed with dashed-lines in Fig.~\ref{fig:spectra} for the six sky regions and in Fig.~\ref{fig:spectra2} for the northern and 
southern parts of the LR42, LR52, LR62, and LR71 regions.
The amplitudes $A^{EE}$ and exponents $\alpha_{XY}$
are listed in Table~\ref{tab:ptes} for the six sky regions. The exponents are also printed in each panel of Figs.~\ref{fig:spectra} and \ref{fig:spectra2}. 
The error bars on $A^{EE}$ include a 3\,\% factor from the 1.5\,\% uncertainty on the 353\,GHz 
polarization efficiency. 

The power laws match the fitted data points well, but not perfectly.  Indeed, for many regions, including the largest ones with the highest signal-to-noise ratios, the
$\chi^2$ values in Table~\ref{tab:ptes} are larger than the number of degrees of freedom, $N_{\rm dof} =24$.
We note that these $\chi^2$ values are calculated for exponents fixed at a common value of $-2.44$.
There is evidence for statistically significant variations of the exponents over sky regions.  Furthermore, there is a difference between the values for the $EE$ and $BB$ spectra, which for the largest sky region are
$\alpha_{EE} = \valpee\pm\valpeeu$ and $\alpha_{BB} = \valpbb\pm\valpbbu$, respectively.

Figures~\ref{fig:spectra} and \ref{fig:spectra2} also show the extrapolation of the power laws to low multipoles, which may be compared to the data points at $\ell < 40$ not used in the fit.  The extrapolation is close to these data points in some cases, but not always.

Dust polarization angular power spectra, like the spectra of synchrotron emission, are related physically to the power spectrum of interstellar magnetic fields. 
Within the phenomenological models of \citet{Ghosh16} and \citet{Vansyngel16}, the exponent of the dust power spectrum is found to be 
close to that of the Gaussian random field used to simulate the turbulent component of the magnetic field.
The spectra are expected to flatten towards low multipoles, when the analysis is of an emitting volume sampling physical scales larger
than the injection scale of turbulence \citep{Cho02}. We do not observe such a flattening, but it might well be hidden by systematic variations of the magnetic field orientation over the
solar neighbourhood.  It will be necessary to extend the work of \citet{Vansyngel16} to low multipoles in order to assess whether 
our new results may be accounted for by statistical variance within their model framework.

\begin{figure}[!htbp]
\includegraphics[width=0.49\textwidth]{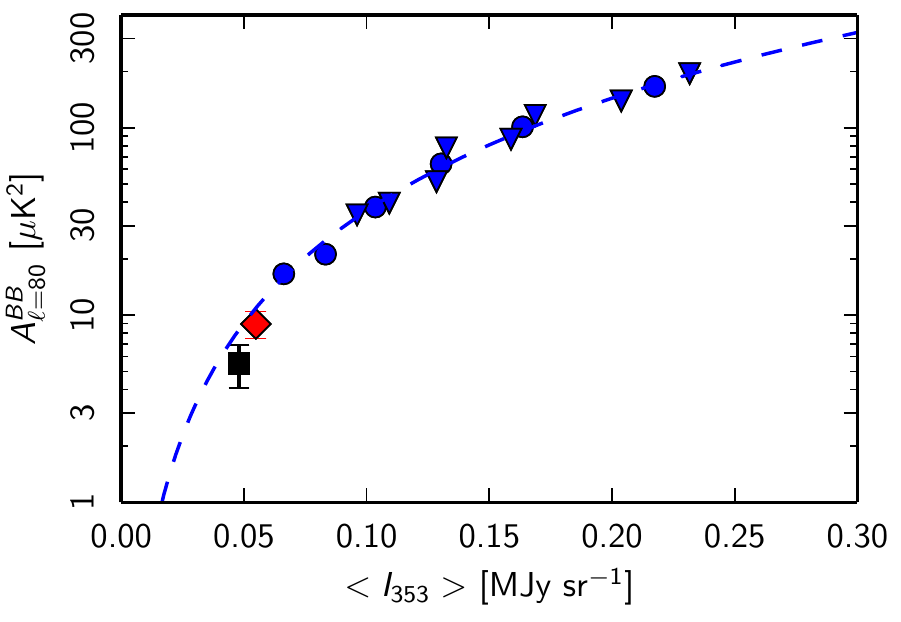}
\caption[]{Scaling of the $BB$ power at $\ell = 80$ versus the mean dust total intensity at 353\,GHz. 
The dashed black line is a power-law fit to values for the six sky regions in our analysis (blue dots) with an exponent of two.
Also shown are the values for the N--S splits of the regions in Fig.~\ref{fig:spectra2} (blue triangles). These results 
are complemented by the measurement (red diamond) over the southern Galactic cap ($\fsky = 8.5$\,\%) by \citet{Ghosh16} and that for the BICEP field (black square) after \citet{PhysRevLett.116.031302}.}
\label{fig:BB_vs_I353}
\end{figure}

\subsection{Scaling of $B$-mode power with total intensity of dust emission}
 \label{subsec:B_power}
 
In Fig.~\ref{fig:BB_vs_I353}, we plot the amplitude $A^{BB} (\ell = 80)$ versus the mean dust total intensity at 353\,GHz, %$\langle I_{353}\rangle$. 
$\ainten$.
The amplitudes are 
well fit by a power-law of the form $\ainten^2$ (the dashed line in Fig.~\ref{fig:BB_vs_I353}), i.e. with the same exponent as that measured for the amplitude of the total dust intensity angular power spectrum in the far-infrared \citep{Miville07}
but slightly greater than the value of 1.9 for dust polarization in PXXX. The fit to our results for the six sky regions also matches the
measurement reported by \citet{Ghosh16} for a region of low \hi\ column density in the southern sky with $\fsky = 8.5$\,\%.
 
We also measured the 353-GHz dust $B$-mode power on the \PR\ maps over the BICEP/Keck field using the mask available on the collaboration website. 
Measurements have been made as well on each of the 300 realizations of the E2E simulations. The dispersion of these measurements provides us with error bars, including both instrumental noise
and uncorrected systematics. We find $(4.4\pm3.4)\,\mu$K$^2$ using half-mission maps to compute cross-spectra, 
and $(0.83\pm3.1)\,\mu$K$^2$ for odd and even surveys. These measurements are consistent with the value 
derived from the correlation of the \Planck\ 353-GHz PR2 maps
with the BICEP/Keck 95- and 150-GHz data in \citet{PhysRevLett.116.031302}; 
to compare that value, $(4.3\pm1.1)\,\mu$K$^2$ at the reference frequency 353\,GHz, with the above results from \Planck\ PSB-only polarization band-integrated data at 353\,GHz, 
we multiply it by the colour correction 1.098$^2$ to obtain 
$(5.2 \pm1.3)\,\mu$K$^2$.
In the BICEP field $\ainten = 0.048\,$MJy\,sr$^{-1}$, for which
extrapolation of the fit to our measurements gives a signal level of approximately $8\,\mu$K$^2$. The difference 
is within the cosmic variance, as estimated by \citet{Ghosh16} using their statistical model of the dust polarization in the southern Galactic cap.

 \subsection{Asymmetry between the power in $E$ and $B$ modes}
 \label{subsec:E_B_ratio}

In Table~\ref{tab:ptes}, for each of the six regions we list the $BB/EE$ ratio of the amplitudes parameterizing the power-law fits.
The weighted mean ratio is $BB/EE = 0.524\pm0.005$, a value consistent with that in PXXX.
For some regions, but not all of them, we find
that the $E/B$ power asymmetry extends to the lowest multipole bins. 
At low multipoles the measured $BB/EE$ power ratio is in the range of about 0.5 to 1. 
%To interpret the variability of the exponents of the power laws and of the $EE/BB$ ratio in the lowest $\ell$ bins is beyond the scope of the paper, we would need to use a statistical model of the dust polarization sky, which extends the \citet{Vansyngel16} model to very low multipoles. 

The weighted mean values of the exponents for the $EE$ and $BB$ power spectra 
are $\alpha_{EE} = \mvalpee$ and $\alpha_{BB} = \mvalpbb$, respectively. 
The weighted dispersions of individual measurements for the six regions are \mvalpeed\ and \mvalpbbd, respectively. 
The exponents measured on the northern and southern parts of the LR42, LR52, LR62, and LR71 regions in Fig.~\ref{fig:spectra2} fit within this statistical characterization of our results 
for the full sky regions.
The exponents that we find are close to the values reported in PXXX. 

However, we find a small difference between the two exponents, which
suggests that the asymmetry changes slightly as a function of multipole.
Such a difference is not unexpected. The filamentary structures in the cold neutral interstellar medium have mainly $E$-mode polarization, due to the 
statistical alignment of the magnetic field orientation with matter \citep{Clark13,planck2015-XXXVIII,Ghosh16}.

\subsection{ The $TE$ correlation}
\label{subsec:TE_correlation}

\citet{planck2015-XXXVIII} related the 
$TE$ correlation to the observed alignment between filamentary structures and the magnetic field in the diffuse ISM,
while \citet{Caldwell17} discussed it theoretically in the context of MHD turbulence.
However, the new data shown here in Figs.~\ref{fig:spectra} and \ref{fig:spectra2} show that the $TE$ correlation extends down to the lowest multipoles, which
characterize dust polarization on angular scales larger than those of interstellar filaments. 
To examine this further,
we performed $\chi^2$ fits of a power law to the $TE$ spectra, as for $EE$ and $BB$, over the multipole range $\ell=40$--600.
The parameters of the fits are listed in Table~\ref{tab:ptes} and displayed in Figs.~\ref{fig:spectra} and \ref{fig:spectra2}.
The data points at $\ell < 40$, not included in the fit, are close to the extrapolation of the power laws to low multipoles. 

These new results show that the filamentary structure of the magnetized interstellar medium
alone cannot account for the observed $TE$ correlation. At least for the lowest multipoles, the correlation must have another origin that will need to be explored in future studies.
One possibility is that the low-$\ell$ TE correlation arises from the correlation between the local structure of the GMF with the geometry of the Local Bubble cavity \citep{Alves18}.

The weighted mean value of the exponent is $\alpha_{TE} = -2.49$, slightly different than $\alpha_{EE} =  \mvalpee$. 
The $TE$ spectrum is shallower (i.e. the absolute value of $\alpha_{TE} $ is smaller) than that measured on average for \hi\ column density maps 
\citep{Miville03,martin2015,Blagrave17}.
However, using line profile decomposition to isolate gas with the lowest velocity dispersion (the cold neutral medium or CNM), \citet{martin2015} and \citet{Ghosh16} provide evidence that the angular power spectrum of the column density of the CNM gas is shallower, in particular with exponent about $-2.4$ in the extended SGC34 region defined by the latter (a 3500 deg$^2$
region comprising 34\,\% of the southern Galactic cap with $f_{\rm sky}^{\rm eff} =0.085$).
As quantified by the modeling in \citet{Ghosh16}, this is in agreement  
with the idea that the $TE$ correlation, and the $E/B$ asymmetry, at $\ell > 40$
are related to the statistical alignment of the magnetic field with filamentary structure in the cold medium \citep{Clark15,planck2015-XXXVIII,Kalberla16}.

\begin{figure}[!htbp]
\includegraphics[width=0.5\textwidth]{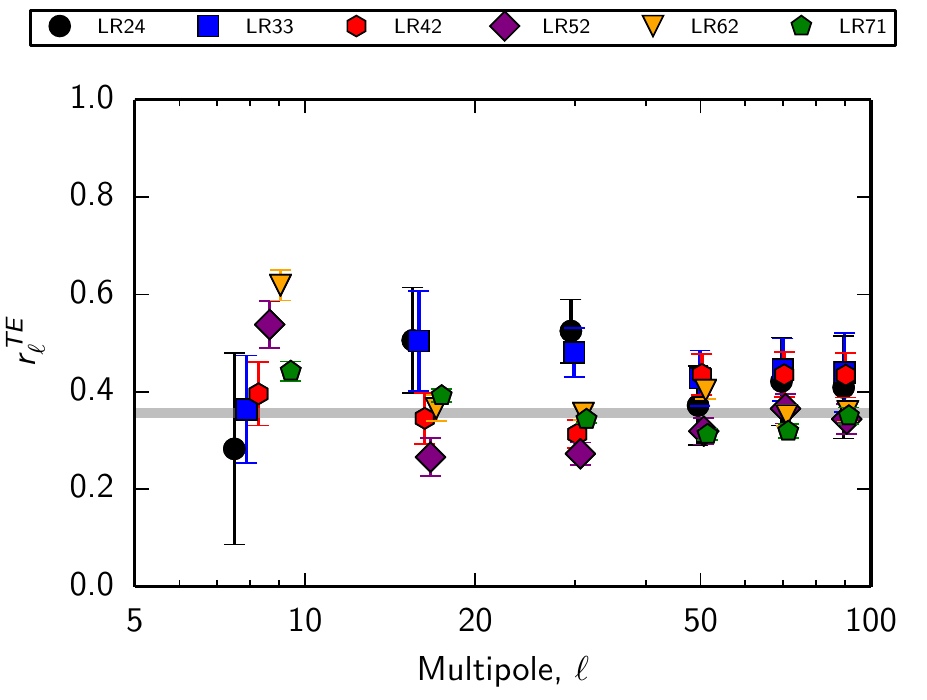} 
\caption[]{$TE$ correlation ratio $r^{TE}_\ell$ versus multipole. The data points are plotted using distinct symbols and colours (see legend at the top) for each of the six sky regions. The error bars
are derived from the E2E simulations.}
\label{fig:correlation_rTE}
\end{figure}

Table~\ref{tab:ptes} gives values of the ratio of the amplitudes of the $TE$ and $EE$ power
spectra.  The weighted mean value of the $TE/EE$ ratios is $2.76\pm0.05$.
We also combine the dust $TE$, $EE$, and $TT$ spectra at 353\,GHz 
to compute the dimensionless correlation ratio $r^{TE}_\ell = {\cal D}_\ell^{TE} /({\cal D}_\ell^{TT} \times {\cal D}_\ell^{EE})^{0.5}$ discussed by \citet{Caldwell17} and introduced in the context of the CMB in appendix~E3 of \citet{planck2014-a13}. The ratio is plotted versus multipole in 
Fig.~\ref{fig:correlation_rTE} for the six regions. The weighted mean of all measurements for all sky regions and multipole bins is $r^{TE}_\ell = 0.357\pm0.003$. The data
show significant scatter, but no systematic dependence on multipole down to the lowest $\ell$ bins or on the sky region.

\begin{figure*}[!htbp]
\includegraphics[width=18cm]{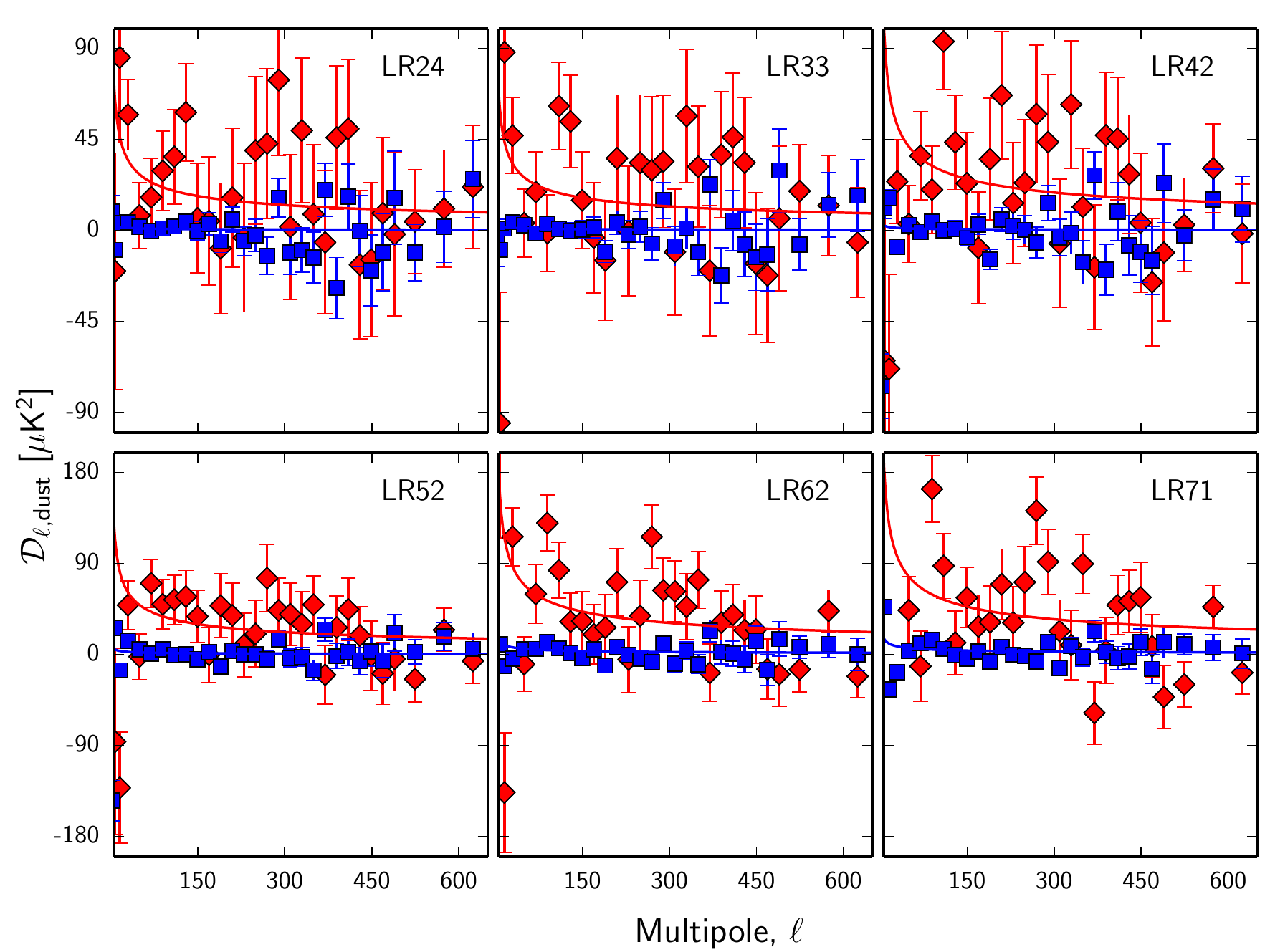}
\caption[]{Power spectra of $TB$ (red diamonds) and $EB$ (blue squares) at 353\,GHz for the six sky regions. The error bars are derived from the E2E simulations.  A power-law fit to the $TB$ data (solid red line) reveals an overall positive $TB$ signal, not seen in the E2E simulations. The $EB$ power (solid blue line fit) is consistent with zero.}
\label{fig:TB_spectra}
\end{figure*}

\subsection{$TB$ and $EB$ power spectra}
\label{subsec:TB_spectra}

The $TB$ and $EB$ angular power spectra are presented in Fig.~\ref{fig:TB_spectra}.
We find a positive $TB$ signal. A similar result was reported using earlier \Planck\ data in PXXX. 
On the largest sky regions providing the best signal-to-noise ratio, the power ratio $TB/TE$ is about 0.1 from a power-law fit (exponent fixed at $-2.44$)
over the $\ell=40$--600 multipole range.  The correlation ratio $r^{TB}_\ell = {\cal D}_\ell^{TB} /({\cal D}_\ell^{TT} \times {\cal D}_\ell^{BB})^{0.5}$, about 0.05, is also much lower than $r^{TE}_\ell$. 
The $EB$ signal is consistent with zero. The $EB/EE$ power ratio is smaller than about 0.03.

The E2E simulations in this paper allow us to check that the $TB$ power does not arise from a known systematic error.  For example, a systematic error in the orientation of the \Planck\ bolometers at $353\,$GHz would induce leakage of the $TE$ power to 
$TB$ and from the $EE$ and $BB$ power to $EB$ \citep{Abitbol16}. To account for a ratio $TB/TE = 0.1$, the error would need to be $3^\circ$, a value that is
one order of magnitude larger than the uncertainties on the orientation of the 
\HFI\ PSBs determined from CMB data analysis for the 100, 143, and 217\,GHz channels \citep[see section A.6 in][]{planck2014-a10}.

We do not see any systematic effects that could produce the $TB$ signal. 
If it is indeed real, this indicates that the dust polarization maps do not satisfy parity invariance. %FB: OK \DS{Adding words here}
Although there is no reason
for Galactic emission to preserve mirror symmetry, to our knowledge there is no straightforward interpretation 
of this observed asymmetry.   The $TB$ signal, at low multipoles, might arise from the structure of the mean magnetic field in the solar neighborhood.  
It might also be related to reference quantities of magnetohydrodynamic turbulence that are not parity invariant, 
such as the magnetic helicity (the volume integral of the scalar product between the vector potential and the magnetic field; see e.g. \citealp{Blackman2015}) and/or the cross-helicity 
(the integral of the scalar product between the gas velocity and the magnetic field; see e.g. \citealp{Yokoi2013}). 
These possible links will need to be explored in further studies. 

Within the context of CMB experiments, as discussed in \citet{Abitbol16} a non-zero dust $TB$ signal can
limit the accuracy to which $TB$ and $EB$ spectra at microwave frequencies may be used to check the orientation of the polarimeter.

\begin{figure}[!htbp]
\includegraphics[width=0.48\textwidth]{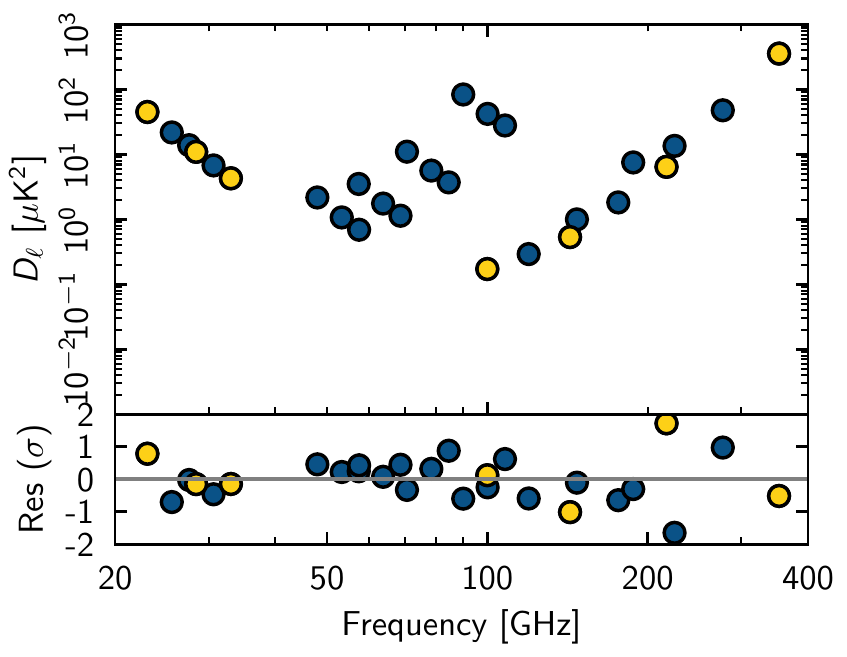}
\includegraphics[width=0.48\textwidth]{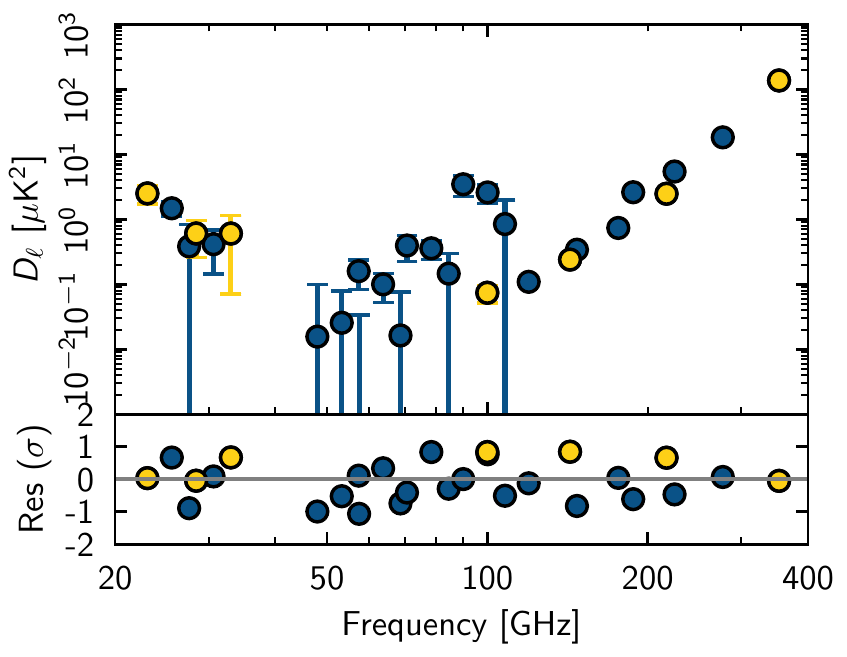}
\caption[]{$BB$ cross-spectra ${\cal D}_\ell (\nu_1 \times \nu_2)$ versus the effective frequency $\nu_{\rm eff} = (\nu_1\times\nu_2)^{0.5}$, 
for the LR62 sky region and two multipole bins: $\ell = 4$-11 (top plot) and 40--59 (bottom). 
Yellow and blue colours represent data values from single and inter-frequency cross spectra, respectively. 
The bottom panel within each plot shows the residuals from the fits normalized to the $1\,\sigma$ uncertainty of each data point.
Lower frequency data (left) points are dominated by the SED of synchrotron polarized emission, while higher frequency (right) data
characterize dust polarized emission, and those at the centre characterize the correlation between the two sources of emission. Differences 
between the two plots illustrate that both the ratio between synchrotron and dust power and the correlation between these two sources of polarized emission
decrease for increasing multipoles.} 
\label{fig:SED-pol}
\end{figure}

%===========================================================
\section{Dust and synchrotron polarized emission at microwave frequencies}
\label{sec:dust_sed}
%============================================================

We now calculate cross-power spectra, build models for them, and compare the foreground signals to the CMB.  Specifically, in Sect.~\ref{subsec:cross_spectra}
using cross-spectra we characterize Galactic polarized emission, including the correlation between dust and synchrotron polarization, as a function of frequency and 
multipole. In Sect.~\ref{subsec:spectral_model}, 
we fit these data with a spectral model and present the parameters determined. Galactic polarized foregrounds as quantified here are 
compared to the CMB primordial $E$- and $B$-mode signals as a function of frequency and multipole in Sect.~\ref{subsec:foregrounds}.

\subsection{Cross-power spectra}
\label{subsec:cross_spectra}

For this study, we consider single and inter-frequency cross-spectra among the four polarized channels 
of \Planck-HFI, at 100, 143, 217, and 353\,GHz, as well as the lowest frequency channel of \Planck-\LFI\ at 30\,GHz, and the two lowest frequencies of \WMAP\ at 23 and 33\,GHz. The three
channels of \LFI\ and \WMAP\ provide the highest signal-to-noise ratio on synchrotron polarization; we use them to estimate the synchrotron contribution to 
the lowest \HFI\ frequencies and characterize the spatial correlation between polarized dust and synchrotron sources of emission. 

Single frequency cross-spectra are computed using maps with independent statistical noise made with data subsets, to avoid noise bias. For \Planck-HFI, 
we use the half-mission maps. 
For \Planck-LFI, we separate data from even and odd years. 
For \WMAP, we combine the first four years on the one hand and the subsequent five years on the other hand. 
For inter-frequency cross-spectra, we consider all the possible combinations among the frequency channels being used. In total, we obtain 21 cross-spectra that combine observations at two distinct frequencies and 7 cross-spectra at a single frequency.  The uncertainties on power spectra are again computed from E2E simulations, as described in Appendix~\ref{appendix:methodology}. 

All 28 spectra are computed for each of the six sky regions described in Sect.~\ref{subsec:dust_power_spectra}, within nine multipole bins in the range $4\leq\ell\leq159$.  The specific multipole bins are top-hat (flat)
in the following ranges: 4--11; 12--19; 20--39; 40--59; 60--79; 80--99; 100--119; 120--139; and 140--159. 
Low signal-to-noise ratios prevent us from deriving meaningful SED parameters at higher multipoles.
Fig.~\ref{fig:SED-pol} presents an example for $B$ modes in the LR62 region for two multipole bins, $\ell = 4$--11 and 40--59.

\subsection{Spectral model}
\label{subsec:spectral_model}

Our SED analysis includes polarized synchrotron emission spatially correlated with polarized thermal dust emission  \citep{kogut2007,page2007,planck2014-XXII,planck2014-a12}.
We use the following spectral model, introduced by \citet{Choi15}:
\begin{align}
\label{eq:SED_model}
{\cal D}_\ell^{XX} (\nu_1 \times \nu_2) &=
 A_{\rm s}^{XX} \left( \frac{\nu_1 \nu_2}{30^2} \right)^{\beta_{\rm s}}
\nonumber \\
&+ A_{\rm d}^{XX} \left(\frac{\nu_1 \nu_2}{353^2} \right)^{\beta_{\rm d}-2} \frac{B_{\nu_1}(T_{\rm d})}{B_{353}(T_{\rm d})} \frac{B_{\nu_2}(T_{\rm d})}{B_{353}(T_{\rm d})} \nonumber \\
+\, \rho^{XX} (A_{\rm s}^{XX} \, A_{\rm d}^{XX})^{0.5} &\,\Bigg[ \left(\frac{\nu_1 } {30} \, \right)^{\beta_{\rm s}} \left(\frac{\nu_2 }{353} \right)^{\beta_{\rm d}-2} \frac{B_{\nu_2}(T_{\rm d})}{B_{353}(T_{\rm d})} \nonumber \\
&+ \left( \frac{\nu_2 } {30} \, \right)^{\beta_{\rm s}} \left(\frac{\nu_1 }{353} \right)^{\beta_{\rm d}-2} \frac{B_{\nu_1}(T_{\rm d})}{B_{353}(T_{\rm d})} \Bigg] \, ,
\end{align}
where $X \in \{E,B\}$ and ${\cal D}_\ell^{XX} (\nu_1 \times \nu_2)$ is the amplitude of the $XX$ cross-spectrum between frequencies $\nu_1$ and $\nu_2$ (expressed in GHz) 
within a given multipole bin $\ell$, expressed in terms of  brightness temperature squared.
The Planck function $B_{\nu}(T_{\rm d})$ is computed for a fixed dust temperature $T_{\rm d} = 19.6\,$K, derived from the fit of the SED of dust total intensity at high Galactic latitude in \citet{planck2014-XXII}.
We use a fixed temperature because, over microwave frequencies, the dust SED depends mainly on the dust spectral index of the 
modified black-body (or MBB) emission law and the temperature 
cannot be determined independently of the spectral index.  As discussed in \citet{planck2014-XXII} and \citet{Choi15}, the cross-correlation between dust and synchrotron polarization might arise
from the magnetic field structure but might also include a contribution from variations of the synchrotron spectral index and anomalous microwave emission (AME) if it is polarized \citep{Hoang16_AME,Draine16,Genova17}. 

%%%%%%%%%%%%%%%%%%%%%%%%%%%%%%%%%%%%
%%%% Table with unit conversion factors and color corrections %%%%%%
\begin{table*}[]
\newdimen\tblskip \tblskip=5pt
\caption{Unit conversion factors and colour corrections}
\label{tab:unit_color} 
\vskip -5mm
%\footnotesize
\setbox\tablebox=\vbox{
 \newdimen\digitwidth
 \setbox0=\hbox{\rm 0}
 \digitwidth=\wd0
 \catcode`*=\active
 \def*{\kern\digitwidth}
 \newdimen\signwidth
 \setbox0=\hbox{+}
 \signwidth=\wd0
 \catcode`!=\active
 \def!{\kern\signwidth}
  \newdimen\dpwidth
  \setbox0=\hbox{.}
  \dpwidth=\wd0
  \catcode`?=\active
  \def?{\kern\dpwidth}
\halign{\tabskip 0pt\hbox to 5.0cm{#\leaderfil}\tabskip 1em&
\hfil#\hfil\tabskip 1.0em& \hfil#\hfil& \hfil#\hfil&
\hfil#\hfil& \hfil#\hfil& \hfil#\hfil&  \hfil#\hfil& \hfil#\hfil& \hfil#\hfil\tabskip 0em\cr
\noalign{\doubleline}
Experiments & \wmap& \LFI& \wmap&  \LFI& \LFI &  \HFI& \HFI& \HFI& \HFI\ \cr
Reference frequencies [GHz] & 23& 28.4& 33& 44.1 & 70.4 & 100& 143& 217& 353\cr
\noalign{\vskip 3pt\hrule\vskip 5pt}
$U$& 0.986  & 0.949  & 0.972  & 0.932 & 0.848 & 0.794  & 0.592  & 0.334  & 0.075\cr
$C$& 1.073  & 1.000  & 1.027 &  1.000 & 0.981 & 1.088  & 1.017  & 1.120  & 1.098 \cr
\noalign{\vskip 3pt\hrule\vskip 5pt}
}}
\endPlancktablewide
\end{table*}

The spectral model has five parameters: the two amplitudes $A_{\rm s}$ and $A_{\rm d}$ and the two spectral indices $\beta_{\rm s}$ and $\beta_{\rm d}$, characterizing the synchrotron and dust 
SEDs, respectively; and the correlation factor $\rho$ quantifying the spatial correlation between synchrotron and dust polarized emission.
In Eq.~(\ref{eq:SED_model}), the  synchrotron and MBB emission are expressed in terms of brightness temperature, whereas the data are in thermodynamic units.  The conversion between the two is accomplished by two factors.  The first, $U$, is a unit conversion from the thermodynamic units to brightness temperature units for some adopted reference spectral dependence, performing the appropriate integrations over the bandpass.  The second, $C$, is a colour correction from the actual spectrum of the model to the adopted reference spectral dependence, again with bandpass integrations.  Accordingly, the spectrum is converted into units of the data by multiplication by $C/U$, and in the application to the fit of the spectral model in Eq.~(\ref{eq:SED_model}) by multiplication by $(C/U)_1 (C/U)_2$.
These factors were computed as in \citet{planck2014-XXII}, for \Planck\ using the procedures  
{\tt hfi\_unit\_conversion} and {\tt hfi\_colour\_correction} (for both \HFI\ and \LFI) and the instrument data files described in the \Planck\ Explanatory 
Supplement,\footnote{\url{http://wiki.cosmos.esa.int/planckpla2015/index.php/Unit\_conversion\_and\_Color\_correction}} and for \wmap\ the formulae and tabulations in \citet{Jarosik03}.  Here, 
for both \HFI\ and \LFI\ the adopted reference spectral dependence is $I_\nu \propto \nu^{-1}$ (see discussion in \citealp{planck2013-p03d} 
and the \Planck\ Explanatory Supplement\footnote{\url{https://wiki.cosmos.esa.int/planckpla2015/index.php/UC_CC_Tables}}),
whereas for \wmap\ it is constant Rayleigh-Jeans temperature.  By construction, the ratio $C/U$ does not depend on the adopted choice.
The conversion factors used are listed in Table~\ref{tab:unit_color}.  These are very close to the factors  in Table~3 of \citet{planck2014-XXII}, though here at 353\,GHz the evaluation is for the PSBs only.
The values of $C$ are evaluated for the following SED.  For the LFI and WMAP channels used, the synchrotron component dominates, for which we assume
$\beta_{\rm s} = -3$, while for the \Planck\ HFI channels the polarized dust MBB spectrum dominates, for which we assume $\beta_{\rm d} = 1.5$ and $T_{\rm d} = 19.6$\,K.  

We fit our spectral model to the $EE$ and $BB$ spectra separately, for each sky region and for each multipole bin independently.
Before fitting, we subtract the amplitude of the CMB power spectrum, estimated from the \Planck\ 2015 $\Lambda$CDM model \citep{planck2014-a15}, from each data point.
The fit is carried out in two steps. First, we fit the model of Eq.~(\ref{eq:SED_model}) using the {\tt MPFIT} code, which
uses the Levenberg-Marquardt algorithm to perform a least-squares fit. 
We then compute the weighted mean and standard deviation of $\beta_{\rm s}$ over the {\tt MPFIT} results for all sky regions and multipole bins, finding $\beta_{\rm s} = -3.13\pm0.13$. 
This value of $\beta_{\rm s}$ is consistent with those obtained by \citet{Fuskeland14} and \citet{Choi15} using all frequency channels of \wmap, as well as that,  
$-3.22 \pm 0.08$, reported by \citet{Krachmalnicoff18} for the frequency range $2.3-33\,$GHz, combining data from the S-band Polarization All-Sky Survey (S-PASS) at $2.3\,$GHz, \wmap, and \Planck.
We use it as a Gaussian prior for a second fit of the same data with the same model. 
This second fit is performed with a Monte Carlo Markov chain (MCMC) algorithm. In both fits we assume that the data points are independent. We checked on the E2E realizations that this is true for
the $B$-mode data. For $E$-mode, the CMB variance introduces a slight correlation that we neglect.
We adopt this two-step procedure because when attempting to fit $\beta_{\rm d}$ without a prior on $\beta_{\rm s}$ we found spurious results for a few combinations of $\ellbin$ and sky regions, when the signal-to-noise ratio 
in the low-frequency channels is too low to constrain the synchrotron SED adequately.

An example is given in Fig.~\ref{fig:SED-pol}, also showing the residuals from the fit.  The $\chi^2$ values for all fits are listed in Tables~\ref{tab:EE_fit} and \ref{tab:BB_fit} for the $EE$ and $BB$ spectra, respectively.
The results obtained on the simulated maps (Fig.~\ref{fig:E2E_sims}) show that the fit parameters match the input values without any bias.

\begin{figure}[!htbp]
\includegraphics[width=0.49\textwidth]{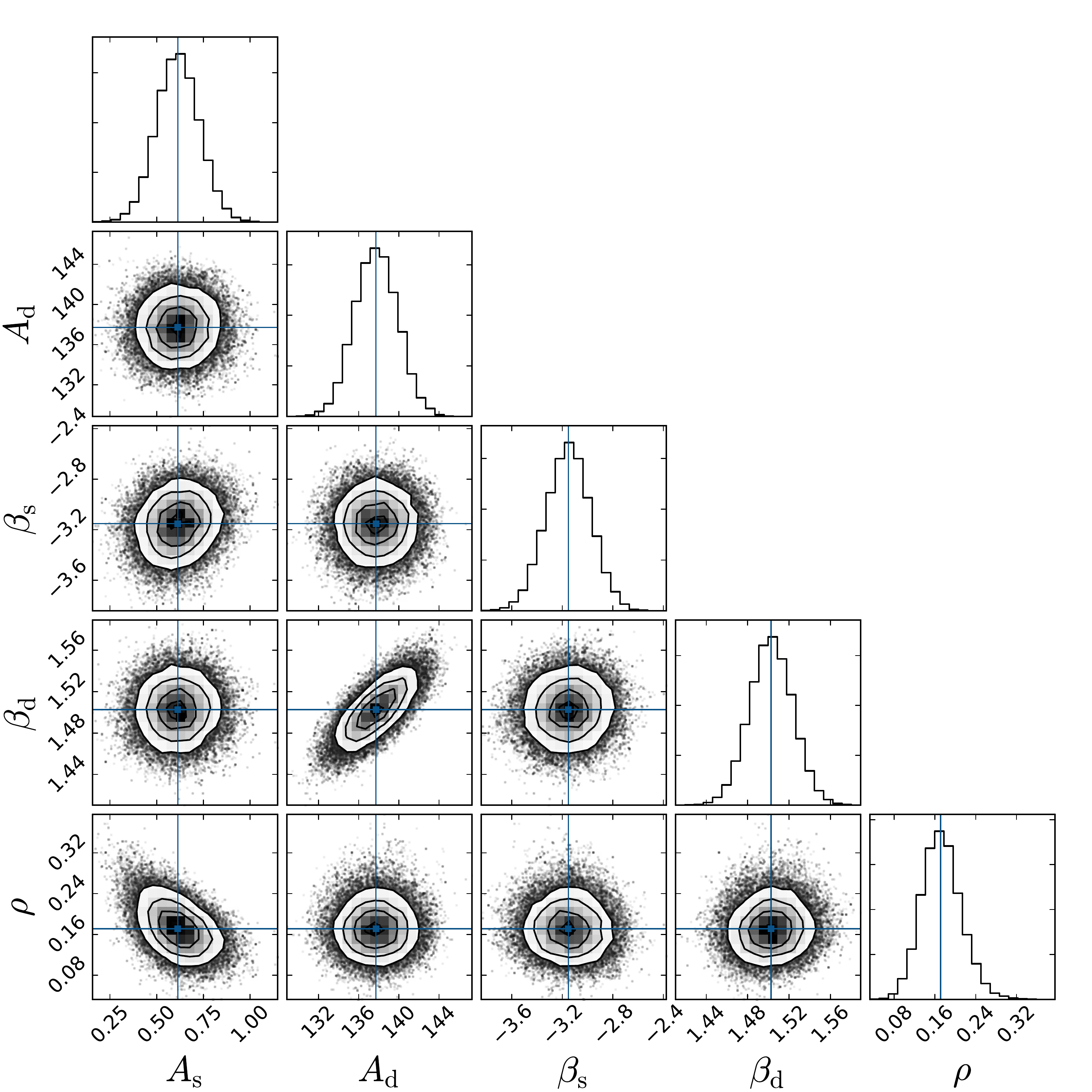}
\caption[]{Posterior distribution for each of the parameters of the spectral model in Eq.~(\ref{eq:SED_model}), as obtained through the MCMC fitting algorithm for $BB$ data points. The MCMC results illustrated here are
for the LR62 region and the multipole bin $\ell=40$--59, one of the two cases shown in Fig.~\ref{fig:SED-pol}. 
The diagonal shows the probability distribution of each parameter.
Median values are $A_{\rm s} = 0.6\pm0.1$, $A_{\rm d} = 137\pm2$, $\beta_{\rm s} = -3.15\pm0.17$, $\beta_{\rm d} = 1.50\pm0.02$, and $\rho = 0.17\pm0.04$.
} 
\label{fig:MCMCexample}
\end{figure}

Continuing the example, Fig.~\ref{fig:MCMCexample} shows the posterior distribution of the model parameters obtained through the MCMC algorithm, for $BB$ data, the LR62 region, and the $\ell = 40$--59 bin.
Best-fit parameters are computed as the median value of the posterior distributions, while errors are obtained from the 16th and 84th percentiles (68\,\% confidence interval).
For all regions and multipole ranges, values for $A_{\rm d}$, $A_{\rm s}$, $\beta_{\rm d}$, $\beta_{\rm s}$, and $\rho$ are listed in Tables~\ref{tab:EE_fit} ($EE$) and \ref{tab:BB_fit} ($BB$).

\begin{figure}[!htbp]
 \includegraphics[width=0.49\textwidth]{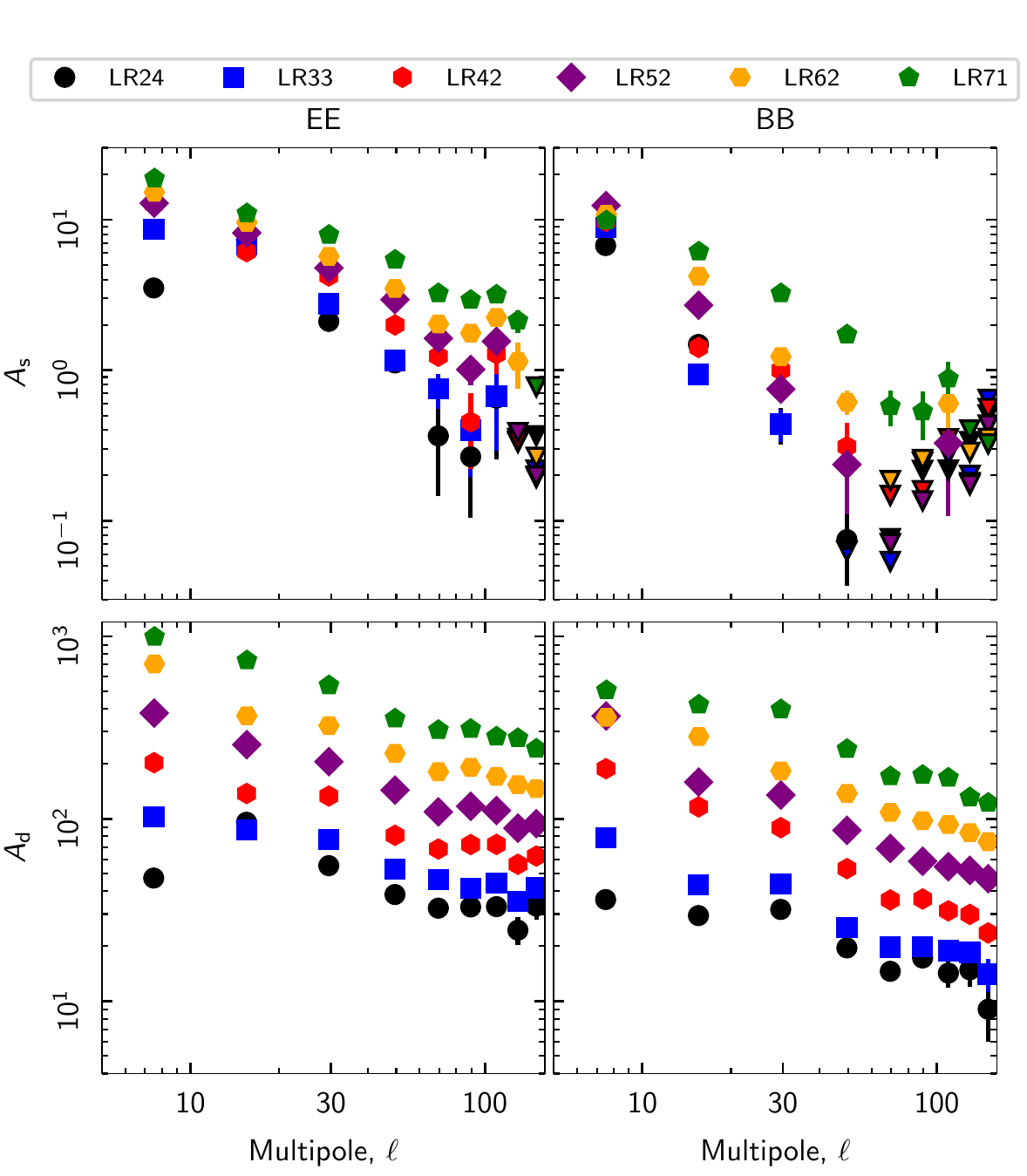}
 \caption[]{Amplitudes of $EE$ and $BB$ power spectra for dust and synchrotron emission at 353 and 30\,GHz, respectively,
shown for each sky region and each multipole bin.  The $A_{\rm s}$ and $A_{\rm d}$ parameters of our spectral model from Eq.~(\ref{eq:SED_model}) are converted from brightness to thermodynamic (CMB) temperature and expressed in $\mu$K$^2$. 
Where the synchrotron amplitude is compatible with zero at the $1\,\sigma$ level, we report an upper limit on $A_{\rm s}$ (68\,\% confidence limit) with triangles pointing down.}
\label{fig:DustSpectra}
\end{figure}

\begin{figure}[!htbp]
\includegraphics[width=0.49\textwidth]{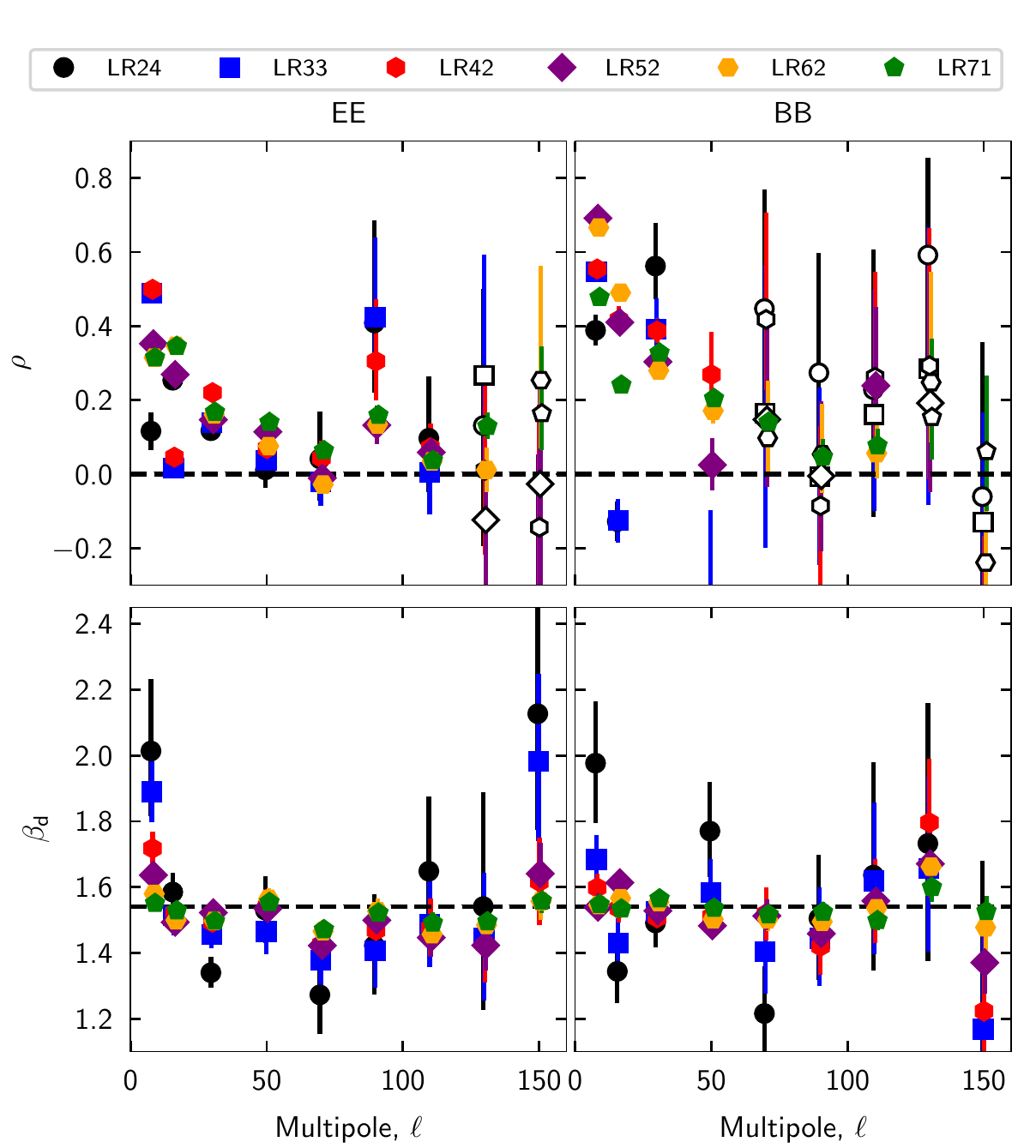}
\caption[]{Fit parameters $\rho$ and $\beta_{\rm d}$ for $E$- and $B$-mode polarization versus multipole.  %Note the linear scale.
Open symbols for $\rho$ represent the cases where the synchrotron amplitude is compatible with zero, making it difficult to measure the correlation.}
\label{fig:TGfits-pol}
\end{figure}

We do not list the amplitudes $A_{\rm d}$ and $A_{\rm s}$ of the dust and synchrotron emission but we note that as expected values of $A_{\rm d}$ are close to the values of the amplitudes $\mathcal{D}_{\ell}^{EE, BB} $ in Table~\ref{tab:dust_spectra}.
In Fig.~\ref{fig:DustSpectra}, $A_{\rm d}$ and $A_{\rm s}$ for $EE$ and $BB$
are plotted versus multipole for the six sky regions. As in the spectra for each region in Fig.~\ref{fig:spectra}, $A_{\rm d}$ has a power-law dependence on $\ell$ and a systematic increase 
with $\feff$ (see e.g. Fig~\ref{fig:BB_vs_I353}) that applies down to lower multipoles beyond $\ell = 40$.
On the other hand, for the multipole bin 4--11 the $B$-mode synchrotron amplitude $A_{\rm s}^{BB}$ is roughly constant over the six sky regions. 
As a corollary, for this multipole bin the ratio between dust and synchrotron $B$-mode polarization increases by about one order of magnitude from the smallest sky region, LR24, to the 
largest one, LR71. We point out that this result is specific to our set of sky regions, which are defined using the dust total intensity map to minimize dust power for a given sky fraction. 

\citet{Krachmalnicoff18} have characterized the synchrotron polarized foreground emission analysing maps of the southern sky from 
S-PASS at $2.3\,$GHz.  Comparison with our synchrotron results in Fig.~\ref{fig:DustSpectra}
is not immediate because power spectra are not measured over the same sky regions. Further, the signal to noise ratio of the S-PASS data for synchrotron emission is 
larger than that of \wmap\ and \Planck, which is a critical advantage in characterizing the faint polarization signal at high Galactic latitude. However,  
contamination by Faraday rotation is likely to be significant for their largest sky regions extending down to Galactic latitude $|b| = 20^\circ$.

Figure~\ref{fig:TGfits-pol} plots the two parameters $\rho$ and $\beta_{\rm d}$ (not $\beta_{\rm s}$ because of the prior applied) for $EE$ and $BB$. 
The top panels show that $\rho$, which quantifies the correlation between dust and synchrotron polarization, decreases with increasing multipole and 
is detected with high confidence only for $\ell \lesssim 40$. The correlation might extend to higher multipoles, but the decreasing signal-to-noise ratio of the synchrotron polarized emission precludes detecting it.
These results are consistent with the analysis done by \citet{Choi15} using all frequency channels of \wmap. 
The bottom panels show that the spectral index $\beta_{\rm d}$ has no systematic dependence on multipole or sky region, except for the lowest multipole bin. 
%The values of the dust spectral indices are discussed in Sect.~\ref{sec:betaT}. 
The dust spectral indices are further discussed in Sect.~\ref{sec:betaT}.

\begin{figure}[!htbp!]
\includegraphics[width=0.48\textwidth]{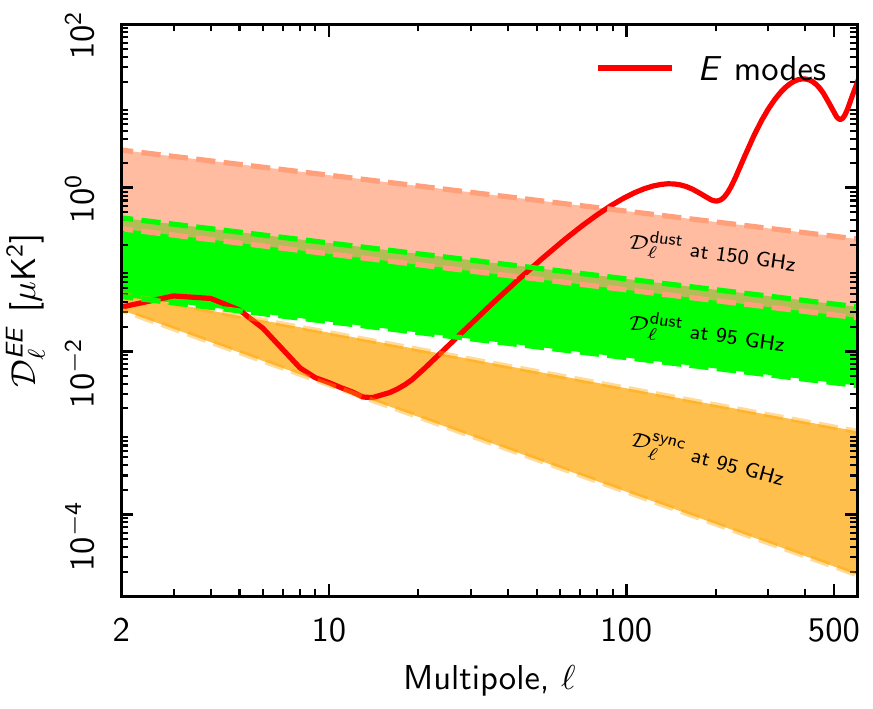}
\caption[]{Dust and synchrotron $E$-mode power versus multipole. 
The dust power at 95 and 150\,GHz and that of synchrotron at 95\,GHz 
are compared with the CMB $E$-mode signal (red-line) computed for the \Planck\ 2015 \LCDM\ model \citep{planck2014-a15} and 
a Thompson scattering optical depth $\tau=0.055$ from \citet{planck2014-a10}.
The coloured bands show the range of power measured from the smallest (LR24) to the largest (LR71) sky regions in our analysis. 
The lower limit of the synchrotron band
is derived from the S-PASS data analysis in \citet{Krachmalnicoff18}.
} 
\label{fig:foregrounds_power_EE}
\end{figure}

\begin{figure}[!htbp]
\includegraphics[width=0.48\textwidth]{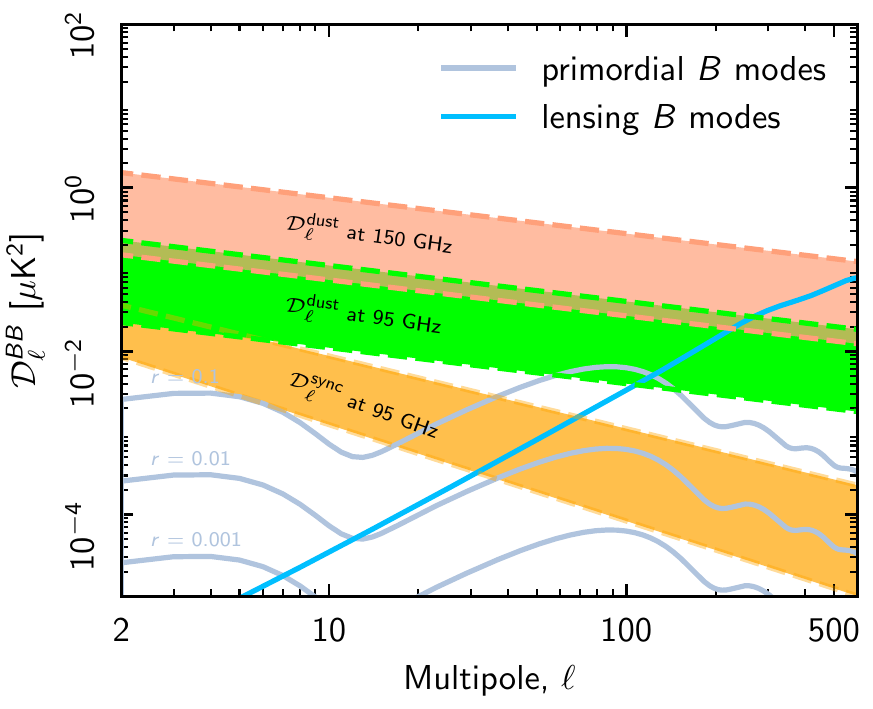}
\caption[]{Dust and synchrotron $B$-mode power versus multipole. % at 95 and 150\,GHz. 
The dust power at 95 and 150\,GHz, and that of synchrotron at 95\,GHz %, at frequencies accessible to ground-based experiments,
are compared with CMB $B$ modes from primordial gravitational waves (grey lines) for three values of the tensor-to-scalar ratio, $r= 0.1$, $0.01$, and $0.001$, 
and from lensing (blue line) for the \Planck\ 2015 \LCDM\ model \citep{planck2014-a15}. %FB: done \DS{Again: use published model.}
The coloured bands show the range of power measured from the smallest (LR24) to the largest (LR71) sky regions in our analysis. 
The lower limit of the synchrotron band
is derived from the S-PASS data analysis in \citet{Krachmalnicoff18}.} 
\label{fig:foregrounds_power_BB}
\end{figure}

\begin{figure}[!htbp]
\includegraphics[width=0.49\textwidth]{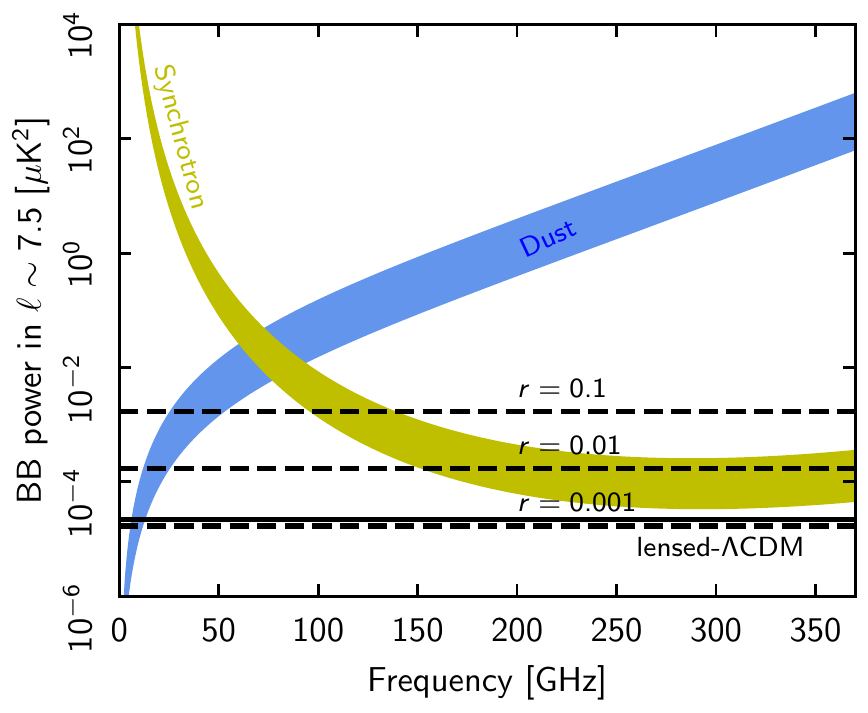}
\includegraphics[width=0.49\textwidth]{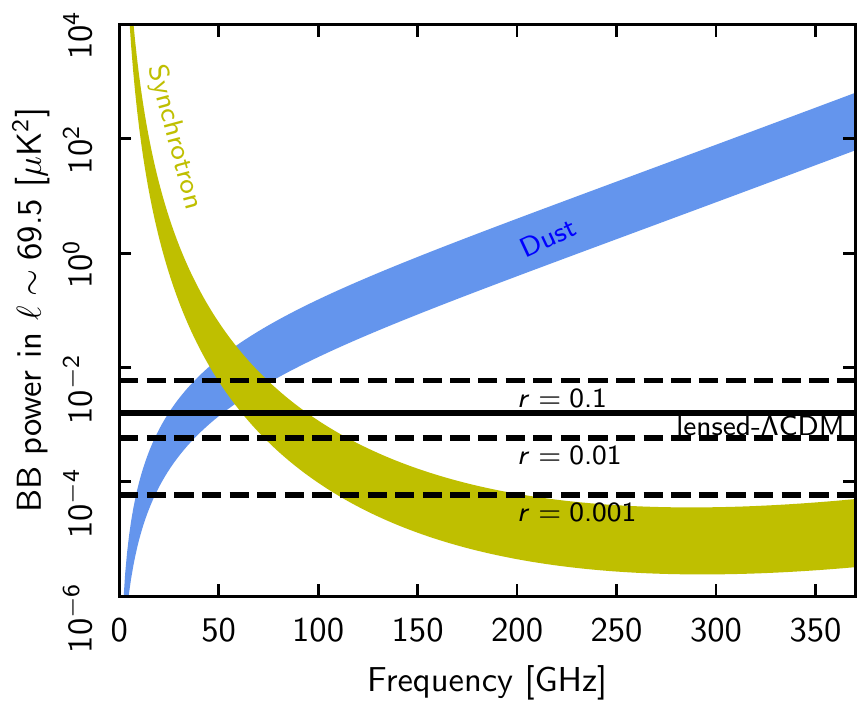}
\caption[]{Dust and synchrotron $B$-mode power versus frequency for two multipole bins: $\ell = 4$--11 (top) and 60--79 (bottom). 
The coloured bands show the range of power measured from the smallest (LR24) to the largest (LR71) sky regions in our analysis. 
The lower limit of the synchrotron band is derived from the S-PASS data analysis in \citet{Krachmalnicoff18}.
The primordial CMB $B$-mode signal, averaged within the appropriate $\ell$ bin, is plotted with dashed lines for three values of the tensor-to-scalar ratio: $r= 0.1$; $10^{-2}$; and $10^{-3}$.
The solid line represents the lensing $B$-mode signal for the \Planck\ 2015 \LCDM\ model \citep{planck2014-a15}. 
}
 \label{fig:foregrounds_seds}
\end{figure}

\subsection{Foregrounds versus CMB polarization}
\label{subsec:foregrounds}

Next, Galactic foregrounds are compared to CMB $E$- and $B$-mode polarization to quantify the challenge of component separation for measuring the low-multipole $E$-mode
CMB signal from reionization (Fig.~\ref{fig:foregrounds_power_EE}), and also for detecting primordial $B$ modes (Figs.~\ref{fig:foregrounds_power_BB} and \ref{fig:foregrounds_seds}).
The results of our spectral analysis allow us to update earlier studies \citep[see e.g.][]{Dunkley09,Krachmalnicoff16,planck2014-a12}.

To prepare Figs.~\ref{fig:foregrounds_power_EE} and \ref{fig:foregrounds_power_BB}, we use the results of our spectral fitting to compute the dust and synchrotron $E$- and $B$-mode power at frequencies 95 and 150\,GHz, which
correspond to the two microwave atmospheric windows providing the best signal-to-noise on the CMB for ground-based observations. 
In both figures, the dust  power is represented by a coloured band that spans the signal range from the smallest (LR24) to the largest (LR71) sky regions in our analysis;
the lower and upper edges of the band represent power-law fits of the values of $A_{\rm d}$  listed in Tables~\ref{tab:EE_fit} and \ref{tab:BB_fit}.
For synchrotron and LR71, we apply the same procedure fitting $A_{\rm s}$ values. For LR24, the signal to noise ratio of our synchrotron results is too low to compute a reliable fit. We choose instead to plot 
the results from the S-PASS data in \citet{Krachmalnicoff18} for their smallest ($|b|>50^\circ$) sky region.  The scaling of the power spectrum amplitude from 2.3 to $95\,$GHz is done using their determination 
of the spectral index ($\beta_{\rm s} = -3.22$). 

The dust $E$-mode power at 95 and 150\,GHz and that of synchrotron at 95\,GHz are compared with the CMB, as a function of multipole, in Fig.~\ref{fig:foregrounds_power_EE}.
Similarly, in Fig.~\ref{fig:foregrounds_power_BB}, the $B$-mode foregrounds at the same two frequencies are compared with the CMB primordial and lensing signals.
The primordial $B$-mode signal has two broad peaks in two multipole ranges, $\ell =2$--8 and 30--200, associated with reionization and recombination, respectively, the amplitude of which
scales linearly with the tensor-to-scalar ratio $r$.
The $E$- and $B$-mode reionization bumps at low multipoles are computed here for a Thompson scattering optical depth $\tau =0.055$ from \citet{planck2014-a10}.

Figure~\ref{fig:foregrounds_power_BB} shows that the synchrotron power decreases more steeply than the dust power with increasing $\ell$. 
Consequently, polarized synchrotron is a more significant foreground for the reionization peak than for the recombination peak. 
 
In Fig.~\ref{fig:foregrounds_seds}, the dust and synchrotron $BB$ power is plotted versus frequency for 
two multipole bins $\ell = 4$--11 (top plot) and 60--79 (bottom plot), which roughly correspond to the reionization and recombination peaks of the primordial $B$-mode
CMB signal, respectively.  The lower and upper edges of the dust band are drawn combining $A_{\rm d}$ values with spectral indices 
 $\beta_{\rm d}$, both listed in Table~\ref{tab:BB_fit}, for the LR24 and LR71 sky regions.  
For synchrotron, as in Figs.~\ref{fig:foregrounds_power_EE} and \ref{fig:foregrounds_power_BB}, we use the 
results from the S-PASS data in \citet{Krachmalnicoff18} for their smallest ($|b|>50^\circ$) sky region to draw the lower edge of the coloured band. 
The two polarized foregrounds have comparable amplitudes at a frequency that depends on the multipole and the sky region.
For average $\ellbin = 7.5$ (top plot) the amplitudes are equal at $\sim 75\,$GHz for both the lower and upper edges of the bands, whereas
 for $\ellbin = 69.5$ (bottom) this equality occurs at a lower frequency $\sim 60\,$GHz.
For higher frequencies, dust quickly dominates synchrotron. For example, for $\ellbin = 69.5$,  the $BB$ dust and synchrotron signals are equal at $60\,$GHz, while at 100\,GHz
the dust and synchrotron powers differ by two orders of magnitude. 
%FB: commented out in this revised version:  , corresponding to the equivalent of $r=0.1$ and $r=10^{-3}$, respectively. 

Our analysis stresses the accuracy with which dust and CMB $B$ modes must be separated to 
search confidently for primordial $B$ modes down to $r =0.01$. 
At this sensitivity level for sub-orbital experiments targeting the recombination peak at 95 and $150\,$GHz, e.g. the BICEP/Keck Array ground-based experiment \citep{PhysRevLett.116.031302} 
and the Spider balloon-borne experiment \citep{Fraisse13}, synchrotron polarization appears not to be a significant foreground over the relevant high latitude southern sky areas at $|b|>50^\circ$ 
used to draw the lower edge of the band.  However, the exact level of contamination will depend in detail on the sky region observed and how the synchrotron power extrapolates from $2.3\,$GHz there.

%===========================================================
\section{Microwave SED of polarized dust emission}
\label{sec:betaT}
%============================================================

This section focusses on the microwave SED of dust emission that is of interest for component separation and as a constraint on dust emission models. 

\subsection{Spectral index of dust polarized emission}
\label{subsec:beta_P}

Within the approximation of an MBB emission law and given a dust temperature, the microwave SED of dust emission is 
determined by the value of the dust spectral index, $\beta_{\rm d}$. 
This index parameterizes the separation of the dust and CMB components and the \Planck\ data constrain it better than ground-based data thanks to \Planck's $353$-GHz channel.  

We compute the mean values $\beta_{\rm d}^{EE}$ and $\beta_{\rm d}^{BB}$ for $E$- and $B$-mode polarization 
from the results of the spectral fitting from Sect.~\ref{sec:dust_sed} in Tables~\ref{tab:EE_fit} and \ref{tab:BB_fit}.
The uncertainty-weighted average of the differences between $\beta_{\rm d}^{BB}$ and $\beta_{\rm d}^{EE} $, computed over all multipole bins and sky regions, 
is  $<\beta_{\rm d}^{BB} - \beta_{\rm d}^{EE}>  \, = \, 0.0150 \pm  0.0053$. We consider the significance of this difference to be marginal because the statistical error-bar assumes 
that the measurements for the different sky regions are independent.
Averaging differences for the LR71 region alone, we find $<\beta_{\rm d}^{BB} - \beta_{\rm d}^{EE}>_{\rm LR71} \, =  \, 0.0180 \pm 0.0069$.

The difference between $\beta_{\rm d}^{EE} $ and $\beta_{\rm d}^{BB}$ is small and so we averaged them. 
Specifically, the uncertainty-weighted average of the fit results for all multipole bins and sky regions 
is $\beta_{\rm d}^P \equiv  0.5\,(\beta_{\rm d}^{EE}  + \beta_{\rm d}^{BB}) = 
\vbetap\pm\vbetapu$,
where the error bar includes the uncertainty from the polarization efficiencies of \HFI\ (Sect.~\ref{sec:data}) and 
the uncertainty from the CMB subtraction, which affects the determination of $\beta_{\rm d}^{EE}$. 
This is the uncertainty of the mean; the weighted dispersions of individual measurements are 0.046 and 0.034 for $E$ and $B$ modes, respectively. 
This value of $\beta_{\rm d}^P$ is lower than the mean polarization index $1.59\pm0.02$ derived from the analysis of earlier (PR2) 
\Planck\ data \citep{planck2014-XXII}. This difference reflects correction of data systematics between the PR2 and PR3 polarization maps. 
We checked that it does not come from the data analysis by analysing  the PR2 data in the same way as the PR3 data in this paper.

\begin{figure}[!htbp]
\includegraphics[width=0.49\textwidth]{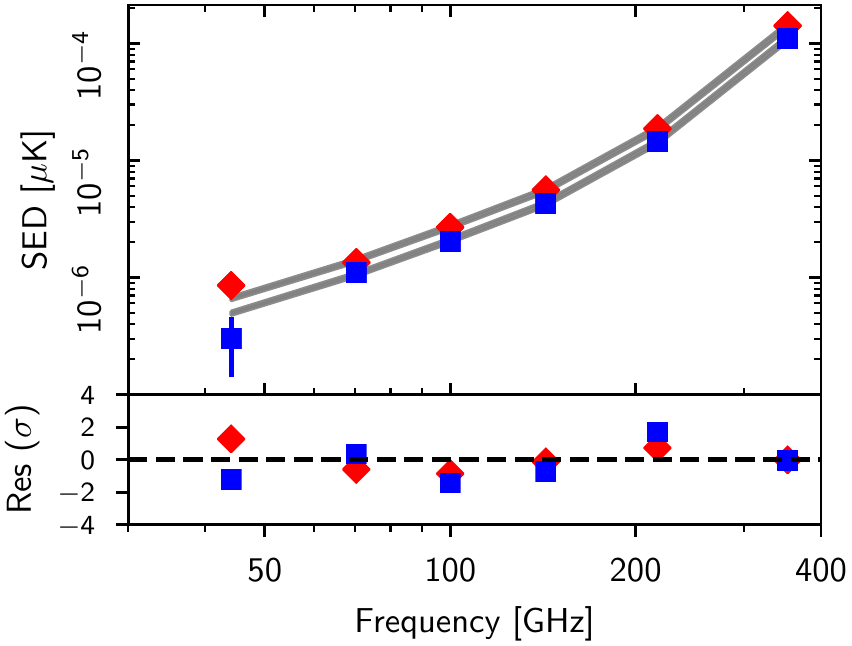}
\caption[]{Dust SEDs for $E$- and $B$-mode 
polarization derived from the {\tt SMICA} component-separation procedure \citep{planck2016-l04}.  
The two grey lines 
represent MBB fits to the $E$- (red diamonds) and $B$-mode (blue squares) data points with a temperature of 19.6\,K. The polarization spectral index
derived from the fits is $\beta_{\rm d}^P = 
\vbetap\pm\vbetapu$.
The residuals to each fit, normalized to the
$1\, \sigma$ data uncertainty, are plotted in the lower part of each panel. 
}
\label{fig:smica}
\end{figure}

\subsection{Dust polarization SED from blind component separation}
\label{subsec:smica}

The dust SEDs for $E$- and $B$-mode polarization were determined jointly with the corresponding synchrotron SEDs using the {\tt SMICA} (Spectral Matching Independent Component Analysis) method 
of blind component separation described in \citet{Cardoso08}, \citet{planck2014-a11} and \citet{planck2016-l04}.
In brief, the method consists of fitting all of the auto- and cross-spectra from 30\,GHz to 353\,GHz to a model consisting of a superposition of the CMB, two foreground emission components, and noise.
The fit is performed under very mild constraints, the free parameters being the angular spectrum of the CMB, the SED of each foreground emission component (assumed independent of angular scale), 
the angular spectra of each foreground emission component and their cross-spectrum, and the noise spectrum at each frequency.  
No prior spectral models of the SEDs are assumed; we do not assume that the dust SED is an MBB or that the synchrotron SED is a power law.

Fitting such a model determines, at the spectral level, a unique
\emph{global} foreground contribution that corresponds to two
underlying templates.  However, because the model allows for an arbitrary
angular correlation between those two templates, as well as an arbitrary
SED for each of them, the templates are linearly degenerate, meaning that each
can be an arbitrary linear combination of synchrotron and dust
emission.  
We choose to resolve this degeneracy by selecting the (essentially unique)
linear combinations, such that one template has no contribution at
353\,GHz while the other has no contribution at 30\,GHz. The latter corresponds to the dust foreground. 

The {\tt SMICA} component separation was performed over the LR71 sky region for comparison with our data analysis. 
The resulting dust SEDs for $E$- and $B$-mode polarization are presented in Fig.~\ref{fig:smica}.
These SEDs, coming from blind component separation, are remarkably close to a single-temperature MBB over the full range of \Planck\ polarization observations, despite the fact that an MBB spectral shape was not a prior assumption.  Performing MBB fits after the fact to the {\tt SMICA} dust spectral data in Fig.~\ref{fig:smica}
(again with $T_{\rm d} = 19.6\,$K and using colour corrections as described in Sect. 4.2),
we find a mean spectral index of $\beta_{\rm d}^P = 1.53\pm0.02$, taking into account the 1.5\,\% uncertainty on the polarization efficiency at 353\,GHz.
The $E$- and $B$-mode data intensities, each normalized to 1 at 353\, GHz, and uncertainties are listed in Table~\ref{tab:smica_sed}. For comparison, 
we also list the corresponding values for a MBB SED with $\beta_{\rm d}^P = 1.53\pm0.02$. 

The fit  is in excellent agreement with our determination in Sect.~\ref{subsec:beta_P}.
This agreement is perhaps not that surprising because our approach to the data analysis is in some aspects quite similar to that used by {\tt SMICA}. 
In both cases, the foreground SEDs are determined by fitting cross-spectra. Both methods allow for correlation between the two foreground components. 
However, the two methods differ in their simplifying assumptions. 
We constrain the dust and synchrotron SEDs to be the MBB and power-law parametric models, while {\tt SMICA} assumes that the SEDs are scale invariant. The agreement of the SEDs is reassuring and a cross-validation of the assumptions, as well as of the technical implementation.

The $BB/EE$ power ratio from {\tt SMICA} is 0.60, whereas we find $BB/EE= 0.53\pm0.01$ (Table~\ref{tab:ptes}).
The slightly higher $BB/EE$ power ratio could result 
from the fact that the $BB/EE$ power ratios in our analysis are determined at $\ell \ge 40$, while {\tt SMICA} includes lower multipoles.  When further constrained to a multipole range approximating ours, the ratio is 0.57.

%%%%%%%%%%%% TABLE 3 %%%%%%%%%%
\begin{table*}[tbp!]
\newdimen\tblskip \tblskip=5pt
\caption{Dust polarization SEDs from {\tt SMICA}  for the LR71 sky region.}
\label{tab:smica_sed}
\vskip -5mm
%\footnotesize
\setbox\tablebox=\vbox{
 \newdimen\digitwidth
 \setbox0=\hbox{\rm 0}
 \digitwidth=\wd0
 \catcode`*=\active
 \def*{\kern\digitwidth}
 \newdimen\signwidth
 \setbox0=\hbox{+}
 \signwidth=\wd0
 \catcode`!=\active
 \def!{\kern\signwidth}
  \newdimen\dpwidth
  \setbox0=\hbox{.}
  \dpwidth=\wd0
  \catcode`?=\active
  \def?{\kern\dpwidth}
\halign{\tabskip 0pt\hbox to 4.0cm{#\leaderfil}\tabskip 1em&
\hfil#\hfil\tabskip 1.0em& \hfil#\hfil& \hfil#\hfil&
\hfil#\hfil& \hfil#\hfil& \hfil#\hfil\tabskip 0em\cr
\noalign{\doubleline}
$\nu$(GHz) & 44.1 & 70.4 & 100 & 143 & 217& 353\cr 
\noalign{\vskip 3pt\hrule\vskip 5pt}
$EE$ SED$^{\rm a}$ & $6.0 \pm 1.1 \times 10^{-3}$  & $9.46 \pm 0.75  \times 10^{-3}$   & $19.0 \pm 0.5  \times 10^{-3} $ & $39.4 \pm 0.7  \times 10^{-3} $ & $13.2 \pm 0.21  \times 10^{-2} $ & 1. \cr
$BB$ SED$^{\rm a}$ & $2.7 \pm 1.4 \times 10^{-3}$  & $9.97 \pm 0.96  \times 10^{-3}$   & $18.3 \pm 0.5  \times 10^{-3} $ & $38.6 \pm 0.7  \times 10^{-3} $ & $13.2 \pm 0.21  \times 10^{-2} $ & 1. \cr
MBB  SED $^{\rm b}$ & $4.6 \pm 0.2 \times 10^{-3}$  & $9.76 \pm 0.31  \times 10^{-3}$   & $19.1 \pm 0.5  \times 10^{-3} $ & $39.1 \pm 0.7  \times 10^{-3} $ & $13.1 \pm 0.13  \times 10^{-2} $ & 1. \cr
\noalign{\vskip 3pt\hrule\vskip 5pt}
}}
\endPlancktablewide
\tablenote {{\rm a}}
Intensities in thermodynamic (CMB) temperature, not colour corrected, normalized to 1 at 353\,GHz. \par % as displayed in Fig~\ref{fig:smica}.\par
\tablenote {{\rm b}}
Corresponding intensities for an MBB SED with  $T_{\rm d} = 19.6\,$K  and $\beta_{\rm d} = 1.53 \pm 0.02$.\par
\end{table*}

%%%%%%%%%%%%%%%%%%%%%%%%

\subsection{Difference between spectral indices for polarization and total intensity}
\label{subsec:betaPandI}

The spectral model in Eq.~(\ref{eq:SED_model}) cannot be applied to the $TT$ spectra because in addition to synchrotron and dust thermal emission there are two other Galactic components, namely
AME and free-free emission, that contribute to the total intensity of the Galactic signal \citep{gold2010,planck2014-a12,planck2014-a31}. 
To compare the SEDs of dust polarization and total intensity, we follow a method similar to that used in \citet{planck2014-XXII}
correlating emission in the 217- and 353-GHz \HFI\ channels.  We work in harmonic space to assess any SED dependence on multipole and 
to be able to compare these results to those from the SED fitting.
In doing this, we implicitly assume that AME and free-free may be neglected at these two frequencies.

We compute the colour ratio,
\begin{equation}
\alpha_\ell^{XX} \, (217,353) \equiv \frac{\mathcal{C}_{\ell}^{XX} (217\times353)}{\mathcal{C}_{\ell}^{XX}(353\times353) }\, ,
\label{eq:color_ratio}
\end{equation}
for the $TT$, $EE$, and $BB$ spectra. The ratios are colour corrected, as described in Sect.~\ref{subsec:spectral_model}. We 
derive the corresponding spectral indices for a dust temperature of 19.6\,K. 
To compute $\alpha_\ell^{TT}$, we subtract CMB anisotropies 
using the map produced with the {\tt SMICA} component-separation method.
The 353-GHz power spectra are computed using half-mission data subsets. 

\begin{figure}[!htbp]
\includegraphics[width=0.49\textwidth]{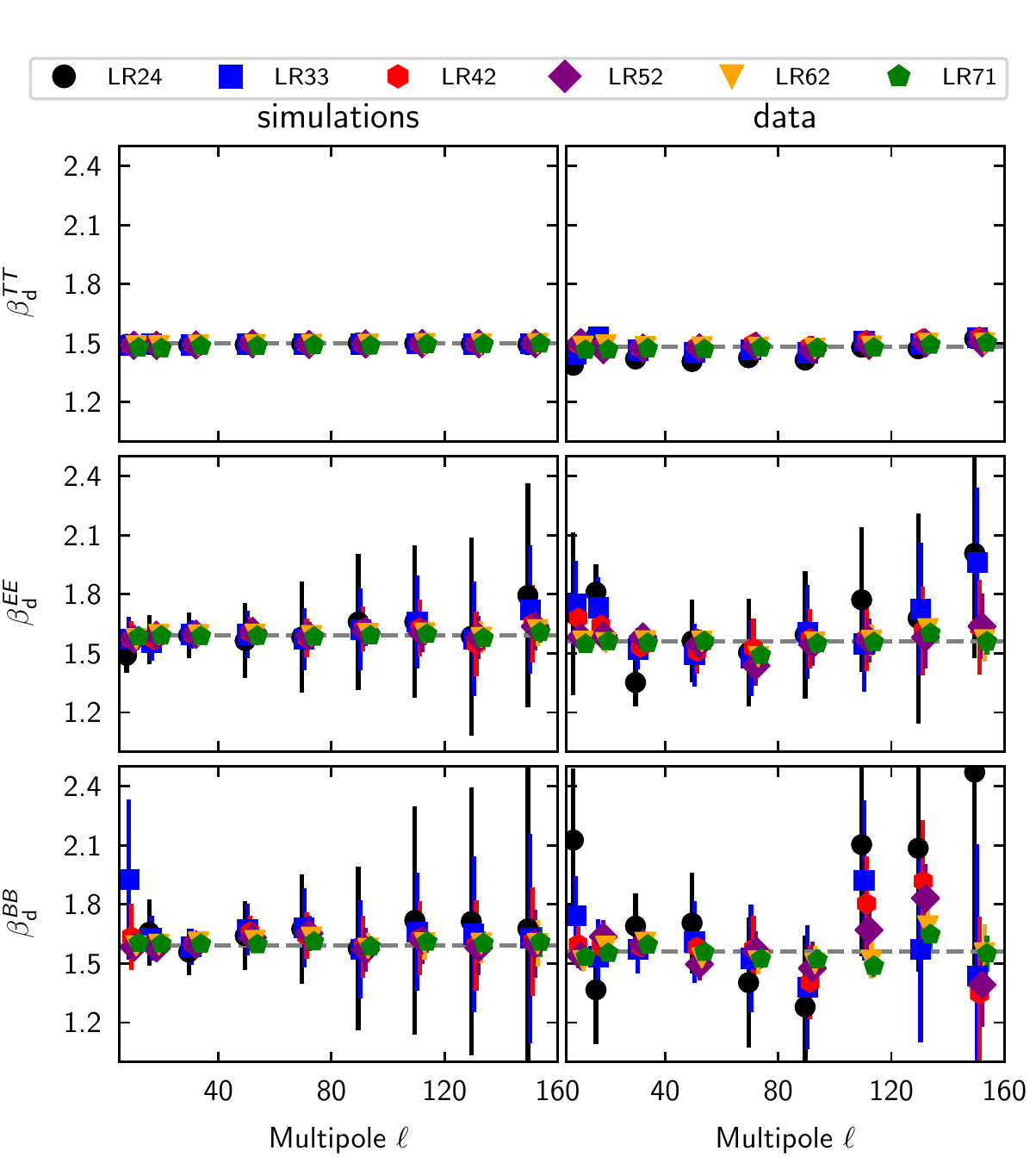}
\caption[]{Comparison of spectral indices of dust polarized emission and total intensity.
The spectral indices are derived from the 353-to-217\,GHz colour ratio. 
Plots to the left show the results obtained from our simulated maps, and the ones to the right are from the \Planck\ data.
Distinct symbols are used to represent each of the six sky regions, as in Fig.~\ref{fig:TGfits-pol}. 
For the simulations, the dashed lines represent the input dust spectral indices ($\beta_{\rm d}^{TT}=1.5$, $\beta_{\rm d}^{EE} = \beta_{\rm d}^{BB}=1.59$). For the data, the dashed lines represent the mean measured dust spectral indices ($\beta_{\rm d}^{TT}=1.48$, $\beta_{\rm d}^{EE} = \beta_{\rm d}^{BB}=1.53$).
}
\label{fig:betaT}
\end{figure}

The spectral indices are listed for each sky region and multipole bin in Table~\ref{tab:betadvaldata} for the \Planck\ \PR\ data.
The results are also presented in Fig.~\ref{fig:betaT}.  
The sky emission model that we use for simulating the total intensity maps includes anisotropies of the cosmic infrared background \citep{planck2013-pip56}. 
For the simulations, we retrieve the dust spectral indices adopted as input (1.50 for the total intensity and 1.59 for polarized intensity) with no bias.

For the \Planck\ maps, the dust spectral index for polarized intensity averaged over all regions and all $\ell$ bins is $\beta_{\rm d}^{P} \equiv 0.5\,(\beta_{\rm d}^{EE} +\beta_{\rm d}^{BB}) = 1.53\pm0.03$, 
taking into account the 1.5\,\% uncertainty on the polarization efficiency at 353\,GHz.
This value agrees well with that inferred from the multi-frequency spectral analysis in Sects.~\ref{subsec:beta_P} and \ref{subsec:smica} above.
The corresponding value for total intensity is $\beta_{\rm d}^{I} \equiv \beta_{\rm d}^{TT} = 1.48$, with much smaller uncertainty\footnote{The difference with the corresponding spectral index 1.51 in
  \citet{planck2014-XXII} follows from a 1.5\% upward photometric calibration change from the PR2 to \PR\ data at 353\,GHz}.
The spectral indices for polarization and total intensity differ by 
$\vdbetapi\pm\vdbetapiu$.
This difference is smaller than that reported in \citet{planck2014-XXII} analysing earlier \Planck\ data. 

\begin{figure}[!htbp]
\includegraphics[width=0.49\textwidth]{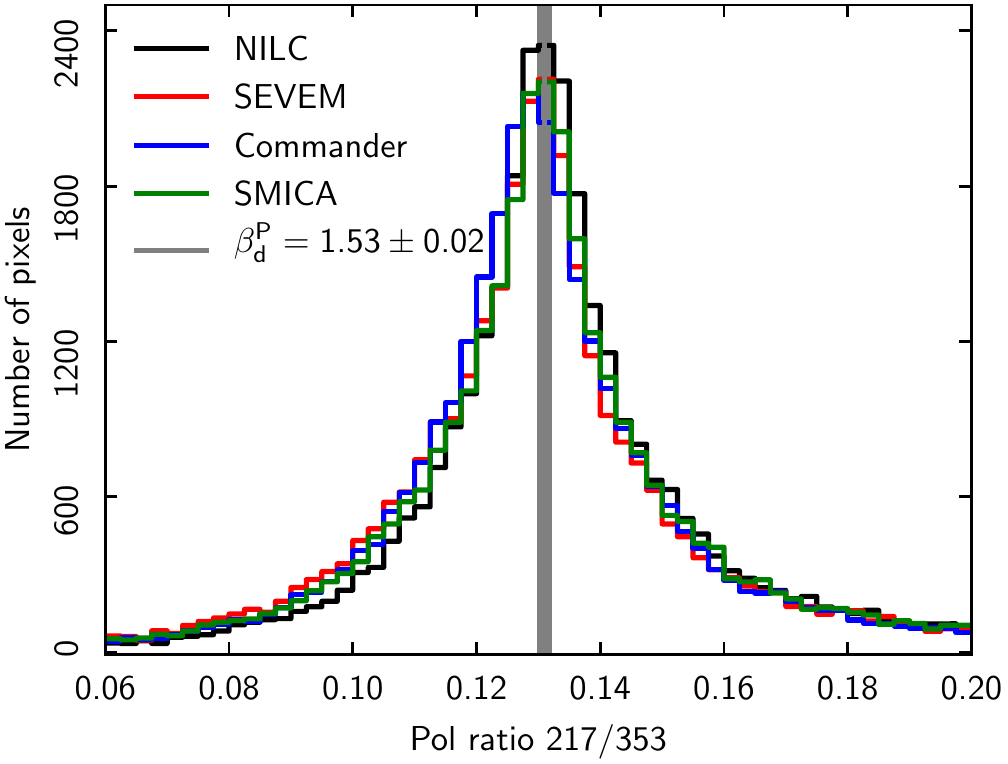}
\caption[]{Illustration of  the consistency between our
analysis and component separation methods. 
The probability distribution of the 217-to-353\,GHz colour ratio for dust polarized intensity, 
computed over the LR71 sky region from \Planck\ CMB-subtracted maps smoothed to a $3^\circ$ beam, 
is plotted for  each of the component separation methods in  \citet{planck2016-l04}.
The vertical line is the value derived from our analysis. For the unit conversion factors and 
color corrections, and our modification of the 217\,GHz  polarization efficiency, it corresponds to
the spectral index $\beta_{\rm d}^{P} = 1.53 \pm 0.02$ from Sects.~\ref{subsec:beta_P} and \ref{subsec:smica}.
The width of the line represents the error bar. }
\label{fig:CompSep_comparaison}
\end{figure}

We checked the consistency of our derivation of the dust spectral index for polarization  
with the component separation methods in \citet{planck2016-l04}, by computing maps of $\beta_{\rm d}^{P}$ from 
\Planck\ 217 and $353\,$GHz CMB-subtracted maps smoothed to a $3^\circ$ beam. This is illustrated in Fig.~\ref{fig:CompSep_comparaison}, where
the probability distribution of the 217-to-353\,GHz colour ratio for dust polarized intensity, computed over the LR71 sky region, is shown for
each of the component separation methods in  \citet{planck2016-l04}. For all methods, the median value of  $\beta_{\rm d}^{P}$, inferred from the colour-ratio for a dust temperature of $19.6\,$K,
is consistent with our estimate $1.53 \pm 0.02$ in Sects.~\ref{subsec:beta_P} and \ref{subsec:smica}.  We point out that the scatter in measured colour-ratios is dominated by data noise.

\subsection{Impact on dust modelling}
\label{subsec:impact}

These results from the spectral fitting of the polarized dust SED provide an additional constraint for dust modelling.
Reviewing the spectral fit in Sect.~\ref{sec:dust_sed},  
for $\ell \le 100$ all of the $\chi^2$ values of the spectral fit (listed in Tables~\ref{tab:EE_fit} and \ref{tab:BB_fit})
are lower than the number of degrees of freedom. Therefore, to the sensitivity of the \planck\ data, a single temperature MBB 
emission law is a satisfactory model
of the polarized dust emission.  This same conclusion is supported by the further analyses in the subsections above.
There is no evidence for a flattening or steepening of the dust SED, which could in principle result from a variation of spectral index with frequency 
as reported from laboratory studies of silicate grains \citep{Demyk17}, or from a significant contribution from magnetic 
dipolar emission from magnetic nano-particles \citep{Draine13}.

Interstellar dust is often modeled
as a mixture of silicates and carbon grains  \citep[e.g.][]{Li01,DraineFraisse2009,Compiegne11,Jones13,Siebenmorgen14,Guillet17}.
A difference between $\beta_{\rm d}^{P}$ and $\beta_{\rm d}^{I}$ might
be evidence that these two dust components have distinct spectral indices and polarization properties.
However, the difference that we have found is small and not of high statistical significance.
This result suggests that the emission from a single grain type dominates the long-wavelength emission in both polarization and total intensity. 
If the emission from silicate grains dominates that of carbon grains in polarization - as it is often assumed \citep{Andersson15} -
this should also hold for the total dust intensity at long-wavelengths.  

The alignment of interstellar silicates may be effective irrespectively of whether the grains contain magnetic inclusions \citep{Lazarian_Hoang07,Andersson15,Hoang16}. 
If silicates do have magnetic inclusions, or if interstellar dust comprises free-flying magnetic grains, the microwave dust emission may  include a significant contribution from
magnetic dipole emission \citep{Draine99,Draine13,Hoang16_AME,Hensley17}. 
The close match between $\beta_{\rm d}^{P}$ and $\beta_{\rm d}^{I}$ constrains this contribution.
More generally, the dust polarization SEDs in Table~\ref{tab:smica_sed} may be used in combination with the dust total intensity SED in  \citet{planck2014-XXII} (corrected for 
the 1.5\% upward photometric calibration change  from the PR2 to PR3 data at 353\,GHz) to test dust emission models. This detailed comparison between data and models is 
beyond the scope of this paper.

%===========================================================
\section{Correlation of dust polarized emission across microwave frequencies}
\label{sec:dust_deco}
%============================================================

Interstellar processes couple the emission properties of dust and grain alignment with the density structure of matter and
that of magnetic fields \citep{Hoang16,Fanciullo17}. Likewise, the cosmic-ray energy spectrum, and thereby the synchrotron emission
spectrum, depend on the magnetic field structure \citep{Strong11}. These physical couplings break the simplest assumption for
component separation, by which the spectral frequency dependence of the Galactic polarization and its angular
structure on the sky are separable \citep{Tassis15,Poh17}. The couplings make polarized foregrounds intrinsically complex, in ways that have
yet to be characterized statistically for optimizing the component separation and taking into account Galactic residuals
in the CMB likelihood function.  This is a critical issue for the analysis of CMB polarization because spatial
variations of the spectral behaviour of polarized dust emission can mistakenly be interpreted as a (false)
detection of primordial CMB $B$ modes.

PL analysed the correlation between the \HFI\ dust polarization maps at 217 and 353\,GHz.  
In Appendix~\ref{appendix:pdf_Rell}, using the new \Planck\ maps, we update and extend the PL analysis (Sect.~\ref{subsec:correlation_ratio}).  
Uncorrected systematics and correlated noise in the data limit how tightly the decorrelation can be constrained.
However, these effects change with frequency and so can potentially be mitigated by analysis across many frequencies.
In Sect.~\ref{subsec:multi_freq_deco}, we present such a multi-frequency correlation analysis, making use of the four polarized \HFI\ channels from 100 to 353\,GHz.
The implications of this new analysis of the \Planck\ data for on-going and future CMB $B$-mode experiments are discussed in Sect.~\ref{subsec:freq_perspective}. 

\subsection{Multi-frequency correlation analysis of dust polarization}
\label{subsec:multi_freq_deco}

The spectral model introduced in Sect.~\ref{subsec:spectral_model} assumes that the dust and synchrotron polarized emission signals are each perfectly correlated across microwave frequencies.
To test this hypothesis, we repeat the spectral fitting with a model modified to allow for a loss of correlation for dust polarization. 
The dust contribution to the amplitude of $BB$ cross-spectra between frequencies $\nu_1$ and $\nu_2$ is
%
%\begin{equation}
\begin{align}
\label{eq:dust_decorrelation}
{\cal D}_\ell^{\rm BB_d} (\nu_1 \times \nu_2) & = 
 A_{\rm d} \, \left(\frac{\nu_1 \nu_2}{353^2} \right)^{\beta_{\rm d}-2} \times  \nonumber \\
& \frac{B_{\nu_1}(T_{\rm d})}{B_{353}(T_{\rm d})}  \frac{B_{\nu_2}(T_{\rm d})}{B_{353}(T_{\rm d})} \, f_{\rm d}(\delta_{\rm d},\nu_1,\nu_2) \, ,
\end{align}
%\end{equation}
%
where the frequencies $\nu_1$ and $\nu_2$ are expressed in GHz and the adopted function $f_{\rm d}$ from appendix~B of \citet{Vansyngel16} is 
\begin{equation}
f_{\rm d}(\delta_{\rm d},\nu_1,\nu_2) =  {\rm exp}\Big\{- \delta_{\rm d} \, \Big[{\rm ln} \left(\nu_1/\nu_2\right)\Big]^2 \Big\} \, .
\label{eq:decorrelation}
\end{equation} 
The loss of correlation introduced by the parameter $\delta_{\rm d}$ increases with the frequency ratio ${\nu_1}/{\nu_2}$. 
From $\delta_{\rm d}$ we also re-express the decorrelation in terms of the spectral correlation ratio $\mathcal{R}_{\ell}^{BB}(217, 353)$ (see Eq.~(\ref{eq:correlation_ratio}))
for comparison with the two-frequency results presented in 
PL and in \citet{Sheehy17}, and for the \PR\ data in Appendix~\ref{appendix:pdf_Rell}.

We fit this model over the four \HFI\ polarized \Planck\ frequencies 100, 143, 217, and 353\,GHz, for the six sky regions LR24 to LR71. 
Synchrotron polarization is ignored because it is negligible in this frequency range (Sect.~\ref{subsec:foregrounds}).
We carry out this analysis for the $BB$ cross-spectra computed from the \Planck\ data and the E2E simulations, for the multipole range 
$\ell = 50$--160, relevant to the search for primordial $B$ modes at the recombination peak.
To allow readers to fit an alternative spectral model,
we list the data values and uncertainties for the LR71 sky region in Table~\ref{tab:SED_decorrelation_LR71}. We also provide 
the corresponding values for the 300 E2E simulations as a FITS Table, which may be used to assess the significance of such an alternative analysis.

\begin{table}[tbp!]
\newdimen\tblskip \tblskip=5pt
\caption{Spectral data constraining the decorrelation of dust polarization for the LR71 sky region}
\label{tab:SED_decorrelation_LR71}
\vskip -5mm
%\footnotesize
\setbox\tablebox=\vbox{
 \newdimen\digitwidth
 \setbox0=\hbox{\rm 0}
 \digitwidth=\wd0
 \catcode`*=\active
 \def*{\kern\digitwidth}
 \newdimen\signwidth
 \setbox0=\hbox{+}
 \signwidth=\wd0
 \catcode`!=\active
 \def!{\kern\signwidth}
  \newdimen\dpwidth
  \setbox0=\hbox{.}
  \dpwidth=\wd0
  \catcode`?=\active
  \def?{\kern\dpwidth}
\halign{\tabskip 0pt\hbox to 5.0cm{#\leaderfil}\tabskip 1em&
\hfil#\hfil\tabskip 1.0em&  \hfil#\hfil\tabskip 0em\cr
\noalign{\doubleline}
%\omit& LR71$^{\rm a}$\cr 
$\nu_1 \times \nu_2$ & ${\cal D}_\ell^{\rm BB} (\nu_1 \times \nu_2)^{\rm a}$\cr
GHz & $\mu$K$^2$ \cr
\noalign{\vskip 3pt\hrule\vskip 5pt}

$100 \times 100$ & $!*0.08\pm0.02$\cr
$ 100 \times 143$  &  $!**0.11    \pm  0.01$ \cr
$100 \times 217 $   & $ !**0.38     \pm    0.02$ \cr
$ 100 \times 353 $  &  $!**2.80    \pm  0.13$ \cr
$ 143 \times 143 $   & $!**0.25    \pm  0.02$ \cr
$ 143 \times 217 $   & $!**0.79    \pm  0.02$ \cr
$ 143 \times 353 $   & $!**6.02     \pm  0.11$ \cr
$ 217 \times 217  $  &  $!**2.78    \pm  0.05$ \cr
$ 217 \times 353 $   & $!*20.92    \pm  0.18$ \cr
$ 353 \times 353 $  &  $!161.46   \pm  1.48$ \cr
\noalign{\vskip 3pt\hrule\vskip 5pt}
}}
\endPlancktable
\tablenote {{\rm a}}
Amplitude of cross-spectra for the multipole range 50--160, not colour corrected.\par
\end{table}

\begin{table*}[tbp!]
\newdimen\tblskip \tblskip=5pt
\caption{Spectral correlation ratio $\mathcal{R}_{\ell}^{BB}(217, 353)$ from multi-frequency MCMC fit for the multipole range 50--160.}
\label{tab:multifreq_Rell_results}
\vskip -5mm
%\footnotesize
\setbox\tablebox=\vbox{
 \newdimen\digitwidth
 \setbox0=\hbox{\rm 0}
 \digitwidth=\wd0
 \catcode`*=\active
 \def*{\kern\digitwidth}
 \newdimen\signwidth
 \setbox0=\hbox{+}
 \signwidth=\wd0
 \catcode`!=\active
 \def!{\kern\signwidth}
  \newdimen\dpwidth
  \setbox0=\hbox{.}
  \dpwidth=\wd0
  \catcode`?=\active
  \def?{\kern\dpwidth}
\halign{\tabskip 0pt\hbox to 5.0cm{#\leaderfil}\tabskip 1em&
\hfil#\hfil\tabskip 1.0em& \hfil#\hfil& \hfil#\hfil&
\hfil#\hfil& \hfil#\hfil& \hfil#\hfil\tabskip 0em\cr
\noalign{\doubleline}
\omit& LR24& LR33& LR42& LR52& LR62& LR71\cr 
\noalign{\vskip 3pt\hrule\vskip 5pt}

\HFI\ data& $0.935\pm0.054$& $0.932\pm0.039$& $0.970\pm0.021$& $0.983\pm0.013$& $0.984\pm0.008$& $0.989\pm0.005$\cr
Mean E2E simulations$^{\rm a}$& $0.976\pm0.043$& $0.988\pm0.026$& $0.993\pm0.016$& $0.993\pm0.011$& $0.995\pm0.008$& $0.997\pm0.005$\cr
E2E lower limits$^{\rm b}$&  0.865& 0.924& 0.963& 0.973& 0.983& 0.991\cr
FFP10 dust model$^{\rm c}$& 0.987& 0.992& 0.994& 0.996& 0.997& 0.998\cr
Two-frequency analysis of data$^{\rm d}$& 0.822& $0.886$& $0.932$& $0.954$& $0.976$& $0.989$\cr
Two-frequency E2E lower limits$^{\rm e}$&  0.756& 0.854& 0.913& 0.949& 0.965& 0.980\cr
\noalign{\vskip 3pt\hrule\vskip 5pt}
}}
\endPlancktablewide
\tablenote {{\rm a}}
Results from MCMC fit to the mean of the E2E simulations.\par
\tablenote {{\rm b}} E2E lower limits on $\mathcal{R}_{\ell}^{BB}$, corresponding
to the 2.5th percentile of the {\tt MPFIT} results on the 300 E2E realizations (Fig.~\ref{fig:Rell_histo}).\par
\tablenote {{\rm c}}
$\mathcal{R}_{\ell}^{BB}$ values measured for the noise-free FFP10 dust
polarization maps (Appendix~\ref{subsec:validation}).\par
\tablenote {{\rm d}}
For comparison, results from the two-frequency analysis of HFI data in Appendix~\ref{appendix:pdf_Rell} (see Table~\ref{tab:Rell_LR} in Appendix~\ref{appendix:tables}).\par
\tablenote {{\rm e}}
For comparison, E2E lower limits from the two-frequency analysis (see Table~\ref{tab:Rell(BB)_limits}).\par
\end{table*}

\begin{figure}[!htbp]
\includegraphics[width=0.5\textwidth]{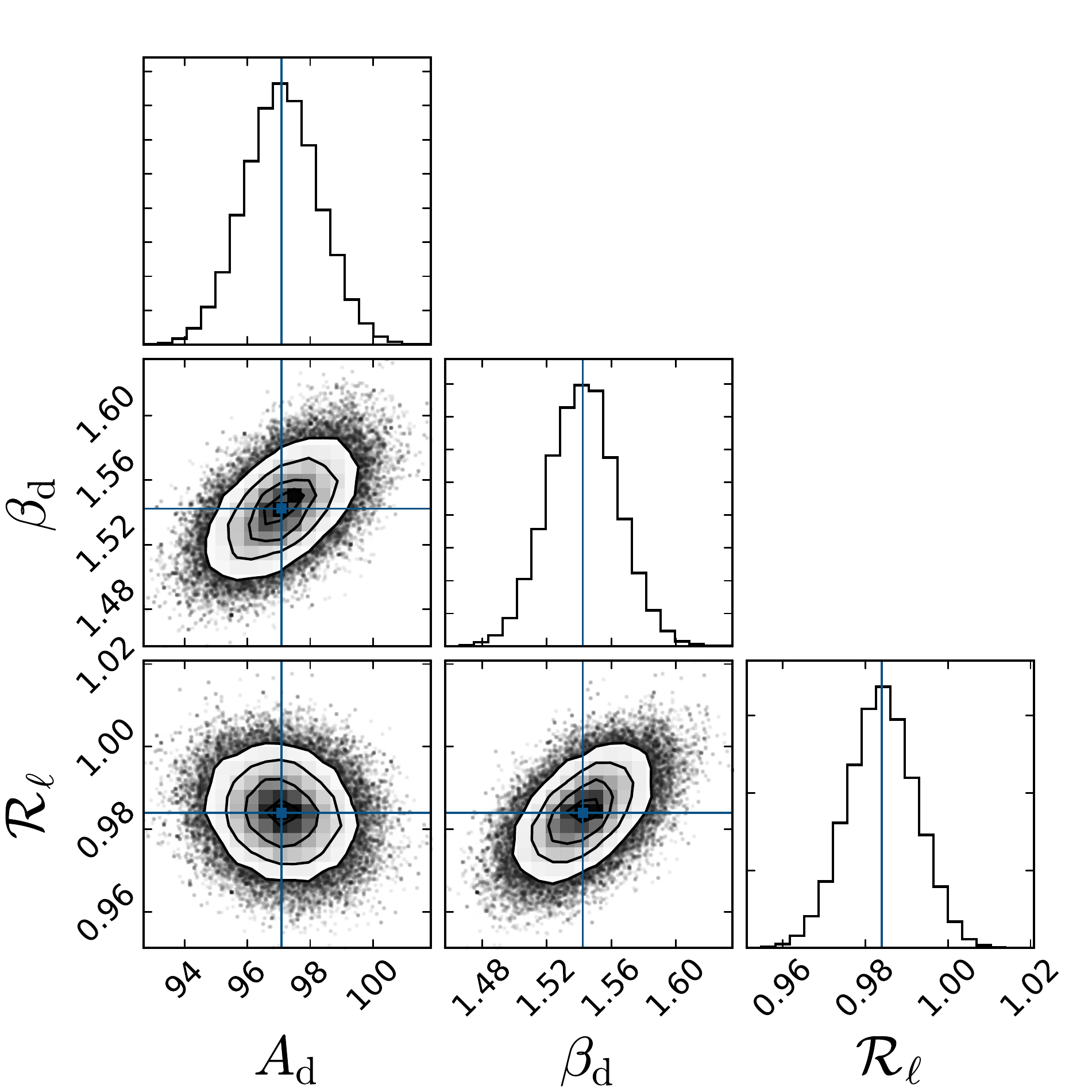}
\caption[]{Posterior distribution for each of the parameters of the spectral model with decorrelation given in Eq.~(\ref{eq:decorrelation}), as obtained through the MCMC fitting algorithm for $BB$ data points. 
The MCMC results illustrated here are for the LR62 region and the multipole range 50--160.  Median values are $A_{\rm d} = 97.1\pm1.2$,  $\beta_{\rm d} = 1.54\pm0.02$, and $\mathcal{R}_{\ell}^{BB}(217, 353) = 0.984\pm0.008$.}
\label{fig:decor_posterior}
\end{figure}

We perform an MCMC fit to the \Planck\ data and to the mean of the E2E simulations computed over the 
300 E2E realizations.  The uncertainties are in both cases inferred from the dispersion of spectra computed with the E2E simulations.
In Fig.~\ref{fig:decor_posterior}, we show for the LR62 region the posterior probability distribution of the model parameters $A_{\rm d}$, $\beta_{\rm d}$, and the correlation ratio $\mathcal{R}_{\ell}^{BB}$ inferred from $\delta_{\rm d}$.
The values of the model parameters are listed in Table~\ref{tab:multifreq_Rell_results} for the data and the mean of the simulations for all six regions.
The dust sky model used in the simulations has a perfect correlation across frequencies (Appendix~\ref{appendix:data_simulations}),  that is for this dust model, $\delta_{\rm d} =0$ and $\mathcal{R}_{\ell}^{BB} = 1$.  
The values of $\mathcal{R}_{\ell}^{BB}$ in Table~\ref{tab:multifreq_Rell_results}, 
inferred from the best-fit value of $\delta_{\rm d}$ for the mean of the 300 E2E realizations, are consistent with 1 within a fraction of the $1\,\sigma$ error bars, for all sky regions. 
This result shows that there is no bias introduced by neglecting the synchrotron contribution at 100\,GHz, even though it is present in the FFP10 sky model (Appendix~\ref{subsec:validation}). 
In this model, the contribution of synchrotron to the $BB$ power at 100\,GHz, in the multipole bin $\ell = 50$--160, rises from 4 to 19\,\% for decreasing $\fsky$ from LR71 to LR24.

\begin{figure}[!htbp]
\includegraphics[width=0.495\textwidth]{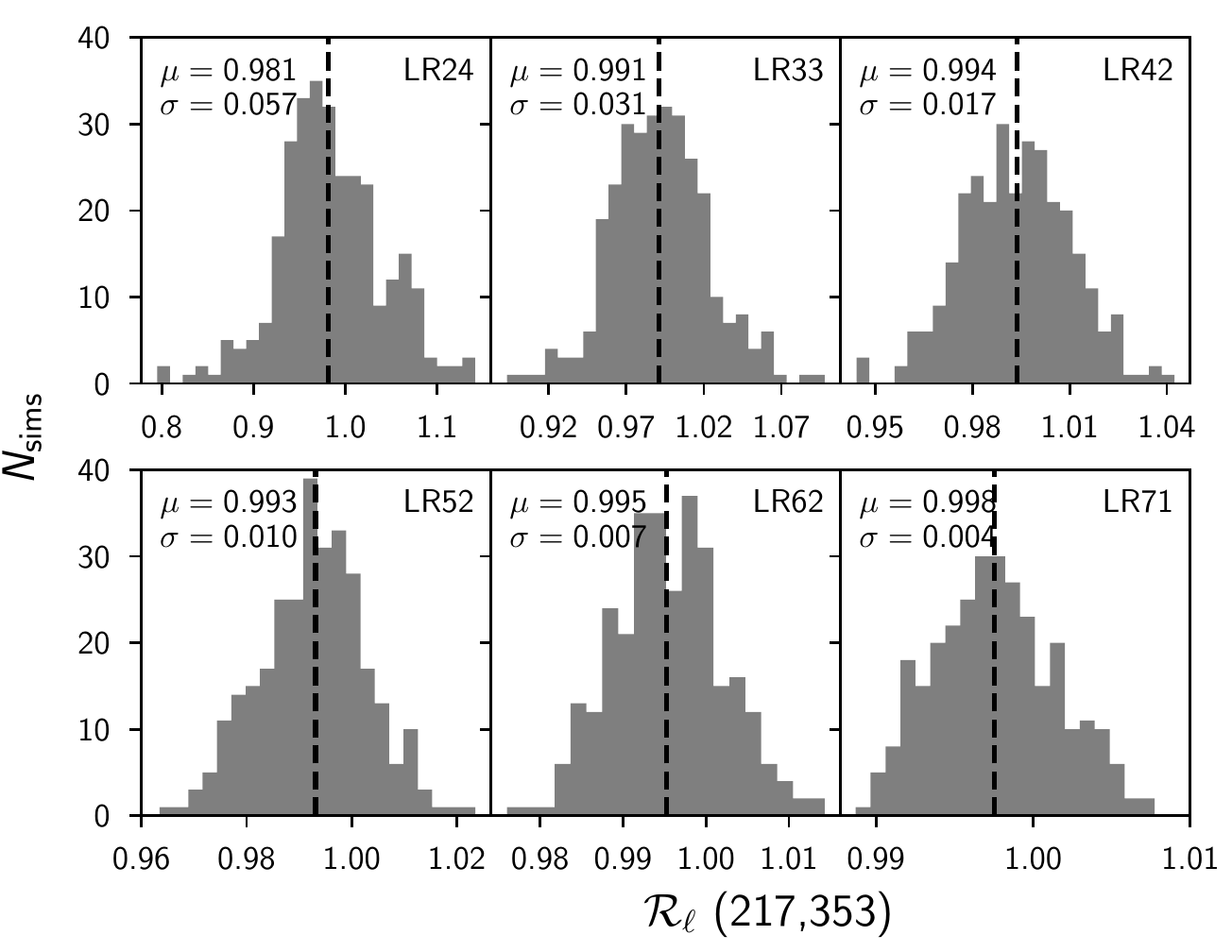}
\caption[]{Distribution of the correlation ratios $\mathcal{R}_{\ell}^{BB}(217, 353)$ inferred from $\delta_{\rm d}$ on the six sky regions for the $\ell$ range 50--160. The histograms are 
computed from the 300 E2E simulations using half-mission data splits. The dashed lines represent the median values on each sky region. This median value, $\mu$, and the standard
deviation, $\sigma$, are printed in the upper right of each panel.
The lower limits on $\mathcal{R}_{\ell}^{BB}$ in Table~\ref{tab:multifreq_Rell_results} are derived from the 2.5th percentile of the distribution for each sky region.}
\label{fig:Rell_histo}
\end{figure}

We obtain histograms of parameter values, 
fitting the spectral model in Eq.~(\ref{eq:decorrelation}) to each of the 300 E2E realizations. To do this, we use the least-squares {\tt MPFIT} algorithm because
the MCMC fit is too computationally-intensive to be run 300 times. We checked that the two methods provide consistent parameter values for the \Planck\ data and for the 
mean of the E2E simulations. 
The probability distributions of $\mathcal{R}_{\ell}^{BB}$ inferred from $\delta_{\rm d}$ values measured on the E2E realizations for each sky region are presented in Fig.~\ref{fig:Rell_histo}.  
Lower limits on $\mathcal{R}_{\ell}^{BB}$ from the E2E simulations are listed in Table~\ref{tab:multifreq_Rell_results}.  These are based on the 95\,\% confidence interval, thus on the 2.5th percentile of the histograms.

The limits from the multi-frequency analysis are tighter than the corresponding ones in Table~\ref{tab:Rell(BB)_limits}, derived from the 217- and 353-GHz correlation alone (see Appendix~\ref{appendix:pdf_Rell} and for convenience reproduced in Table~\ref{tab:multifreq_Rell_results}). 
However, it is important to keep in mind that the limits derived from our multi-frequency analysis depend on an assumption of the applicability of the spectral model in Eq.~(\ref{eq:decorrelation}), while
the two-frequency results are model independent.

The multi-frequency analysis shows no evidence for a loss of correlation, within the limits provided by the analysis of the E2E simulations.  As discussed, these new limits are much tighter 
than those obtained from the 217- and 353-GHz correlation ratio in Appendix~\ref{appendix:pdf_Rell}. However, current limits are still consistent with (i.e. still allow the presence of) significant variations of the dust spectral
index over the sky. 
To illustrate this statement quantitatively, we have computed $\mathcal{R}_{\ell}^{BB}$ 
for the noise-free FFP10 dust polarization maps (Appendix~\ref{subsec:validation}), built from 353-GHz polarization templates computed using the \citet{Vansyngel16} model.  These 
353-GHz templates were scaled to other frequencies using maps of 
dust temperature and spectral index that were derived from the analysis of dust total intensity maps in \citet{planck2016-XLVIII} and \citet{planck2013-p06b}. 
The standard deviations of the dust spectral index for our six sky regions, measured using the 217- to 353-GHz colour ratio of model maps smoothed to a $1^\circ$ resolution, 
are in the range $\sigma (\beta_d) = 0.092\pm 0.005$. 
Nevertheless, the values of $\mathcal{R}_{\ell}^{BB}$ that we obtained, listed in Table~\ref{tab:multifreq_Rell_results}, are within
the lower limits inferred from the E2E simulations. 

Frequency decorrelation might result from variations of the spectral index both across the sky and along the line of sight. In the FFP10 maps, only the former is taken into account, and thus
polarization angles do not vary with frequency \citep[][PL]{Tassis15}. We also note again that the results of our multi-frequency analysis depend in detail on the adopted spectral model (Eq.~(\ref{eq:dust_decorrelation})).

\begin{figure}[!htbp]
\includegraphics[width=0.49\textwidth]{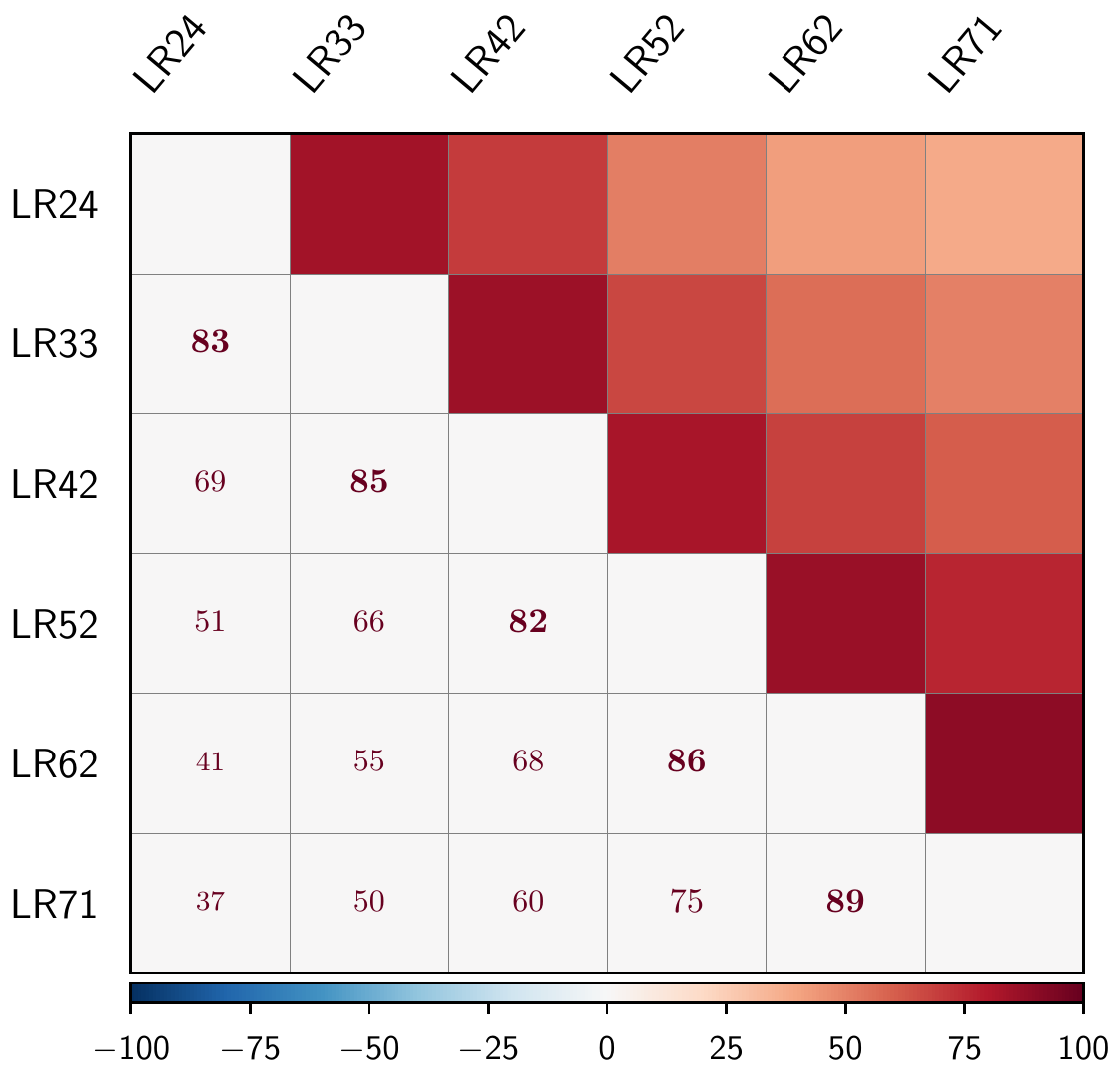}
\caption[]{Cross-correlation factor $\mathcal{R}_{\ell}^{BB}$ between sky regions, determined from the 
multi-frequency fits expressed as percentages.  As found for the correlation analysis between the 217- and 353-GHz data presented in Appendix~\ref{appendix:pdf_Rell}, the results from the 
multi-frequency fit are correlated between the six nested sky regions. }
\label{fig:correlation_LRmulti}
\end{figure}

We have also used the 300 E2E realizations to compute the cross-correlation of $\mathcal{R}_{\ell}^{BB}$ measured for our six sky regions (Fig.~\ref{fig:correlation_LRmulti}).
As found for the correlation analysis between the 217- and 353-GHz data 
discussed in Appendix~\ref{appendix:pdf_Rell} and by \citet{Sheehy17}, 
the results from the multi-frequency fit are also correlated between sky regions, which makes sense, of course, because they are nested. 

\subsection{Perspective for on-going and future CMB experiments}
\label{subsec:freq_perspective}

Here, we discuss the implications of our multi-frequency analysis of \Planck\ dust polarization for on-going and future CMB experiments that are designed to search for primordial $B$ modes. 
A somewhat comforting view concerning the complexity of dust polarization as a CMB foreground is suggested by two of our results. 
First, the data show no departure from a one-parameter MBB emission law (with a single fixed temperature) for the 
dust polarization SED spanning from 353\,GHz to below $70\,$GHz. Second, the data do not provide evidence for frequency decorrelation.

For our largest sky region, LR71, our lower limit on the correlation ratio $\mathcal{R}_{\ell}^{BB}(217, 353)$ is tightest and quite close to unity for the multipole range relevant to the search for primordial $B$ modes at the recombination peak. 
Using Eq.~(\ref{eq:decorrelation}), this limit translates to limits on the correlation ratio between the frequencies (95, 150, 220\,GHz) used in ground-based and balloon-borne experiments, namely 0.996 and 0.979 for $\mathcal{R}_{\ell}^{BB}(150, 220)$ and $\mathcal{R}_{\ell}^{BB}(95, 220)$, respectively. 

Sub-orbital CMB experiments typically target smaller sky regions, for which frequency decorrelation is less constrained by the \Planck\ data.
If the LR71 \Planck\ limit applies to these cleaner sky regions, then frequency decorrelation of dust polarization might not be a problem for CMB experiments aiming at a primordial $B$-mode detection limit of $r\simeq0.01$ at the recombination peak.
To quantify this, we consider LR24 as representative of clean sky regions
used by sub-orbital experiments for their CMB studies.
For this region, at $\ell =80$, and for 150 and 95\,GHz, the dust power is 45 and 7 times the primordial $B$-mode signal for $r=0.01$, respectively. 
Combining these factors with the corresponding values of $\mathcal{R}_{\ell}^{BB} (150, 220)$ and $\mathcal{R}_{\ell}^{BB}(95, 220)$ for LR71, the potential bias from decorrelation when combining data at 95, 150, and 220\,GHz could be smaller than the CMB $B$-mode signal for $r=0.01$. 

We caution that this view might be too optimistic because the frequency decorrelation might not be homogeneous over the sky. 
Indeed, the tight limit that \Planck\ data provide for LR71 might not apply to smaller sky regions. In particular, 
decorrelation could have a large statistical variance for small sky regions, where there is not a large amount of averaging over relevant scales in the ISM. 
Furthermore, the \Planck\ limits in Table~\ref{tab:multifreq_Rell_results} are dependent on the adopted model and so
frequency decorrelation of dust polarization at microwave frequencies could be larger than we have estimated.

It is worth stressing that the \Planck\ sensitivity precludes identifying how difficult the component separation will be
for more ambitious experiments -- CMB Stage IV \citep{CMBS4ScienceBook} and a future space mission, for example LiteBIRD \citep{Ishino16} -- that are targeting a $B$-mode detection limit $r\simeq0.001$ 
using the recombination and reionization peaks. Given the limitations of available data, it is essential to continue to make progress in assessing the component-separation problem 
by using increasingly realistic models developed in relation to the astrophysics of the dusty magnetized ISM.

%============================================================
\section{Conclusions}
\label{sec:summary}
%============================================================

Using the \Planck\ \PR\ polarization maps, this paper has extended the characterization of
Galactic dust polarized emission that is foreground to CMB polarization.  Our data analysis 
is validated using E2E simulations, where the mapmaking pipeline is run on simulated data derived by combining fixed maps of polarized sky emission with independent realizations
of the data noise and systematics. This final section summarizes the main results of our study.

% power spectra Section 3
% 3.1
The power spectra of dust polarized emission ($EE$, $BB$, $TE$, $TB$, and $EB$) are measured for six nested high-Galactic latitude regions, covering a range of sky fractions $\feff$ from 
24 to 71\,\%, over the multipole range ($\ell \le 600$) relevant to the analysis of $E$- and $B$-mode CMB polarization associated with 
% in \ell order
reionization, recombination, and lensing.
% power-law fits 3.2
We present power-law fits to the angular power spectra that reveal
statistically significant variations of the exponents over sky regions and a difference between the values for the $EE$ and $BB$ spectra, which for the largest sky region are
$\alpha_{EE} = \valpee\pm\valpeeu$ and $\alpha_{BB} = \valpbb\pm\valpbbu$, respectively.
The difference persists in the weighted mean values,
$\alpha_{EE} = \mvalpee$ and $\alpha_{BB} = \mvalpbb$. 
The small difference between the two exponents is not unexpected because 
the filamentary structures in the cold neutral interstellar medium have mainly $E$-mode polarization, due to the 
statistical alignment of the magnetic field orientation with matter.
%3.3
The $BB$ power scales as the square of the mean intensity of the region, $\ainten$.
% 3.5 and 3.4
The spectra show that the $TE$ correlation and the $E/B$ power asymmetry discovered by \Planck\ extend to low multipoles, which were not included in the
earlier \Planck\ polarization papers due to residual data systematics. 
% 3.5
The weighted mean value of $TE/EE$ is $2.76\pm0.05$.
The mean $TE$ correlation ratio is $r^{TE}_\ell = 0.357\pm0.003$,
with a scatter of around 0.1 between measured values, but no systematic dependence on multipole down to the lowest $\ell$ bins or on sky region.
% 3.6
We also report a significant $TB$ signal with 
a $TB/TE$ ratio of approximately 0.1 and correlation ratio $r^{TB}_\ell$ about 0.05.

% Sect 4. foreground spectral model dust and synchrotron
Combining data from \Planck\ and \WMAP, we characterize the mean SED of polarized Galactic foregrounds 
for the six sky regions as a function of multipole, for $\ell\,{<}\,160$. 
Our spectral model takes into account polarized synchrotron emission and its correlation with polarized dust emission.
The results of this analysis quantify the challenge of the component-separation procedure required for measuring the low-$\ell$ CMB $E$-mode reionization signal and detecting 
the reionization and recombination peaks of primordial CMB $B$ modes.

% Sect 5
% 5.1
In our analysis, we do not find systematic variations of the polarized dust SED with multipole value or with sky region.
The mean dust spectral index is $\beta_{\rm d}^P= \vbetap\pm\vbetapu$ for a dust temperature of $19.6$\,K.
The systematic error follows from the uncertainties on the polarization efficiencies of \HFI\ and 
includes the uncertainty from the CMB subtraction. 
%5.2
The dust SED in polarization from blind component separation is remarkably well fit by a single temperature MBB 
emission law from 353 to 44\,GHz with a similar index.
% 5.3
The difference between the indices for polarization and total intensity is small and not of high statistical significance, $\beta_{\rm d}^P - \beta_{\rm d}^I
= \vdbetapi\pm\vdbetapiu$. 
%5.4
This result suggests that the emission from a single grain type dominates the long-wavelength emission in both polarization and total intensity.
It constrains dust models involving multiple dust components (e.g. separate carbon and silicate grains), 
magnetic dipole emission and variations of the spectral index of the dust emissivity with frequency. Detailed modelling, beyond the scope of this paper,
is required to quantify these constraints.

% Sect 6
% 6.1
We analyse the correlation of the dust polarization maps across microwave frequencies by fitting cross-spectra between \Planck\ data at 100, 143, 217, and 353\,GHz. 
We find no evidence for a loss of correlation with frequency, within limits provided by the analysis of the E2E simulations. These new results provide tighter lower limits to the correlation ratio than we obtain from the comparison of the 217- and 353-GHz maps alone.
%6.2
If the \Planck\ limit on decorrelation for the largest sky region applies to the smaller sky regions observed by sub-orbital experiments, 
then frequency decorrelation of dust polarization might not be a problem for CMB experiments aiming at a primordial $B$-mode detection limit at the $r\simeq0.01$ level using the recombination peak.
However, the sensitivity of \Planck\ prevents us drawing any conclusions about how difficult the component-separation problem will be
for more ambitious experiments targeting lower levels for $r$.

\begin{acknowledgements}
The Planck Collaboration acknowledges the support of: ESA; CNES, and
CNRS/INSU-IN2P3-INP (France); ASI, CNR, and INAF (Italy); NASA and DoE
(USA); STFC and UKSA (UK); CSIC, MINECO, JA, and RES (Spain); Tekes, AoF,
and CSC (Finland); DLR and MPG (Germany); CSA (Canada); DTU Space
(Denmark); SER/SSO (Switzerland); RCN (Norway); SFI (Ireland);
FCT/MCTES (Portugal); ERC and PRACE (EU). A description of the Planck
Collaboration and a list of its members, indicating which technical
or scientific activities they have been involved in, can be found at
\href{http://www.cosmos.esa.int/web/planck/planck-collaboration}{\texttt{http://www.cosmos.esa.int/web/planck/planck-collaboration}}.
This research has received funding by the Agence Nationale de la Recherche (ANR-17-CE31-0022).
\end{acknowledgements}

\bibliographystyle{aa}
\bibliography{Planck_bib,CPP_L11}

\begin{thebibliography}{111}
\expandafter\ifx\csname natexlab\endcsname\relax\def\natexlab#1{#1}\fi

\bibitem[{{Abazajian} {et~al.}(2016){Abazajian}, {Adshead}, {Ahmed}, {Allen},
  {Alonso}, {Arnold}, {Baccigalupi}, {Bartlett}, {Battaglia}, {Benson},
  {Bischoff}, {Borrill}, {Buza}, {Calabrese}, {Caldwell}, {Carlstrom}, {Chang},
  {Crawford}, {Cyr-Racine}, {De Bernardis}, {de Haan}, {di Serego Alighieri},
  {Dunkley}, {Dvorkin}, {Errard}, {Fabbian}, {Feeney}, {Ferraro}, {Filippini},
  {Flauger}, {Fuller}, {Gluscevic}, {Green}, {Grin}, {Grohs}, {Henning},
  {Hill}, {Hlozek}, {Holder}, {Holzapfel}, {Hu}, {Huffenberger}, {Keskitalo},
  {Knox}, {Kosowsky}, {Kovac}, {Kovetz}, {Kuo}, {Kusaka}, {Le Jeune}, {Lee},
  {Lilley}, {Loverde}, {Madhavacheril}, {Mantz}, {Marsh}, {McMahon},
  {Meerburg}, {Meyers}, {Miller}, {Munoz}, {Nguyen}, {Niemack}, {Peloso},
  {Peloton}, {Pogosian}, {Pryke}, {Raveri}, {Reichardt}, {Rocha}, {Rotti},
  {Schaan}, {Schmittfull}, {Scott}, {Sehgal}, {Shandera}, {Sherwin}, {Smith},
  {Sorbo}, {Starkman}, {Story}, {van Engelen}, {Vieira}, {Watson}, {Whitehorn},
  \& {Kimmy Wu}}]{CMBS4ScienceBook}
{Abazajian}, K.~N., {Adshead}, P., {Ahmed}, Z., {et~al.} 2016, ArXiv:1610.02743

\bibitem[{{Abitbol} {et~al.}(2016){Abitbol}, {Hill}, \& {Johnson}}]{Abitbol16}
{Abitbol}, M.~H., {Hill}, J.~C., \& {Johnson}, B.~R. 2016, \mnras, 457, 1796

\bibitem[{{Alves} {et~al.}(2018){Alves}, {Boulanger}, {Ferri{\`e}re}, \&
  {Montier}}]{Alves18}
{Alves}, M.~I.~R., {Boulanger}, F., {Ferri{\`e}re}, K., \& {Montier}, L. 2018,
  \aap, 611, L5

\bibitem[{{Andersson} {et~al.}(2015){Andersson}, {Lazarian}, \&
  {Vaillancourt}}]{Andersson15}
{Andersson}, B.-G., {Lazarian}, A., \& {Vaillancourt}, J.~E. 2015, \araa, 53,
  501

\bibitem[{{Arnold} {et~al.}(2014){Arnold}, {Stebor}, {Ade}, {Akiba}, {Anthony},
  {Atlas}, {Barron}, {Bender}, {Boettger}, {Borrill}, {Chapman}, {Chinone},
  {Cukierman}, {Dobbs}, {Elleflot}, {Errard}, {Fabbian}, {Feng}, {Gilbert},
  {Goeckner-Wald}, {Halverson}, {Hasegawa}, {Hattori}, {Hazumi}, {Holzapfel},
  {Hori}, {Inoue}, {Jaehnig}, {Jaffe}, {Katayama}, {Keating}, {Kermish},
  {Keskitalo}, {Kisner}, {Le Jeune}, {Lee}, {Leitch}, {Linder}, {Matsuda},
  {Matsumura}, {Meng}, {Miller}, {Morii}, {Myers}, {Navaroli}, {Nishino},
  {Okamura}, {Paar}, {Peloton}, {Poletti}, {Raum}, {Rebeiz}, {Reichardt},
  {Richards}, {Ross}, {Rotermund}, {Schenck}, {Sherwin}, {Shirley}, {Sholl},
  {Siritanasak}, {Smecher}, {Steinbach}, {Stompor}, {Suzuki}, {Suzuki},
  {Takada}, {Takakura}, {Tomaru}, {Wilson}, {Yadav}, \& {Zahn}}]{Arnold14}
{Arnold}, K., {Stebor}, N., {Ade}, P.~A.~R., {et~al.} 2014, in \procspie, Vol.
  9153, Millimeter, Submillimeter, and Far-Infrared Detectors and
  Instrumentation for Astronomy VII, 91531F

\bibitem[{{Ashton} {et~al.}(2018){Ashton}, {Ade}, {Angil{\`e}}, {Benton},
  {Devlin}, {Dober}, {Fissel}, {Fukui}, {Galitzki}, {Gandilo}, {Klein},
  {Korotkov}, {Li}, {Martin}, {Matthews}, {Moncelsi}, {Nakamura},
  {Netterfield}, {Novak}, {Pascale}, {Poidevin}, {Santos}, {Savini}, {Scott},
  {Shariff}, {Soler}, {Thomas}, {Tucker}, {Tucker}, \&
  {Ward-Thompson}}]{Ashton2017}
{Ashton}, P.~C., {Ade}, P.~A.~R., {Angil{\`e}}, F.~E., {et~al.} 2018, \apj,
  857, 10

\bibitem[{{Austermann} {et~al.}(2012){Austermann}, {Aird}, {Beall}, {Becker},
  {Bender}, {Benson}, {Bleem}, {Britton}, {Carlstrom}, {Chang}, {Chiang},
  {Cho}, {Crawford}, {Crites}, {Datesman}, {de Haan}, {Dobbs}, {George},
  {Halverson}, {Harrington}, {Henning}, {Hilton}, {Holder}, {Holzapfel},
  {Hoover}, {Huang}, {Hubmayr}, {Irwin}, {Keisler}, {Kennedy}, {Knox}, {Lee},
  {Leitch}, {Li}, {Lueker}, {Marrone}, {McMahon}, {Mehl}, {Meyer}, {Montroy},
  {Natoli}, {Nibarger}, {Niemack}, {Novosad}, {Padin}, {Pryke}, {Reichardt},
  {Ruhl}, {Saliwanchik}, {Sayre}, {Schaffer}, {Shirokoff}, {Stark}, {Story},
  {Vanderlinde}, {Vieira}, {Wang}, {Williamson}, {Yefremenko}, {Yoon}, \&
  {Zahn}}]{Austermann12}
{Austermann}, J.~E., {Aird}, K.~A., {Beall}, J.~A., {et~al.} 2012, in
  \procspie, Vol. 8452, Millimeter, Submillimeter, and Far-Infrared Detectors
  and Instrumentation for Astronomy VI, 84521E

\bibitem[{{Baumann}(2009)}]{Baumann09}
{Baumann}, D. 2009, arXiv:0907.5424

\bibitem[{{Bennett} {et~al.}(2013){Bennett}, {Larson}, {Weiland}, {Jarosik},
  {Hinshaw}, {Odegard}, {Smith}, {Hill}, {Gold}, {Halpern}, {Komatsu}, {Nolta},
  {Page}, {Spergel}, {Wollack}, {Dunkley}, {Kogut}, {Limon}, {Meyer}, {Tucker},
  \& {Wright}}]{bennett2012}
{Bennett}, C.~L., {Larson}, D., {Weiland}, J.~L., {et~al.} 2013, \apjs, 208, 20

\bibitem[{{BICEP2 and Keck Array
  Collaborations}(2016)}]{PhysRevLett.116.031302}
{BICEP2 and Keck Array Collaborations}. 2016, Phys. Rev. Lett., 116, 031302

\bibitem[{{BICEP2/Keck Array and Planck Collaborations}(2015)}]{pb2015}
{BICEP2/Keck Array and Planck Collaborations}. 2015, \prl, 114, 101301

\bibitem[{{Blackman}(2015)}]{Blackman2015}
{Blackman}, E.~G. 2015, \ssr, 188, 59

\bibitem[{{Blagrave} {et~al.}(2017){Blagrave}, {Martin}, {Joncas}, {Kothes},
  {Stil}, {Miville-Desch{\^e}nes}, {Lockman}, \& {Taylor}}]{Blagrave17}
{Blagrave}, K., {Martin}, P.~G., {Joncas}, G., {et~al.} 2017, \apj, 834, 126

\bibitem[{{Caldwell} {et~al.}(2017){Caldwell}, {Hirata}, \&
  {Kamionkowski}}]{Caldwell17}
{Caldwell}, R.~R., {Hirata}, C., \& {Kamionkowski}, M. 2017, \apj, 839, 91

\bibitem[{{Cardoso} {et~al.}(2008){Cardoso}, {Le Jeune}, {Delabrouille},
  {Betoule}, \& {Patanchon}}]{Cardoso08}
{Cardoso}, J.-F., {Le Jeune}, M., {Delabrouille}, J., {Betoule}, M., \&
  {Patanchon}, G. 2008, IEEE Journal of Selected Topics in Signal Processing,
  2, 735

\bibitem[{{Chluba} {et~al.}(2017){Chluba}, {Hill}, \& {Abitbol}}]{Chluba17}
{Chluba}, J., {Hill}, J.~C., \& {Abitbol}, M.~H. 2017, \mnras, 472, 1195

\bibitem[{{Cho} \& {Lazarian}(2002)}]{Cho02}
{Cho}, J. \& {Lazarian}, A. 2002, \apjl, 575, L63

\bibitem[{{Choi} \& {Page}(2015)}]{Choi15}
{Choi}, S.~K. \& {Page}, L.~A. 2015, \jcap, 12, 020

\bibitem[{{Clark} {et~al.}(2015){Clark}, {Hill}, {Peek}, {Putman}, \&
  {Babler}}]{Clark15}
{Clark}, S.~E., {Hill}, J.~C., {Peek}, J.~E.~G., {Putman}, M.~E., \& {Babler},
  B.~L. 2015, Physical Review Letters, 115, 241302

\bibitem[{{Clark} {et~al.}(2014){Clark}, {Peek}, \& {Putman}}]{Clark13}
{Clark}, S.~E., {Peek}, J.~E.~G., \& {Putman}, M.~E. 2014, \apj, 789, 82

\bibitem[{{Compi{\`e}gne} {et~al.}(2011){Compi{\`e}gne}, {Verstraete}, {Jones},
  {Bernard}, {Boulanger}, {Flagey}, {Le Bourlot}, {Paradis}, \&
  {Ysard}}]{Compiegne11}
{Compi{\`e}gne}, M., {Verstraete}, L., {Jones}, A., {et~al.} 2011, \aap, 525,
  A103

\bibitem[{{Demyk} {et~al.}(2017){Demyk}, {Meny}, {Lu}, {Papatheodorou},
  {Toplis}, {Leroux}, {Depecker}, {Brubach}, {Roy}, {Nayral}, {Ojo}, {Delpech},
  {Paradis}, \& {Gromov}}]{Demyk17}
{Demyk}, K., {Meny}, C., {Lu}, X.-H., {et~al.} 2017, \aap, 600, A123

\bibitem[{{Draine} \& {Fraisse}(2009)}]{DraineFraisse2009}
{Draine}, B.~T. \& {Fraisse}, A.~A. 2009, \apj, 696, 1

\bibitem[{{Draine} \& {Hensley}(2013)}]{Draine13}
{Draine}, B.~T. \& {Hensley}, B. 2013, \apj, 765, 159

\bibitem[{{Draine} \& {Hensley}(2016)}]{Draine16}
{Draine}, B.~T. \& {Hensley}, B.~S. 2016, \apj, 831, 59

\bibitem[{{Draine} \& {Lazarian}(1999)}]{Draine99}
{Draine}, B.~T. \& {Lazarian}, A. 1999, \apj, 512, 740

\bibitem[{{Dunkley} {et~al.}(2009){Dunkley}, {Amblard}, {Baccigalupi},
  {Betoule}, {Chuss}, {Cooray}, {Delabrouille}, {Dickinson}, {Dobler},
  {Dotson}, {Eriksen}, {Finkbeiner}, {Fixsen}, {Fosalba}, {Fraisse}, {Hirata},
  {Kogut}, {Kristiansen}, {Lawrence}, {Magalha\~{}Es}, {Miville-Deschenes},
  {Meyer}, {Miller}, {Naess}, {Page}, {Peiris}, {Phillips}, {Pierpaoli},
  {Rocha}, {Vaillancourt}, \& {Verde}}]{Dunkley09}
{Dunkley}, J., {Amblard}, A., {Baccigalupi}, C., {et~al.} 2009, in American
  Institute of Physics Conference Series, Vol. 1141, American Institute of
  Physics Conference Series, ed. S.~{Dodelson}, D.~{Baumann}, A.~{Cooray},
  J.~{Dunkley}, A.~{Fraisse}, M.~G. {Jackson}, A.~{Kogut}, L.~{Krauss},
  M.~{Zaldarriaga}, \& K.~{Smith}, 222--264

\bibitem[{{Errard} {et~al.}(2016){Errard}, {Feeney}, {Peiris}, \&
  {Jaffe}}]{Errard16}
{Errard}, J., {Feeney}, S.~M., {Peiris}, H.~V., \& {Jaffe}, A.~H. 2016, \jcap,
  3, 052

\bibitem[{{Essinger-Hileman} {et~al.}(2014){Essinger-Hileman}, {Ali}, {Amiri},
  {Appel}, {Araujo}, {Bennett}, {Boone}, {Chan}, {Cho}, {Chuss}, {Colazo},
  {Crowe}, {Denis}, {D{\"u}nner}, {Eimer}, {Gothe}, {Halpern}, {Harrington},
  {Hilton}, {Hinshaw}, {Huang}, {Irwin}, {Jones}, {Karakla}, {Kogut}, {Larson},
  {Limon}, {Lowry}, {Marriage}, {Mehrle}, {Miller}, {Miller}, {Moseley},
  {Novak}, {Reintsema}, {Rostem}, {Stevenson}, {Towner}, {U-Yen}, {Wagner},
  {Watts}, {Wollack}, {Xu}, \& {Zeng}}]{Essinger-Hileman14}
{Essinger-Hileman}, T., {Ali}, A., {Amiri}, M., {et~al.} 2014, in \procspie,
  Vol. 9153, Millimeter, Submillimeter, and Far-Infrared Detectors and
  Instrumentation for Astronomy VII, 91531I

\bibitem[{{Fanciullo} {et~al.}(2017){Fanciullo}, {Guillet}, {Boulanger}, \&
  {Jones}}]{Fanciullo17}
{Fanciullo}, L., {Guillet}, V., {Boulanger}, F., \& {Jones}, A.~P. 2017, \aap,
  602, A7

\bibitem[{{Fraisse} {et~al.}(2013){Fraisse}, {Ade}, {Amiri}, {Benton}, {Bock},
  {Bond}, {Bonetti}, {Bryan}, {Burger}, {Chiang}, {Clark}, {Contaldi}, {Crill},
  {Davis}, {Dor{\'e}}, {Farhang}, {Filippini}, {Fissel}, {Gandilo}, {Golwala},
  {Gudmundsson}, {Hasselfield}, {Hilton}, {Holmes}, {Hristov}, {Irwin},
  {Jones}, {Kuo}, {MacTavish}, {Mason}, {Montroy}, {Morford}, {Netterfield},
  {O'Dea}, {Rahlin}, {Reintsema}, {Ruhl}, {Runyan}, {Schenker}, {Shariff},
  {Soler}, {Trangsrud}, {Tucker}, {Tucker}, {Turner}, \& {Wiebe}}]{Fraisse13}
{Fraisse}, A.~A., {Ade}, P.~A.~R., {Amiri}, M., {et~al.} 2013, \jcap, 4, 047

\bibitem[{{Fuskeland} {et~al.}(2014){Fuskeland}, {Wehus}, {Eriksen}, \&
  {N{\ae}ss}}]{Fuskeland14}
{Fuskeland}, U., {Wehus}, I.~K., {Eriksen}, H.~K., \& {N{\ae}ss}, S.~K. 2014,
  \apj, 790, 104

\bibitem[{{Gandilo} {et~al.}(2016){Gandilo}, {Ade}, {Angil{\`e}}, {Ashton},
  {Benton}, {Devlin}, {Dober}, {Fissel}, {Fukui}, {Galitzki}, {Klein},
  {Korotkov}, {Li}, {Martin}, {Matthews}, {Moncelsi}, {Nakamura},
  {Netterfield}, {Novak}, {Pascale}, {Poidevin}, {Santos}, {Savini}, {Scott},
  {Shariff}, {Diego Soler}, {Thomas}, {Tucker}, {Tucker}, \&
  {Ward-Thompson}}]{Gandilo2016}
{Gandilo}, N.~N., {Ade}, P.~A.~R., {Angil{\`e}}, F.~E., {et~al.} 2016, \apj,
  824, 84

\bibitem[{{G{\'e}nova-Santos} {et~al.}(2017){G{\'e}nova-Santos},
  {Rubi{\~n}o-Mart{\'{\i}}n}, {Pel{\'a}ez-Santos}, {Poidevin}, {Rebolo},
  {Vignaga}, {Artal}, {Harper}, {Hoyland}, {Lasenby},
  {Mart{\'{\i}}nez-Gonz{\'a}lez}, {Piccirillo}, {Tramonte}, \&
  {Watson}}]{Genova17}
{G{\'e}nova-Santos}, R., {Rubi{\~n}o-Mart{\'{\i}}n}, J.~A.,
  {Pel{\'a}ez-Santos}, A., {et~al.} 2017, \mnras, 464, 4107

\bibitem[{{Ghosh} {et~al.}(2017){Ghosh}, {Boulanger}, {Martin}, {Bracco},
  {Vansyngel}, {Aumont}, {Bock}, {Dor{\'e}}, {Haud}, {Kalberla}, \&
  {Serra}}]{Ghosh16}
{Ghosh}, T., {Boulanger}, F., {Martin}, P.~G., {et~al.} 2017, \aap, 601, A71

\bibitem[{{Gold} {et~al.}(2011){Gold}, {Odegard}, {Weiland}, {Hill}, {Kogut},
  {Bennett}, {Hinshaw}, {Chen}, {Dunkley}, {Halpern}, {Jarosik}, {Komatsu},
  {Larson}, {Limon}, {Meyer}, {Nolta}, {Page}, {Smith}, {Spergel}, {Tucker},
  {Wollack}, \& {Wright}}]{gold2010}
{Gold}, B., {Odegard}, N., {Weiland}, J.~L., {et~al.} 2011, \apjs, 192, 15

\bibitem[{{Grayson} {et~al.}(2016){Grayson}, {Ade}, {Ahmed}, {Alexander},
  {Amiri}, {Barkats}, {Benton}, {Bischoff}, {Bock}, {Boenish}, {Bowens-Rubin},
  {Buder}, {Bullock}, {Buza}, {Connors}, {Filippini}, {Fliescher}, {Halpern},
  {Harrison}, {Hilton}, {Hristov}, {Hui}, {Irwin}, {Kang}, {Karkare}, {Karpel},
  {Kefeli}, {Kernasovskiy}, {Kovac}, {Kuo}, {Leitch}, {Lueker}, {Megerian},
  {Monticue}, {Namikawa}, {Netterfield}, {Nguyen}, {O'Brient}, {Ogburn},
  {Pryke}, {Reintsema}, {Richter}, {Schwarz}, {Sorenson}, {Sheehy},
  {Staniszewski}, {Steinbach}, {Teply}, {Thompson}, {Tolan}, {Tucker},
  {Turner}, {Vieregg}, {Wandui}, {Weber}, {Wiebe}, {Willmert}, {Wu}, \&
  {Yoon}}]{Grayson16}
{Grayson}, J.~A., {Ade}, P.~A.~R., {Ahmed}, Z., {et~al.} 2016, in \procspie,
  Vol. 9914, Millimeter, Submillimeter, and Far-Infrared Detectors and
  Instrumentation for Astronomy VIII, 99140S

\bibitem[{{Guillet} {et~al.}(2018){Guillet}, {Fanciullo}, {Verstraete},
  {Boulanger}, {Jones}, {Miville-Desch{\^e}nes}, {Ysard}, {Levrier}, \&
  {Alves}}]{Guillet17}
{Guillet}, V., {Fanciullo}, L., {Verstraete}, L., {et~al.} 2018, \aap, 610, A16

\bibitem[{{Guth}(1981)}]{Guth81}
{Guth}, A.~H. 1981, \prd, 23, 347

\bibitem[{{Hennebelle}(2013)}]{Hennebelle13}
{Hennebelle}, P. 2013, \aap, 556, A153

\bibitem[{{Hensley} \& {Bull}(2018)}]{Hensley17}
{Hensley}, B.~S. \& {Bull}, P. 2018, \apj, 853, 127

\bibitem[{{Hildebrand}(1988)}]{Hildebrand88}
{Hildebrand}, R.~H. 1988, \qjras, 29, 327

\bibitem[{{Hoang} \& {Lazarian}(2016{\natexlab{a}})}]{Hoang16}
{Hoang}, T. \& {Lazarian}, A. 2016{\natexlab{a}}, \apj, 831, 159

\bibitem[{{Hoang} \& {Lazarian}(2016{\natexlab{b}})}]{Hoang16_AME}
{Hoang}, T. \& {Lazarian}, A. 2016{\natexlab{b}}, \apj, 821, 91

\bibitem[{{Inoue} \& {Inutsuka}(2016)}]{Inoue16}
{Inoue}, T. \& {Inutsuka}, S.-i. 2016, \apj, 833, 10

\bibitem[{{Ishino} {et~al.}(2016){Ishino}, {Akiba}, {Arnold}, {Barron},
  {Borrill}, {Chendra}, {Chinone}, {Cho}, {Cukierman}, {de Haan}, {Dobbs},
  {Dominjon}, {Dotani}, {Elleflot}, {Errard}, {Fujino}, {Fuke}, {Funaki},
  {Goeckner-Wald}, {Halverson}, {Harvey}, {Hasebe}, {Hasegawa}, {Hattori},
  {Hattori}, {Hazumi}, {Hidehira}, {Hill}, {Hilton}, {Holzapfel}, {Hori},
  {Hubmayr}, {Ichiki}, {Imada}, {Inatani}, {Inoue}, {Inoue}, {Irie}, {Irwin},
  {Ishitsuka}, {Jeong}, {Kanai}, {Karatsu}, {Kashima}, {Katayama}, {Kawano},
  {Kawasaki}, {Keating}, {Kernasovskiy}, {Keskitalo}, {Kibayashi}, {Kida},
  {Kimura}, {Kimura}, {Kisner}, {Kohri}, {Komatsu}, {Komatsu}, {Kuo},
  {Kuromiya}, {Kusaka}, {Lee}, {Li}, {Linder}, {Maki}, {Matsuhara},
  {Matsumura}, {Matsuoka}, {Matsuura}, {Mima}, {Minami}, {Mitsuda}, {Nagai},
  {Nagasaki}, {Nagata}, {Nakajima}, {Nakamura}, {Namikawa}, {Naruse},
  {Nishibori}, {Nishijo}, {Nishino}, {Noda}, {Noguchi}, {Ogawa}, {Ogburn},
  {Oguri}, {Ohta}, {Okada}, {Okamoto}, {Okamura}, {Otani}, {Pisano}, {Rebeiz},
  {Richards}, {Sakai}, {Sakurai}, {Sato}, {Sato}, {Segawa}, {Sekiguchi},
  {Sekimoto}, {Sekine}, {Seljak}, {Sherwin}, {Shimizu}, {Shinozaki}, {Shu},
  {Stompor}, {Sugai}, {Sugita}, {Suzuki}, {Suzuki}, {Suzuki}, {Tajima},
  {Takada}, {Takakura}, {Takano}, {Takatori}, {Takei}, {Tanabe}, {Tomaru},
  {Tomita}, {Turin}, {Uozumi}, {Utsunomiya}, {Uzawa}, {Wada}, {Watanabe},
  {Westbrook}, {Whitehorn}, {Yamada}, {Yamamoto}, {Yamasaki}, {Yamashita},
  {Yoshida}, {Yoshida}, \& {Yotsumoto}}]{Ishino16}
{Ishino}, H., {Akiba}, Y., {Arnold}, K., {et~al.} 2016, in \procspie, Vol.
  9904, Space Telescopes and Instrumentation 2016: Optical, Infrared, and
  Millimeter Wave, 99040X

\bibitem[{{Jarosik} {et~al.}(2003){Jarosik}, {Bennett}, {Halpern}, {Hinshaw},
  {Kogut}, {Limon}, {Meyer}, {Page}, {Pospieszalski}, {Spergel}, {Tucker},
  {Wilkinson}, {Wollack}, {Wright}, \& {Zhang}}]{Jarosik03}
{Jarosik}, N., {Bennett}, C.~L., {Halpern}, M., {et~al.} 2003, \apjs, 145, 413

\bibitem[{{Jones} {et~al.}(2013){Jones}, {Fanciullo}, {K{\"o}hler},
  {Verstraete}, {Guillet}, {Bocchio}, \& {Ysard}}]{Jones13}
{Jones}, A.~P., {Fanciullo}, L., {K{\"o}hler}, M., {et~al.} 2013, \aap, 558,
  A62

\bibitem[{{Kalberla} {et~al.}(2016){Kalberla}, {Kerp}, {Haud}, {Winkel}, {Ben
  Bekhti}, {Fl{\"o}er}, \& {Lenz}}]{Kalberla16}
{Kalberla}, P.~M.~W., {Kerp}, J., {Haud}, U., {et~al.} 2016, \apj, 821, 117

\bibitem[{{Kandel} {et~al.}(2017){Kandel}, {Lazarian}, \&
  {Pogosyan}}]{Kandel17}
{Kandel}, D., {Lazarian}, A., \& {Pogosyan}, D. 2017, \mnras, 472, L10

\bibitem[{{Kandel} {et~al.}(2018){Kandel}, {Lazarian}, \&
  {Pogosyan}}]{Kandel18}
{Kandel}, D., {Lazarian}, A., \& {Pogosyan}, D. 2018, \mnras, 478, 530

\bibitem[{{Kogut} {et~al.}(2007){Kogut}, {Dunkley}, {Bennett}, {Dor{\'e}},
  {Gold}, {Halpern}, {Hinshaw}, {Jarosik}, {Komatsu}, {Nolta}, {Odegard},
  {Page}, {Spergel}, {Tucker}, {Weiland}, {Wollack}, \& {Wright}}]{kogut2007}
{Kogut}, A., {Dunkley}, J., {Bennett}, C.~L., {et~al.} 2007, \apj, 665, 355

\bibitem[{{Kogut} {et~al.}(2011){Kogut}, {Fixsen}, {Chuss}, {Dotson}, {Dwek},
  {Halpern}, {Hinshaw}, {Meyer}, {Moseley}, {Seiffert}, {Spergel}, \&
  {Wollack}}]{Kogut11}
{Kogut}, A., {Fixsen}, D.~J., {Chuss}, D.~T., {et~al.} 2011, \jcap, 7, 025

\bibitem[{{Krachmalnicoff} {et~al.}(2016){Krachmalnicoff}, {Baccigalupi},
  {Aumont}, {Bersanelli}, \& {Mennella}}]{Krachmalnicoff16}
{Krachmalnicoff}, N., {Baccigalupi}, C., {Aumont}, J., {Bersanelli}, M., \&
  {Mennella}, A. 2016, \aap, 588, A65

\bibitem[{{Krachmalnicoff} {et~al.}(2018){Krachmalnicoff}, {Carretti},
  {Baccigalupi}, {Bernardi}, {Brown}, {Gaensler}, {Haverkorn}, {Kesteven},
  {Perrotta}, {Poppi}, \& {Staveley-Smith}}]{Krachmalnicoff18}
{Krachmalnicoff}, N., {Carretti}, E., {Baccigalupi}, C., {et~al.} 2018, \aap,
  618, A166

\bibitem[{Kritsuk {et~al.}(2018)Kritsuk, Flauger, \& Ustyugov}]{Kritsuk17}
Kritsuk, A.~G., Flauger, R., \& Ustyugov, S.~D. 2018, Phys. Rev. Lett., 121,
  021104

\bibitem[{{Lazarian} \& {Hoang}(2007)}]{Lazarian_Hoang07}
{Lazarian}, A. \& {Hoang}, T. 2007, \mnras, 378, 910

\bibitem[{{Lazarian} {et~al.}(2018){Lazarian}, {Yuen}, {Ho}, {Chen},
  {Lazarian}, {Lu}, {Yang}, \& {Hu}}]{Lazarian18}
{Lazarian}, A., {Yuen}, K.~H., {Ho}, K.~W., {et~al.} 2018, \apj, 865, 46

\bibitem[{{Li} \& {Draine}(2001)}]{Li01}
{Li}, A. \& {Draine}, B.~T. 2001, \apj, 554, 778

\bibitem[{{Linde}(1982)}]{Linde82}
{Linde}, A.~D. 1982, Physics Letters B, 108, 389

\bibitem[{{Martin}(2007)}]{Martin07}
{Martin}, P.~G. 2007, in EAS Publications Series, Vol.~23, EAS Publications
  Series, ed. M.-A. {Miville-Desch{\^e}nes} \& F.~{Boulanger}, 165--188

\bibitem[{{Martin} {et~al.}(2015){Martin}, {Blagrave}, {Lockman}, {Pinheiro
  Gon{\c c}alves}, {Boothroyd}, {Joncas}, {Miville-Desch{\^e}nes}, \&
  {Stephan}}]{martin2015}
{Martin}, P.~G., {Blagrave}, K.~P.~M., {Lockman}, F.~J., {et~al.} 2015, \apj,
  809, 153

\bibitem[{{McClure-Griffiths} {et~al.}(2006){McClure-Griffiths}, {Dickey},
  {Gaensler}, {Green}, \& {Haverkorn}}]{McClure06}
{McClure-Griffiths}, N.~M., {Dickey}, J.~M., {Gaensler}, B.~M., {Green}, A.~J.,
  \& {Haverkorn}, M. 2006, \apj, 652, 1339

\bibitem[{{Mennella} {et~al.}(2011){Mennella}, {Butler}, {Curto}, {Cuttaia},
  {Davis}, {Dick}, {Frailis}, {Galeotta}, {Gregorio}, {Kurki-Suonio},
  {Lawrence}, {Leach}, {Leahy}, {Lowe}, {Maino}, {Mandolesi}, {Maris},
  {Mart{\'{\i}}nez-Gonz{\'a}lez}, {Meinhold}, {Morgante}, {Pearson},
  {Perrotta}, {Polenta}, {Poutanen}, {Sandri}, {Seiffert}, {Suur-Uski},
  {Tavagnacco}, {Terenzi}, {Tomasi}, {Valiviita}, {Villa}, {Watson},
  {Wilkinson}, {Zacchei}, {Zonca}, {Aja}, {Artal}, {Baccigalupi}, {Banday},
  {Barreiro}, {Bartlett}, {Bartolo}, {Battaglia}, {Bennett}, {Bonaldi},
  {Bonavera}, {Borrill}, {Bouchet}, {Burigana}, {Cabella}, {Cappellini},
  {Chen}, {Colombo}, {Cruz}, {Danese}, {D'Arcangelo}, {Davies}, {de Gasperis},
  {de Rosa}, {de Zotti}, {Dickinson}, {Diego}, {Donzelli}, {Efstathiou},
  {En{\ss}lin}, {Eriksen}, {Falvella}, {Finelli}, {Foley}, {Franceschet},
  {Franceschi}, {Gaier}, {G{\'e}nova-Santos}, {George}, {G{\'o}mez},
  {Gonz{\'a}lez-Nuevo}, {G{\'o}rski}, {Gruppuso}, {Hansen}, {Herranz},
  {Herreros}, {Hoyland}, {Hughes}, {Jewell}, {Jukkala}, {Juvela},
  {Kangaslahti}, {Keih{\"a}nen}, {Keskitalo}, {Kilpia}, {Kisner}, {Knoche},
  {Knox}, {Laaninen}, {L{\"a}hteenm{\"a}ki}, {Lamarre}, {Leonardi},
  {Le{\'o}n-Tavares}, {Leutenegger}, {Lilje}, {L{\'o}pez-Caniego}, {Lubin},
  {Malaspina}, {Marinucci}, {Massardi}, {Matarrese}, {Matthai}, {Melchiorri},
  {Mendes}, {Miccolis}, {Migliaccio}, {Mitra}, {Moss}, {Natoli}, {Nesti},
  {N{\o}rgaard-Nielsen}, {Pagano}, {Paladini}, {Paoletti}, {Partridge},
  {Pasian}, {Pettorino}, {Pietrobon}, {Pospieszalski}, {Pr{\'e}zeau}, {Prina},
  {Procopio}, {Puget}, {Quercellini}, {Rachen}, {Rebolo}, {Reinecke},
  {Ricciardi}, {Robbers}, {Rocha}, {Roddis}, {Rubino-Mart{\'{\i}}n},
  {Savelainen}, {Scott}, {Silvestri}, {Simonetto}, {Sjoman}, {Smoot}, {Sozzi},
  {Stringhetti}, {Tauber}, {Tofani}, {Toffolatti}, {Tuovinen}, {T{\"u}rler},
  {Umana}, {Valenziano}, {Varis}, {Vielva}, {Vittorio}, {Wade}, {Watson},
  {White}, \& {Winder}}]{planck2011-1.4}
{Mennella}, A., {Butler}, R.~C., {Curto}, A., {et~al.} 2011, \aap, 536, A3

\bibitem[{{Miville-Desch{\^e}nes} {et~al.}(2003){Miville-Desch{\^e}nes},
  {Joncas}, {Falgarone}, \& {Boulanger}}]{Miville03}
{Miville-Desch{\^e}nes}, M.-A., {Joncas}, G., {Falgarone}, E., \& {Boulanger},
  F. 2003, \aap, 411, 109

\bibitem[{{Miville-Desch{\^e}nes} {et~al.}(2007){Miville-Desch{\^e}nes},
  {Lagache}, {Boulanger}, \& {Puget}}]{Miville07}
{Miville-Desch{\^e}nes}, M.-A., {Lagache}, G., {Boulanger}, F., \& {Puget},
  J.-L. 2007, \aap, 469, 595

\bibitem[{{Naess} {et~al.}(2014){Naess}, {Hasselfield}, {McMahon}, {Niemack},
  {Addison}, {Ade}, {Allison}, {Amiri}, {Battaglia}, {Beall}, {de Bernardis},
  {Bond}, {Britton}, {Calabrese}, {Cho}, {Coughlin}, {Crichton}, {Das},
  {Datta}, {Devlin}, {Dicker}, {Dunkley}, {D{\"u}nner}, {Fowler}, {Fox},
  {Gallardo}, {Grace}, {Gralla}, {Hajian}, {Halpern}, {Henderson}, {Hill},
  {Hilton}, {Hilton}, {Hincks}, {Hlozek}, {Ho}, {Hubmayr}, {Huffenberger},
  {Hughes}, {Infante}, {Irwin}, {Jackson}, {Muya Kasanda}, {Klein}, {Koopman},
  {Kosowsky}, {Li}, {Louis}, {Lungu}, {Madhavacheril}, {Marriage}, {Maurin},
  {Menanteau}, {Moodley}, {Munson}, {Newburgh}, {Nibarger}, {Nolta}, {Page},
  {Pappas}, {Partridge}, {Rojas}, {Schmitt}, {Sehgal}, {Sherwin}, {Sievers},
  {Simon}, {Spergel}, {Staggs}, {Switzer}, {Thornton}, {Trac}, {Tucker},
  {Uehara}, {Van Engelen}, {Ward}, \& {Wollack}}]{Naess14}
{Naess}, S., {Hasselfield}, M., {McMahon}, J., {et~al.} 2014, \jcap, 10, 007

\bibitem[{{Page} {et~al.}(2007){Page}, {Hinshaw}, {Komatsu}, {Nolta},
  {Spergel}, {Bennett}, {Barnes}, {Bean}, {Dor{\'e}}, {Dunkley}, {Halpern},
  {Hill}, {Jarosik}, {Kogut}, {Limon}, {Meyer}, {Odegard}, {Peiris}, {Tucker},
  {Verde}, {Weiland}, {Wollack}, \& {Wright}}]{page2007}
{Page}, L., {Hinshaw}, G., {Komatsu}, E., {et~al.} 2007, \apjs, 170, 335

\bibitem[{{Planck HFI Core Team}(2011)}]{planck2011-1.5}
{Planck HFI Core Team}. 2011, \aap, 536, A4

\bibitem[{{\sorthelp{Planck Collaboration 2014A}}{Planck Collaboration
  I}(2014)}]{planck2013-p01}
{\sorthelp{Planck Collaboration 2014A}}{Planck Collaboration I}. 2014, \aap,
  571, A1

\bibitem[{{\sorthelp{Planck Collaboration 2014I}}{Planck Collaboration
  IX}(2014)}]{planck2013-p03d}
{\sorthelp{Planck Collaboration 2014I}}{Planck Collaboration IX}. 2014, \aap,
  571, A9

\bibitem[{{\sorthelp{Planck Collaboration 2014K}}{Planck Collaboration
  XI}(2014)}]{planck2013-p06b}
{\sorthelp{Planck Collaboration 2014K}}{Planck Collaboration XI}. 2014, \aap,
  571, A11

\bibitem[{{\sorthelp{Planck Collaboration 2014ZE}}{Planck Collaboration
  XXX}(2014)}]{planck2013-pip56}
{\sorthelp{Planck Collaboration 2014ZE}}{Planck Collaboration XXX}. 2014, \aap,
  571, A30

\bibitem[{{\sorthelp{Planck Collaboration 2015H}}{Planck Collaboration
  VIII}(2016)}]{planck2014-a09}
{\sorthelp{Planck Collaboration 2015H}}{Planck Collaboration VIII}. 2016, \aap,
  594, A8

\bibitem[{{\sorthelp{Planck Collaboration 2015I}}{Planck Collaboration
  IX}(2016)}]{planck2014-a11}
{\sorthelp{Planck Collaboration 2015I}}{Planck Collaboration IX}. 2016, \aap,
  594, A9

\bibitem[{{\sorthelp{Planck Collaboration 2015J}}{Planck Collaboration
  X}(2016)}]{planck2014-a12}
{\sorthelp{Planck Collaboration 2015J}}{Planck Collaboration X}. 2016, \aap,
  594, A10

\bibitem[{{\sorthelp{Planck Collaboration 2015K}}{Planck Collaboration
  XI}(2016)}]{planck2014-a13}
{\sorthelp{Planck Collaboration 2015K}}{Planck Collaboration XI}. 2016, \aap,
  594, A11

\bibitem[{{\sorthelp{Planck Collaboration 2015L}}{Planck Collaboration
  XII}(2016)}]{planck2014-a14}
{\sorthelp{Planck Collaboration 2015L}}{Planck Collaboration XII}. 2016, \aap,
  594, A12

\bibitem[{{\sorthelp{Planck Collaboration 2015M}}{Planck Collaboration
  XIII}(2016)}]{planck2014-a15}
{\sorthelp{Planck Collaboration 2015M}}{Planck Collaboration XIII}. 2016, \aap,
  594, A13

\bibitem[{{\sorthelp{Planck Collaboration 2015Y}}{Planck Collaboration
  XXV}(2016)}]{planck2014-a31}
{\sorthelp{Planck Collaboration 2015Y}}{Planck Collaboration XXV}. 2016, \aap,
  594, A25

\bibitem[{{\sorthelp{Planck Collaboration 2018A}}{Planck Collaboration
  I}(2018)}]{planck2016-l01}
{\sorthelp{Planck Collaboration 2018A}}{Planck Collaboration I}. 2018, \aap,
  submitted, arXiv:1807.06205

\bibitem[{{\sorthelp{Planck Collaboration 2018B}}{Planck Collaboration
  II}(2018)}]{planck2016-l02}
{\sorthelp{Planck Collaboration 2018B}}{Planck Collaboration II}. 2018, \aap,
  submitted, arXiv:1807.06206

\bibitem[{{\sorthelp{Planck Collaboration 2018C}}{Planck Collaboration
  III}(2018)}]{planck2016-l03}
{\sorthelp{Planck Collaboration 2018C}}{Planck Collaboration III}. 2018, \aap,
  accepted, arXiv:1807.06207

\bibitem[{{\sorthelp{Planck Collaboration 2018D}}{Planck Collaboration
  IV}(2018)}]{planck2016-l04}
{\sorthelp{Planck Collaboration 2018D}}{Planck Collaboration IV}. 2018, \aap,
  submitted, arXiv:1807.06208

\bibitem[{{\sorthelp{Planck Collaboration 2018L}}{Planck Collaboration
  XII}(2018)}]{planck2016-l11B}
{\sorthelp{Planck Collaboration 2018L}}{Planck Collaboration XII}. 2018, \aap,
  submitted, arXiv:1807.06212

\bibitem[{{\sorthelp{Planck Collaboration IntQ}}{Planck Collaboration Int.
  XVII}(2014)}]{planck2013-XVII}
{\sorthelp{Planck Collaboration IntQ}}{Planck Collaboration Int. XVII}. 2014,
  \aap, 566, A55

\bibitem[{{\sorthelp{Planck Collaboration IntS}}{Planck Collaboration Int.
  XIX}(2015)}]{planck2014-XIX}
{\sorthelp{Planck Collaboration IntS}}{Planck Collaboration Int. XIX}. 2015,
  \aap, 576, A104

\bibitem[{{\sorthelp{Planck Collaboration IntT}}{Planck Collaboration Int.
  XX}(2015)}]{planck2014-XX}
{\sorthelp{Planck Collaboration IntT}}{Planck Collaboration Int. XX}. 2015,
  \aap, 576, A105

\bibitem[{{\sorthelp{Planck Collaboration IntU}}{Planck Collaboration Int.
  XXI}(2015)}]{planck2014-XXI}
{\sorthelp{Planck Collaboration IntU}}{Planck Collaboration Int. XXI}. 2015,
  \aap, 576, A106

\bibitem[{{\sorthelp{Planck Collaboration IntV}}{Planck Collaboration Int.
  XXII}(2015)}]{planck2014-XXII}
{\sorthelp{Planck Collaboration IntV}}{Planck Collaboration Int. XXII}. 2015,
  \aap, 576, A107

\bibitem[{{\sorthelp{Planck Collaboration IntZE}}{Planck Collaboration Int.
  XXX}(2016)}]{planck2014-XXX}
{\sorthelp{Planck Collaboration IntZE}}{Planck Collaboration Int. XXX}. 2016,
  \aap, 586, A133

\bibitem[{{\sorthelp{Planck Collaboration IntZG}}{Planck Collaboration Int.
  XXXII}(2016)}]{planck2014-XXXII}
{\sorthelp{Planck Collaboration IntZG}}{Planck Collaboration Int. XXXII}. 2016,
  \aap, 586, A135

\bibitem[{{\sorthelp{Planck Collaboration IntZJ}}{Planck Collaboration Int.
  XXXV}(2016)}]{planck2015-XXXV}
{\sorthelp{Planck Collaboration IntZJ}}{Planck Collaboration Int. XXXV}. 2016,
  \aap, 586, A138

\bibitem[{{\sorthelp{Planck Collaboration IntZM}}{Planck Collaboration Int.
  XXXVIII}(2016)}]{planck2015-XXXVIII}
{\sorthelp{Planck Collaboration IntZM}}{Planck Collaboration Int. XXXVIII}.
  2016, \aap, 586, A141

\bibitem[{{\sorthelp{Planck Collaboration IntZS}}{Planck Collaboration Int.
  XLIV}(2016)}]{planck2016-XLIV}
{\sorthelp{Planck Collaboration IntZS}}{Planck Collaboration Int. XLIV}. 2016,
  \aap, 596, A105

\bibitem[{{\sorthelp{Planck Collaboration IntZU}}{Planck Collaboration Int.
  XLVI}(2016)}]{planck2014-a10}
{\sorthelp{Planck Collaboration IntZU}}{Planck Collaboration Int. XLVI}. 2016,
  \aap, 596, A107

\bibitem[{{\sorthelp{Planck Collaboration IntZW}}{Planck Collaboration Int.
  XLVIII}(2016)}]{planck2016-XLVIII}
{\sorthelp{Planck Collaboration IntZW}}{Planck Collaboration Int. XLVIII}.
  2016, \aap, 596, A109

\bibitem[{{\sorthelp{Planck Collaboration IntZY}}{Planck Collaboration Int.
  L}(2017)}]{planck2016-L}
{\sorthelp{Planck Collaboration IntZY}}{Planck Collaboration Int. L}. 2017,
  \aap, 599, A51

\bibitem[{{Poh} \& {Dodelson}(2017)}]{Poh17}
{Poh}, J. \& {Dodelson}, S. 2017, \prd, 95, 103511

\bibitem[{{Remazeilles} {et~al.}(2018){Remazeilles}, {Banday}, {Baccigalupi},
  {Basak}, {Bonaldi}, {De Zotti}, {Delabrouille}, {Dickinson}, {Eriksen},
  {Errard}, {Fernandez-Cobos}, {Fuskeland}, {Herv{\'{\i}}as-Caimapo},
  {L{\'o}pez-Caniego}, {Martinez-Gonz{\'a}lez}, {Roman}, {Vielva}, {Wehus},
  {Achucarro}, {Ade}, {Allison}, {Ashdown}, {Ballardini}, {Banerji},
  {Bartlett}, {Bartolo}, {Baumann}, {Bersanelli}, {Bonato}, {Borrill},
  {Bouchet}, {Boulanger}, {Brinckmann}, {Bucher}, {Burigana}, {Buzzelli},
  {Cai}, {Calvo}, {Carvalho}, {Castellano}, {Challinor}, {Chluba}, {Clesse},
  {Colantoni}, {Coppolecchia}, {Crook}, {D'Alessandro}, {de Bernardis}, {de
  Gasperis}, {Diego}, {Di Valentino}, {Feeney}, {Ferraro}, {Finelli},
  {Forastieri}, {Galli}, {Genova-Santos}, {Gerbino}, {Gonz{\'a}lez-Nuevo},
  {Grandis}, {Greenslade}, {Hagstotz}, {Hanany}, {Handley},
  {Hernandez-Monteagudo}, {Hills}, {Hivon}, {Kiiveri}, {Kisner}, {Kitching},
  {Kunz}, {Kurki-Suonio}, {Lamagna}, {Lasenby}, {Lattanzi}, {Lesgourgues},
  {Lewis}, {Liguori}, {Lindholm}, {Luzzi}, {Maffei}, {Martins}, {Masi},
  {Matarrese}, {McCarthy}, {Melin}, {Melchiorri}, {Molinari}, {Monfardini},
  {Natoli}, {Negrello}, {Notari}, {Paiella}, {Paoletti}, {Patanchon}, {Piat},
  {Pisano}, {Polastri}, {Polenta}, {Pollo}, {Poulin}, {Quartin},
  {Rubino-Martin}, {Salvati}, {Tartari}, {Tomasi}, {Tramonte}, {Trappe},
  {Trombetti}, {Tucker}, {Valiviita}, {Van de Weijgaert}, {van Tent}, {Vennin},
  {Vittorio}, {Young}, \& {Zannoni}}]{Remazeilles17}
{Remazeilles}, M., {Banday}, A.~J., {Baccigalupi}, C., {et~al.} 2018, \jcap, 4,
  023

\bibitem[{{Remazeilles} {et~al.}(2011){Remazeilles}, {Delabrouille}, \&
  {Cardoso}}]{Remazeilles11}
{Remazeilles}, M., {Delabrouille}, J., \& {Cardoso}, J.-F. 2011, \mnras, 418,
  467

\bibitem[{{Sheehy} \& {Slosar}(2018)}]{Sheehy17}
{Sheehy}, C. \& {Slosar}, A. 2018, \prd, 97, 043522

\bibitem[{{Siebenmorgen} {et~al.}(2014){Siebenmorgen}, {Voshchinnikov}, \&
  {Bagnulo}}]{Siebenmorgen14}
{Siebenmorgen}, R., {Voshchinnikov}, N.~V., \& {Bagnulo}, S. 2014, \aap, 561,
  A82

\bibitem[{{Soler} \& {Hennebelle}(2017)}]{Soler17}
{Soler}, J.~D. \& {Hennebelle}, P. 2017, \aap, 607, A2

\bibitem[{{Starobinski{\v i}}(1979)}]{Starobinsky79}
{Starobinski{\v i}}, A.~A. 1979, Soviet Journal of Experimental and Theoretical
  Physics Letters, 30, 682

\bibitem[{{Stein}(1966)}]{Stein66}
{Stein}, W. 1966, \apj, 144, 318

\bibitem[{{Strong} {et~al.}(2011){Strong}, {Orlando}, \& {Jaffe}}]{Strong11}
{Strong}, A.~W., {Orlando}, E., \& {Jaffe}, T.~R. 2011, \aap, 534, A54

\bibitem[{{Tassis} \& {Pavlidou}(2015)}]{Tassis15}
{Tassis}, K. \& {Pavlidou}, V. 2015, \mnras, 451, L90

\bibitem[{{Tristram} {et~al.}(2005){Tristram}, {Mac{\'{\i}}as-P{\'e}rez},
  {Renault}, \& {Santos}}]{Tristram05}
{Tristram}, M., {Mac{\'{\i}}as-P{\'e}rez}, J.~F., {Renault}, C., \& {Santos},
  D. 2005, \mnras, 358, 833

\bibitem[{{Vansyngel} {et~al.}(2017){Vansyngel}, {Boulanger}, {Ghosh},
  {Wandelt}, {Aumont}, {Bracco}, {Levrier}, {Martin}, \&
  {Montier}}]{Vansyngel16}
{Vansyngel}, F., {Boulanger}, F., {Ghosh}, T., {et~al.} 2017, \aap, 603, A62

\bibitem[{{Yokoi}(2013)}]{Yokoi2013}
{Yokoi}, N. 2013, Geophysical and Astrophysical Fluid Dynamics, 107, 114

\end{thebibliography}

%%%%%%%%%%%%%%%%%%%%%%%%%%%%%%%%%%%%%%%%%%%%%%%%%%%%%%%%%%%%%%%%%%%%%%%%%%%%%%%%%%%%
%%%%%%%%%%%%%%%%%%%%%%%%%%%%%%   Appendices   %%%%%%%%%%%%%%%%%%%%%%%%%%%%%%%%%%%%%%%%%%%%%%%
%%%%%%%%%%%%%%%%%%%%%%%%%%%%%%%%%%%%%%%%%%%%%%%%%%%%%%%%%%%%%%%%%%%%%%%%%%%%%%%%%%%%

\appendix
\section{Data simulations and uncertainties}
\label{appendix:methodology}

In this Appendix, we detail how the uncertainties in the data are assessed and propagated to the power spectra used in Sects.~\ref{sec:power_spectra} 
and \ref{sec:dust_sed}.
End-to-end (E2E) simulations are used throughout the paper
in order to check for potential biases in our data analysis procedure and evaluate uncertainties in our results. 

\subsection{Maps of data uncertainties}
\label{subsec:validation}
%============================================================

We use E2E simulations to build uncertainty maps for the \PR\ \HFI\ polarization data, which include both residual systematics and instrumental noise. 
Simulations of \Planck\ timelines are computed by combining the scanning strategy with models of known systematic effects and models of the sky emission (Appendix~\ref{appendix:data_simulations}).
The systematic effects included in the simulations are described in \citet{planck2016-l03}. One single sky model, referred to as the FFP10 sky model in this paper,  is used.
For all of the sky components except dust polarization, in particular polarized synchrotron emission, we used the latest version of the \Planck\ sky model described in
\citet{planck2014-a14}.  

The dust model maps are built as follows. 
The Stokes $I$ map at 353\,GHz is the dust total intensity \Planck\ map obtained by applying the Generalized Needlet Internal Linear Combination (GNILC) method of \citet{Remazeilles11} to the 2015 release of \Planck\ 
\HFI\ maps (PR2), as described in \citet{planck2016-XLVIII}, and subtracting the monopole of the cosmic infrared background \citep{planck2014-a09}.
For the the Stokes $Q$ and $U$ maps at 353\,GHz, we started with one realization of the statistical model of \citet{Vansyngel16}.
The portions of the simulated Stokes $Q$ and $U$ maps near Galactic plane were replaced by the \Planck\ 353\,GHz \PR\ data. The transition between data and the simulations was made using a Galactic mask with a 
$5^\circ$ apodisation,\footnote{From the set of \Planck\ common Galactic masks available in the \Planck\ Explanatory Supplement (\url{http://wiki.cosmos.esa.int/planckpla2015/index.php/Frequency\_Maps}).} which leaves 68\,\% of the sky unmasked at high latitude.  Furthermore, on the full sky, the large angular scales in the simulated Stokes $Q$ and $U$ maps were replaced by the \Planck\ data. 
Specifically, the first ten multipoles were the \Planck\ data, while over $\ell=10$ to $\ell=20$ the simulations were introduced smoothly using the function $\left(1+\sin\left[\pi\left(15-\ell\right)/10\right]\right)/2$. 
The resulting 353\,GHz Stokes maps were scaled to other frequencies using the maps of dust temperature and spectral index, coming from fitting
the SED of dust total intensity in \citet{planck2016-XLVIII} and \citet{planck2013-p06b}, respectively, which introduces significant spectral variations as discussed in Sect.~\ref{subsec:multi_freq_deco}.
Hereafter, we refer to these simulated maps as the FFP10 dust polarization maps.

Independent realizations of the detector noise and data systematics are computed for each simulation of the \HFI\ timelines, 
keeping the sky emission components (including the CMB) the same.  The mapmaking pipeline is run on the simulated timelines to produce simulated maps. This process is repeated to obtain
300 realizations of simulated maps at each of the four \HFI\ polarized channels, at 100, 143, 217, and 353$\,$GHz.
We compute difference maps by subtracting the input sky model from each of the simulated maps.
The 300 difference maps are independent E2E realizations of \HFI\ uncertainty maps, including both detector noise and systematic effects. 

\begin{figure}[!htbp]
\includegraphics[width=0.49\textwidth]{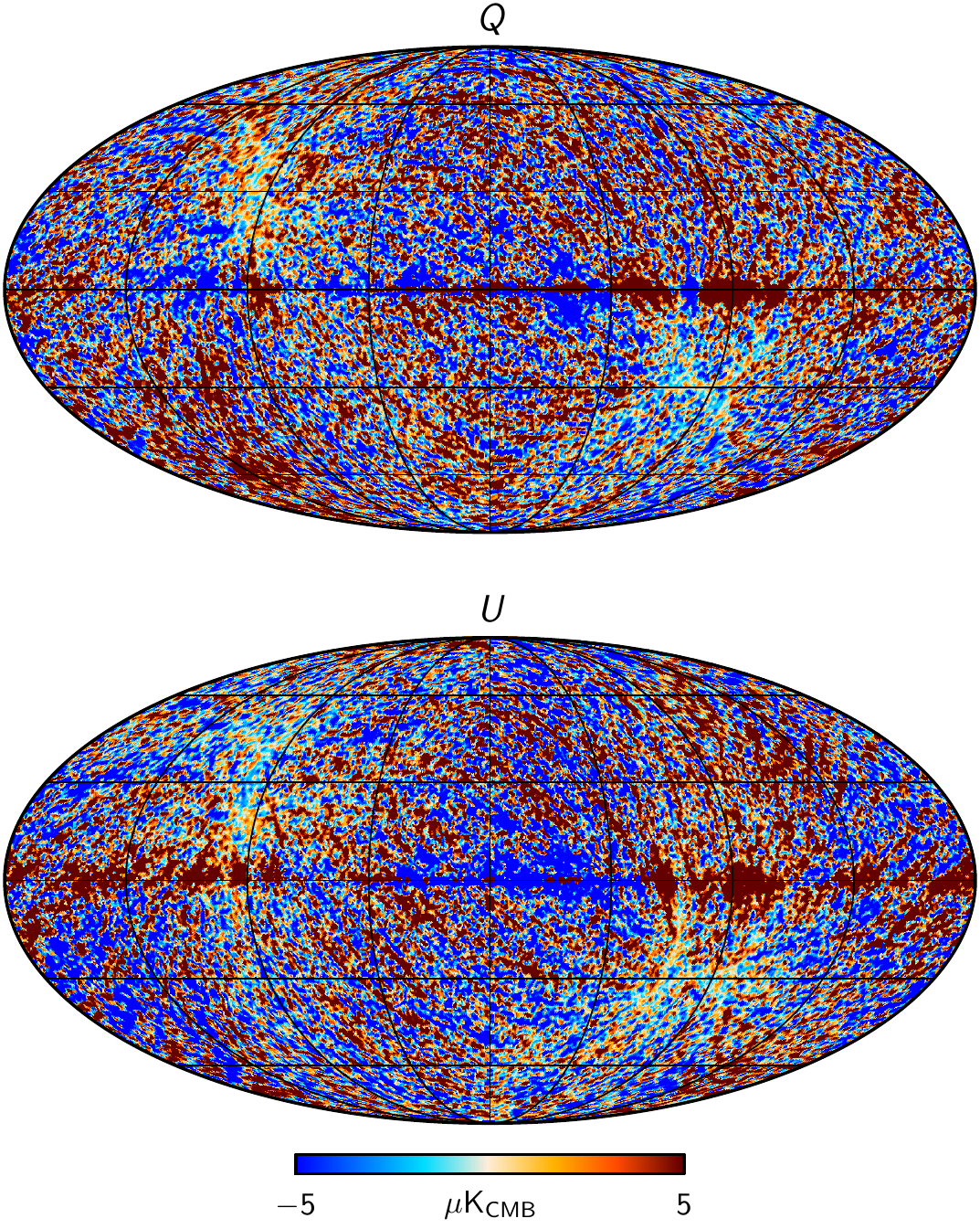}
\caption[]{Difference between the Stokes $Q$ and $U$ output maps and the sky model inputs at $1^\circ$ resolution, for one E2E simulation at $353\,$GHz. Such pairs 
of difference maps are used to quantify statistical and systematic uncertainties in our analysis of \Planck\ \HFI\ data. }
\label{fig:difmaps}
\end{figure}

\begin{figure*}[!htbp]
\begin{tabular}{c}
\includegraphics[width=1.\textwidth]{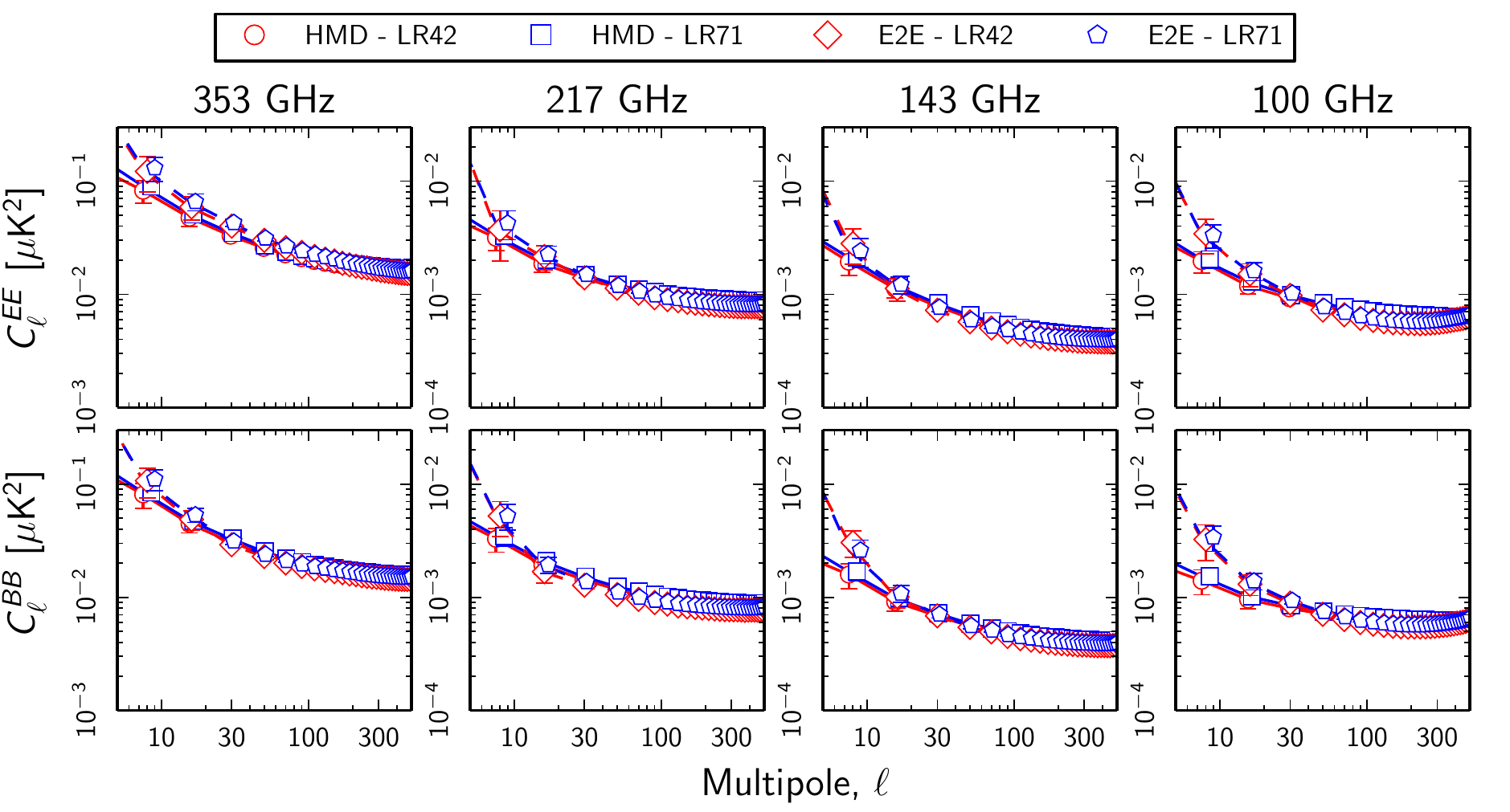}
\end{tabular}
\caption[]{
$EE$ (left) and $BB$ (right) power spectra from the \HFI\ uncertainty maps described in Appendix~\ref{subsec:validation}. The data points and error bars represent the mean and standard deviation of the power in each multipole bin computed over the 300 realizations for the LR42 (red) and LR71 (blue) sky regions. 
The dashed lines represent analytical fits (a power law plus a constant) to the data points. 
For comparison, the plots also present the spectra of the difference between half-mission maps (solid lines, labelled ``HMD''). }
\label{fig:E2E_noise_spectra}
\end{figure*}

\begin{figure*}[!htbp]
\begin{tabular}{c}
\includegraphics[width=1.\textwidth]{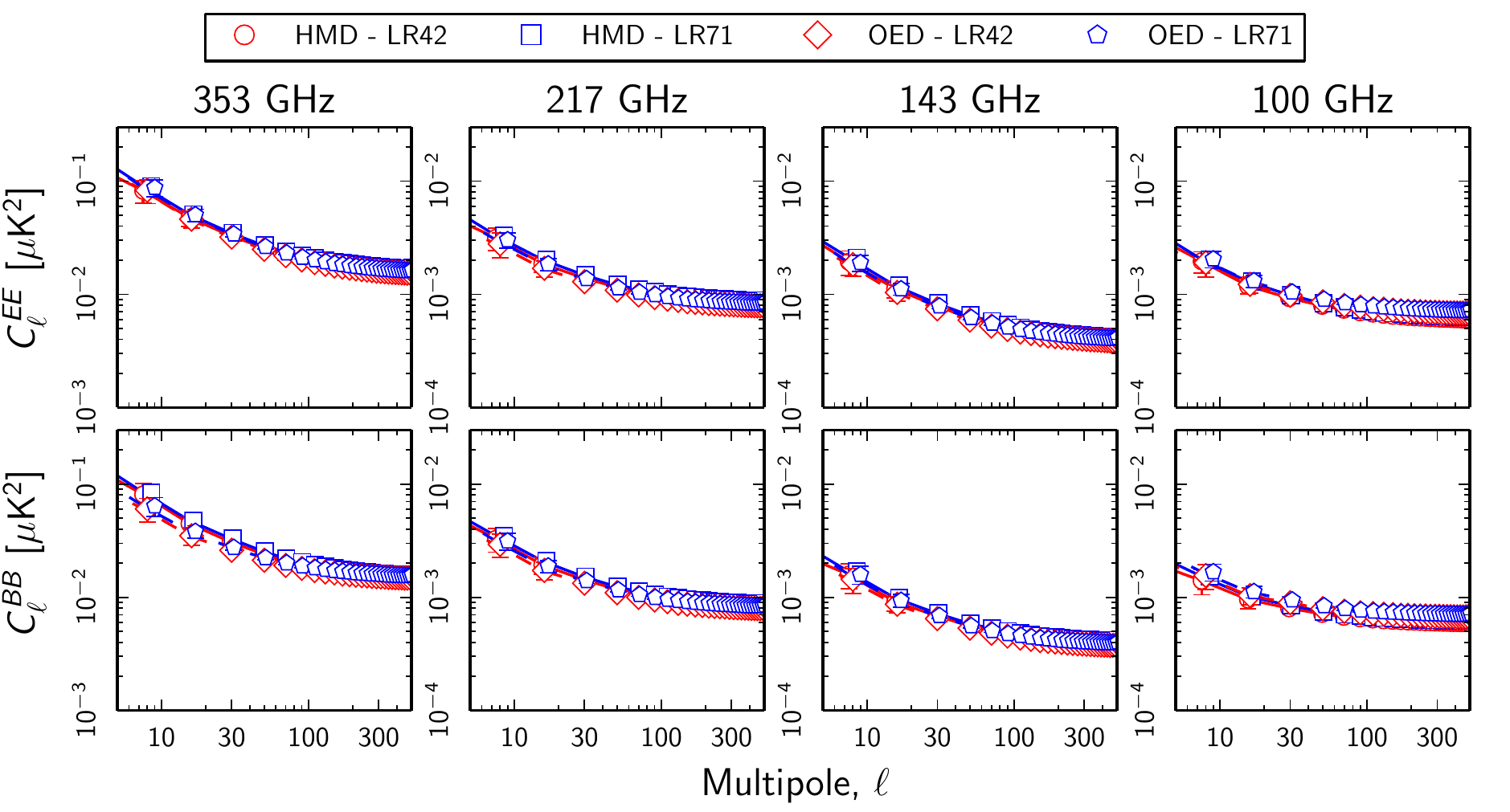} 
\end{tabular}
\caption[]{
Like Fig.~\ref{fig:E2E_noise_spectra}, but comparing 
$EE$ and $BB$ power spectra of the data noise estimated from the half-differences between odd and even survey maps (OED data points and dashed lines) with that estimated from the half-mission maps (solid lines and HMD data points, as in Fig.~\ref{fig:E2E_noise_spectra}).}
\label{fig:HM_OE_noise_spectra}
\end{figure*}

For illustration, one difference map obtained for one random realization of the E2E simulations at 353$\,$GHz is presented in Fig.~\ref{fig:difmaps}.
Figure~\ref{fig:E2E_noise_spectra} presents the mean $EE$ and $BB$ power spectra of the residual maps, computed over the set of 300 simulations. 
These spectra are compared with those computed with the half-difference of the two half-mission \PR\ maps, which quantify the instrumental noise in the data. 
The noise spectra computed from the half-differences between half-mission and odd-even survey maps are very similar, as shown in Fig.~\ref{fig:HM_OE_noise_spectra}. 
The excess power in the E2E spectra at low multipoles corresponds to residual uncorrected systematics, which are taken into account in the \HFI\ E2E uncertainty maps.

We use \LFI\ and \WMAP\ maps to separate dust and synchrotron polarized emission and quantify the
correlation between the two sources of emission. Because E2E simulations are not available for these data, we compute maps of uncertainties from Gaussian realizations of the data noise.
Power spectra of this noise are derived from the half-difference of half-mission \Planck\ \LFI\ maps 
and the difference of year maps
for \WMAP. We note that it is easy to produce a large number (1000 or more) of data realizations with Gaussian data noise,
while only 300 E2E realizations are available for \HFI.

\subsection{E2E simulated maps}
\label{appendix:data_simulations}

The uncertainty maps for all four \HFI\ frequencies described in the previous section are combined with a simple model of dust and synchrotron polarized emission to produce what we call
E2E simulated maps.  We use these to quantify error bars on the power spectra and validate our data analysis. The same single set of 300 E2E simulated maps is used for all sky regions throughout the paper. 

A single sky model is used for all simulations. Stokes maps of the dust polarization at 353\,GHz are computed using the model of \citet{Vansyngel16}. These model maps are 
scaled to the other frequencies with a single SED for all sky pixels, namely an MBB emission law with a spectral index of 1.59 and a temperature of 19.6\,K based on the \Planck\ 
data analysis in \citet{planck2013-XVII,planck2014-XXII}. This model, unlike the FFP10 dust polarization maps (Appendix~\ref{subsec:validation}), has no spectral variations.
The spatial template for the synchrotron polarized emission is derived from the \Planck\ sky model, as in \citet{planck2014-a14}. 
The SED of the synchrotron polarized emission is a power law with a spectral index of $-3.0$ for all sky pixels. Independent realizations of the CMB polarization maps, computed
from the \Planck\ 2015 \LCDM\ model \citep{planck2014-a15}, are added to the Stokes maps of the dust and synchrotron emission. 
By combining the uncertainty maps from
the E2E simulations (Appendix~\ref{subsec:validation}) with Galactic polarization maps
and CMB realizations distinct from those used in the simulations (we note that the Stokes I dust map is unchanged), 
we erase potential correlations between data systematics and the 
polarized sky emission. Such correlations have been shown to have negligible impact on the CMB data analysis \citep{planck2014-a10,planck2016-l03}. We also checked 
that the correlation between dust polarization and residual systematics is not a dominant uncertainty at $353\,$GHz. 

\subsection{Uncertainties propagated to power spectra in the data analysis}
\label{appendix:uncertainties_power_spectra}

\begin{figure}[!htbp]
\includegraphics[width=0.49\textwidth]{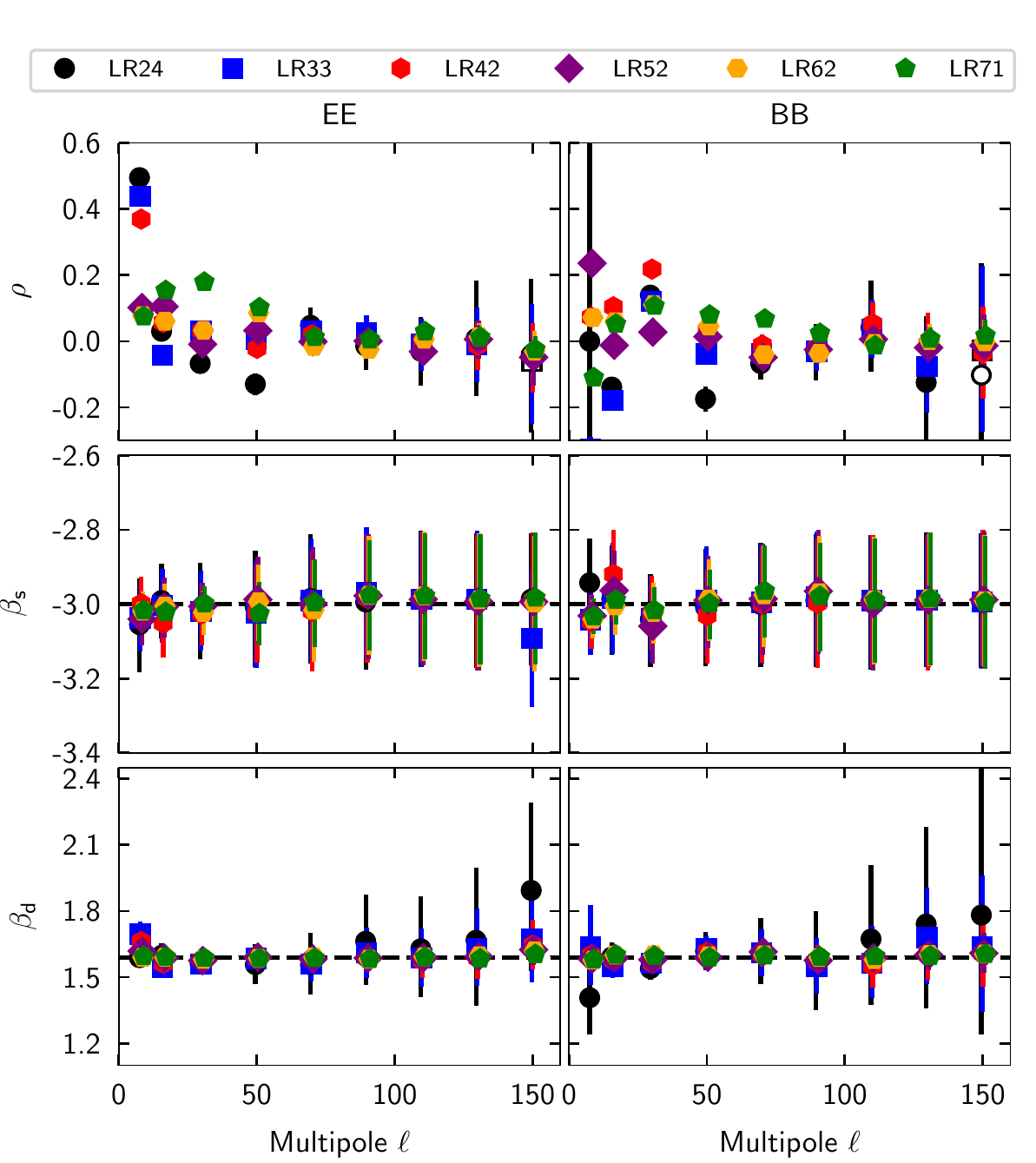}
\caption[]{Results of the spectral fit to simulated data with uncertainties derived from the E2E simulated maps. Distinct symbols and colours represent the different sky regions (see top row).
The dashed lines in plots of $\beta_{\rm s}$ and $\beta_{\rm d}$ are the input values of our sky model. We retrieve these input values without any bias. 
The sky model has a non-zero correlation between dust and synchrotron polarization at low $\ell$, which is consistent with the value of $\rho$ found by fitting the simulated data.}
\label{fig:E2E_sims}
\end{figure}

We compute $EE$ and $BB$ power spectra of a number of E2E simulated maps, which combine the sky model (including CMB polarization)
with independent realizations of the statistical Gaussian noise for \LFI\ and \WMAP\ or the 
E2E uncertainty maps for \HFI. The spectra are computed with \XPol\ \citep{Tristram05} 
for the same multipole bins and sky regions used for the data analysis. 
The CMB contribution is subtracted from the power spectra using the \Planck\ 2015 \LCDM\ model \citep{planck2014-a15}.

The dispersion of values computed over a set of 300 realizations define the statistical uncertainties that we adopt for the corresponding $\mathcal{C}_\ell$ coefficient measured
on the data maps. Because each set of Stokes $I$, $Q$, and $U$ E2E simulated maps includes a different realization of the CMB polarization, the uncertainties include the cosmic
variance of the CMB. These uncertainties are used in the power-law fits in Sect.~\ref{subsec:power_law_fits} and the fits to the 
spectral model in Sect.~\ref{subsec:spectral_model}. 

The data simulations are also used to check for potential biases in the fit of the spectral model. For example, the results presented in 
Fig.~\ref{fig:E2E_sims} for the parameters of the spectral fit to the E2E simulated \HFI\ maps show that for all multipole bins
we recover the input spectral indices of dust and synchrotron polarized emission ($\beta_{\rm d}$ and $\beta_{\rm s}$) 
of the sky model without any bias. We point out that this validation has been done using the least-squares {\tt MPFIT} fitting routine and not the more computationally-intensive MCMC code used in Sect.~\ref{subsec:spectral_model} to fit the data. 
We have checked that the two methods produce consistent determinations of the model parameters when applied to the same data.

\section{Two-frequency analysis of the spectral correlation ratio between 217 and 353\,GHz}
\label{appendix:pdf_Rell}

PL computed the spectral correlation ratio $\mathcal{R}_{\ell}^{BB}$, defined as
\begin{equation}
\mathcal{R}_{\ell}^{BB} \equiv \frac{\mathcal{C}^{BB}_{\ell} (217\times353)}
{\sqrt{\mathcal{C}^{BB}_{\ell}(353\times353) \,
 \mathcal{C}^{BB}_{\ell} (217\times217)}} \, .
\label{eq:correlation_ratio}
\end{equation}
If the $B$-mode emission is perfectly 
correlated between the two frequencies, then $\mathcal{R}_{\ell}^{BB}= 1$, whereas a value lower than 1 is indicative of a correlation that is only partial. PL interpreted their results as evidence for decorrelation
and spatial variations of dust polarization 
between 217 and $353\,$GHz, over multipoles relevant to the search for 
primordial $B$ modes at the recombination peak.
In the course of our analysis of the \PR\ data we found that this conclusion was compromized as described below, so that the significance of the decorrelation was overstated.
\citet{Sheehy17} discovered this independently.

In this Appendix, using the new \Planck\ maps, we update and extend the PL analysis (Sect.~\ref{subsec:correlation_ratio}).  The E2E simulations are used to assess the uncertainty of $\mathcal{R}_{\ell}^{BB}$ and the statistical significance of the results (Sect.~\ref{subsec:correlation_statistics}).  We 
note that $\mathcal{R}_{\ell}^{BB}$ does not depend on the 1.5\,\% uncertainty on the $353\,$GHz polarization efficiency.

\subsection{Measured ratios on \PR\ polarizations maps}
\label{subsec:correlation_ratio}

Here, we compute 
$\mathcal{R}_{\ell}^{BB}$ using the \PR\ \Planck\ data for five broad ranges of multipoles, namely $\ell = 4$--11, 11--50, 50--160, 160--320,
and 320--500. The three last $\ell$ bins are common to the analysis reported by PL using the PR2 \Planck\ data.  The lowest two $\ell$ bins are new.

\begin{figure*}[!htbp]
\begin{tabular}{c}
\includegraphics[width=0.95\textwidth]{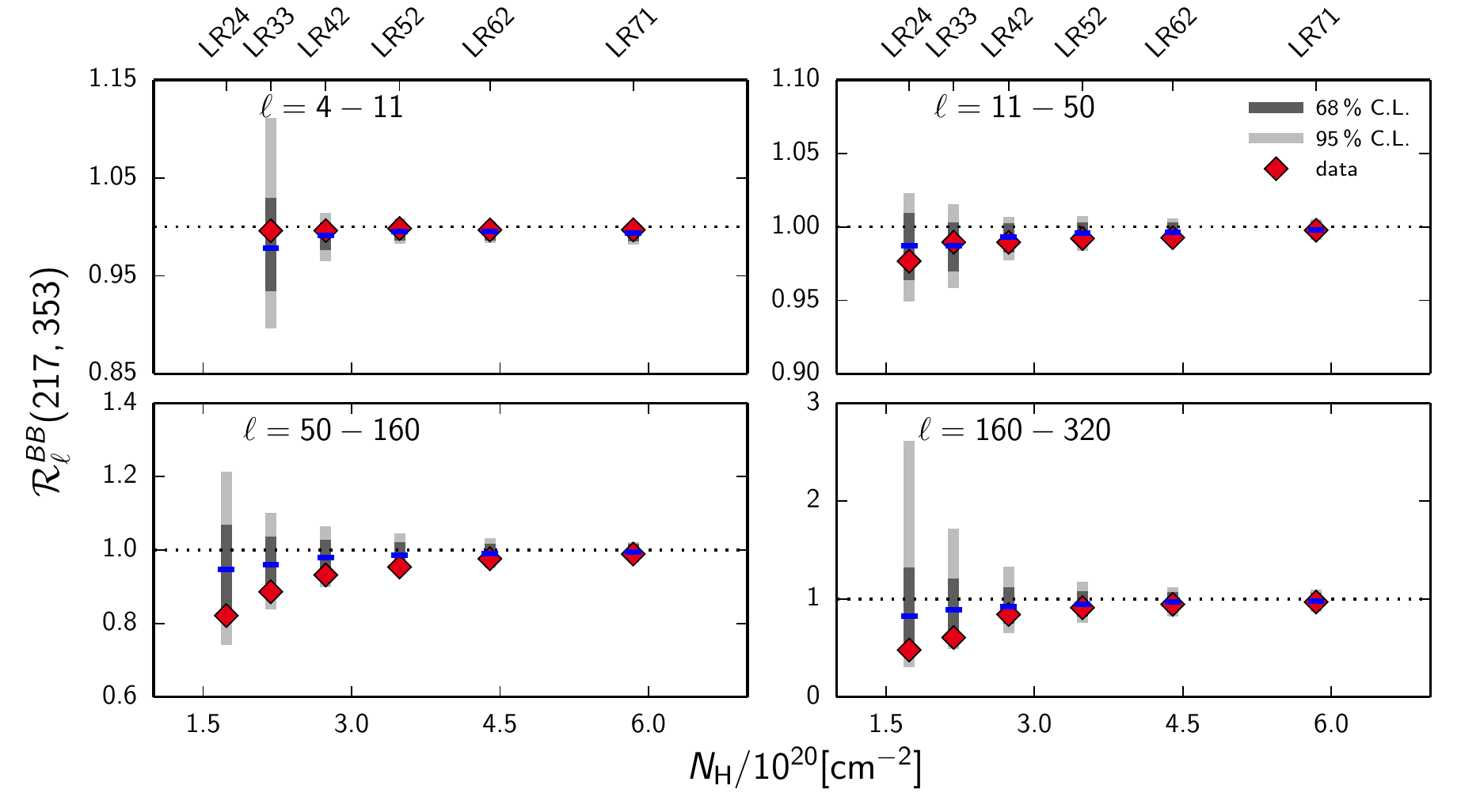}
\end{tabular}
\caption[]{Spectral correlation ratio $\mathcal{R}_{\ell}^{BB}$ versus 
mean hydrogen column density for the six sky regions. Each panel presents the results 
for one of the $\ell$ bins, $\ell = 4$--11, 11--50, 50--160, and 160--320. %from the upper left to the bottom right.
The dark and light grey bars represent the 68\,\% and 95\,\% confidence intervals computed over the 300 E2E simulations, while the blue segments mark the median values. 
The bottom left panel for the $\ell=50$--160 bin is directly comparable to the corresponding plot from the PL analysis, their figure~3. }
\label{fig:Rell_vs_NH}
\end{figure*}

\begin{figure}[!htbp]
\includegraphics[width=0.48\textwidth]{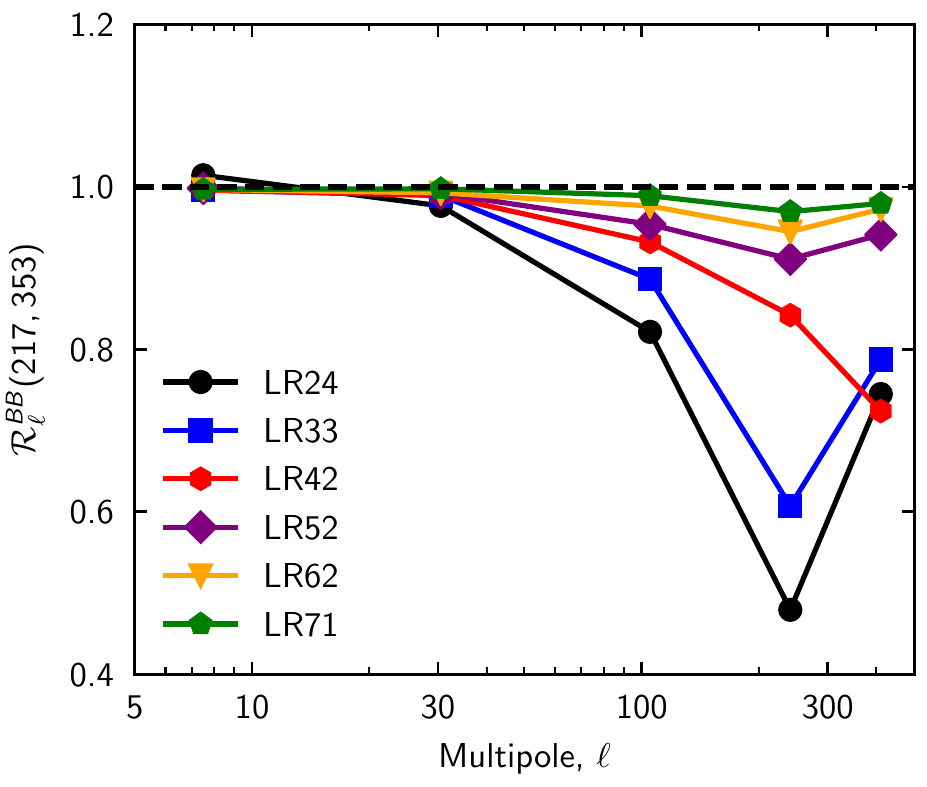}
\caption[]{Spectral correlation ratio $\mathcal{R}_{\ell}^{BB}$ versus multipole for the six sky regions. 
Like in PL, we find that the apparent decorrelation of the dust polarization measured by the difference $1-\mathcal{R}_{\ell}^{BB}$ 
increases for decreasing $\feff$, that is towards low-brightness regions at high Galactic latitude.}
\label{fig:decorrelation_Rll}
\end{figure}

The values of $\mathcal{R}_{\ell}^{BB}$ are listed for the six sky regions, LR24 to LR71, in Table~\ref{tab:Rell_LR}. 
In Fig.~\ref{fig:Rell_vs_NH}, the ratios measured for the first four $\ell$ bins are plotted versus the mean hydrogen column density, $N_{\rm H}$.
The bottom left panel, for the bin $\ell=50$--160, is directly comparable to the corresponding plot from the PL analysis of the PR2 \Planck\ data in their figure~3. 
For this common $\ell$ bin, we find results consistent with the earlier analysis in PL, that the departure from 1 (i.e. the apparent spectral decorrelation) increases with decreasing column density. 
As in PL, we also find that the apparent spectral decorrelation increases with increasing multipole, as illustrated in Fig.~\ref{fig:decorrelation_Rll}. However, the two new lowest $\ell$ bins
show very little decorrelation.

% versus NH

% histograms
\begin{figure*}[!htbp]
\includegraphics[width=0.88\textwidth]{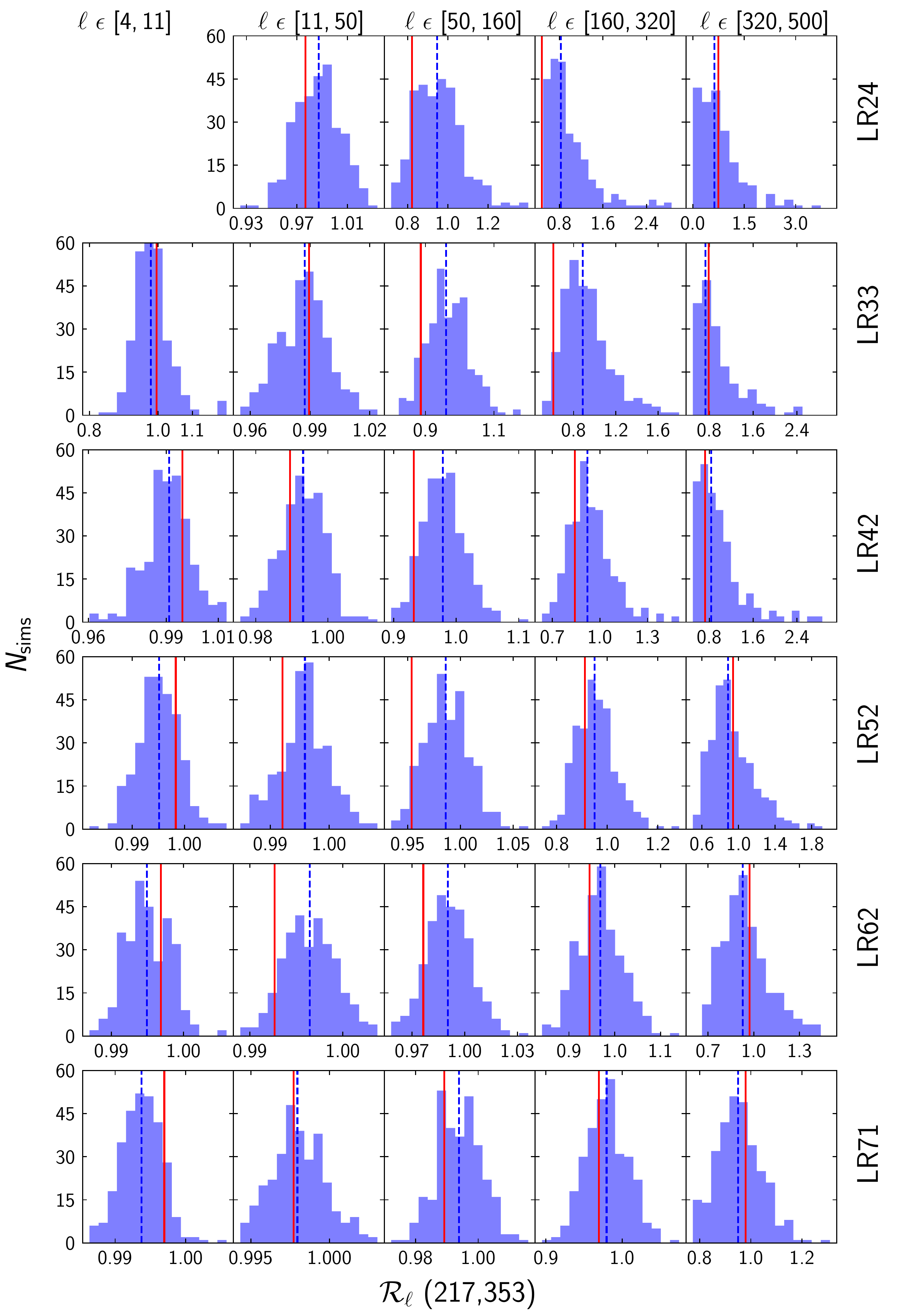}
\caption{Distribution of the correlation ratios $\mathcal{R}_{\ell}^{BB}$ on the six sky regions for each of the five $\ell$ bins. The histograms are 
computed from the 300 E2E simulations using half-mission data splits. Histograms computed using odd-even surveys give similar results.
The values derived from the data are displayed as vertical lines. The dashed lines represent the median value for the simulations.}
\label{fig:Rell_E2E_all_bins}
\end{figure*}

\begin{figure}[!htbp]
\includegraphics[width=0.49\textwidth]{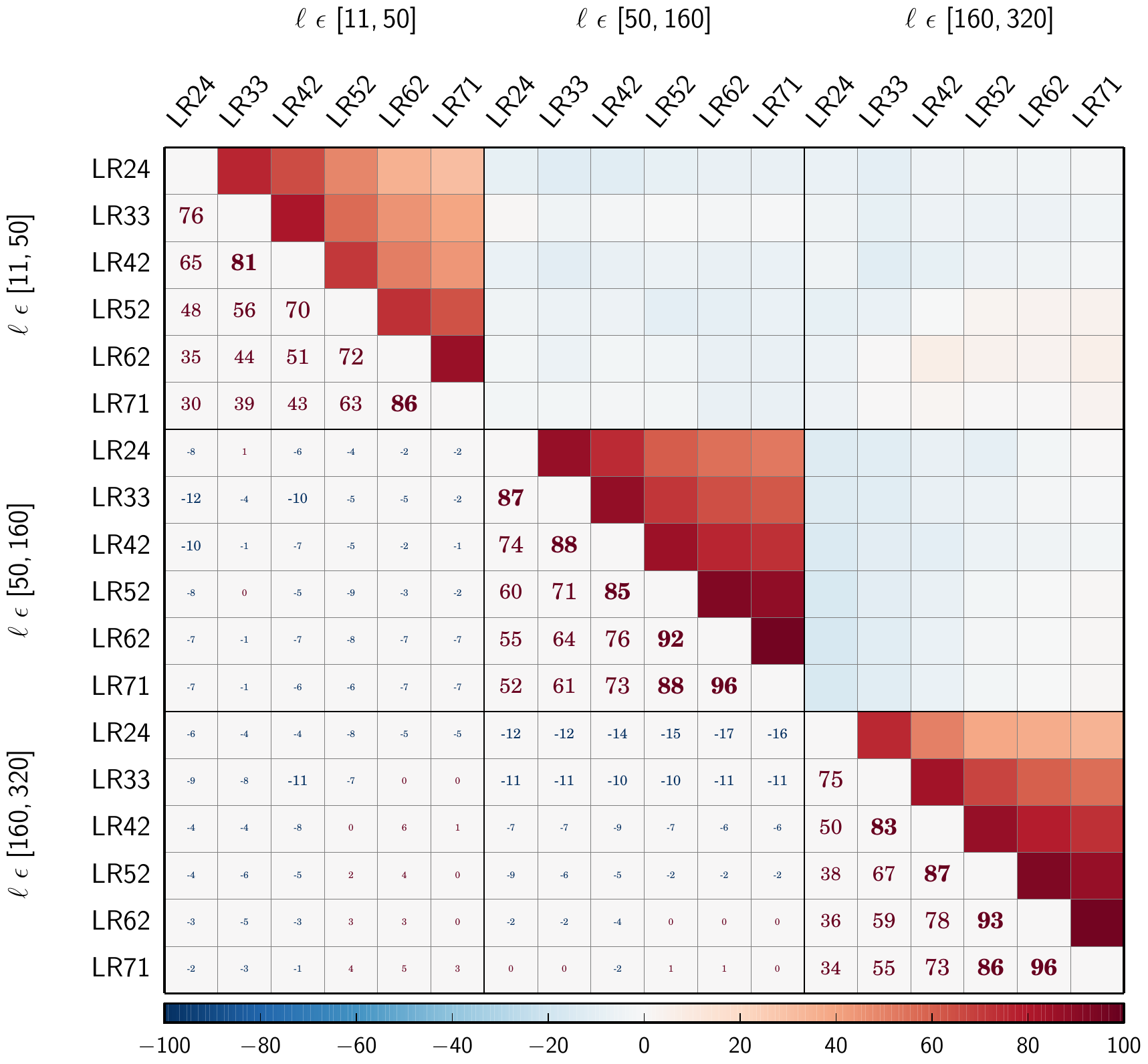}
\caption[]{$\mathcal{R}_{\ell}^{BB}$ cross-correlation factor between sky regions for three $\ell$ bins, expressed as percentages.}
\label{fig:correlation_LR}
\end{figure}

\subsection{Statistical significance of the \Planck\ results}
\label{subsec:correlation_statistics}

%%%%%%%%%%%%%%% Histograms %%%%%%%%%%
We have used the E2E simulated \HFI\ maps described in Appendix~\ref{appendix:data_simulations} to compute probability distribution functions of the $\mathcal{R}_{\ell}^{BB}$ ratio and thereby
quantify the statistical significance of the values found from the \Planck\ data. 
The histograms are shown for the five bins, $\ell =4$--11, 11--50, 50--160, 160--320, and 320--500, in Fig.~\ref{fig:Rell_E2E_all_bins}.
The dust polarization model used in the simulations has the same SED for each sky pixel, i.e. a perfect correlation. 
The plots in Fig.~\ref{fig:Rell_vs_NH} show the 
68\,\% and 95\,\% confidence intervals 
for the values measured on the 300 E2E simulations. 
The shift of the median values, marked by the dashed lines from $\mathcal{R}_{\ell}^{BB}=1$, is a bias due to data uncertainties. The bias and the width of the distribution are both larger in this analysis (based on the E2E simulations) than in PL (based on uncertainty maps derived from Gaussian noise realizations). These differences result from the fact that E2E simulations take into account 
uncorrected systematics, ignored in Gaussian noise realizations.
For our two lowest $\ell$ bins, 
the data values are within the 
68\,\% interval of the simulation results. 
For the $\ell = 50$--160 and 160--320 bins, the values measured on the \Planck\ data are intermediate between 
the  68\,\% and 98\,\% intervals of 
the simulation results, i.e. between the 1 and $2\,\sigma$ limits for a Gaussian distribution.\footnote{We note, however, that the distributions are not Gaussian.}

The E2E simulations show that for a given $\ell$ bin the $\mathcal{R}_{\ell}^{BB}$ values are highly correlated over sky regions. 
The measured cross-correlation ratios between regions are displayed in Fig.~\ref{fig:correlation_LR} for all our $\ell$ bins. Values range from 50\,\% to more than 90\,\%.
This correlation follows from the fact that the sky regions we use are nested.
The data points in Fig.~\ref{fig:Rell_vs_NH} are not independent, a fact also identified by \citet{Sheehy17}. % but missed in the PL analysis.

The $\mathcal{R}_{\ell}^{BB}$ values in independent $\ell$ bins are uncorrelated, but noise introduces a systematic trend, where $\mathcal{R}_{\ell}^{BB}$ decreases for 
increasing multipole, which accounts for the systematic trend in Fig.~\ref{fig:decorrelation_Rll}.

\begin{figure*}[!htbp]
\begin{tabular}{c}
\includegraphics[width=0.95\textwidth]{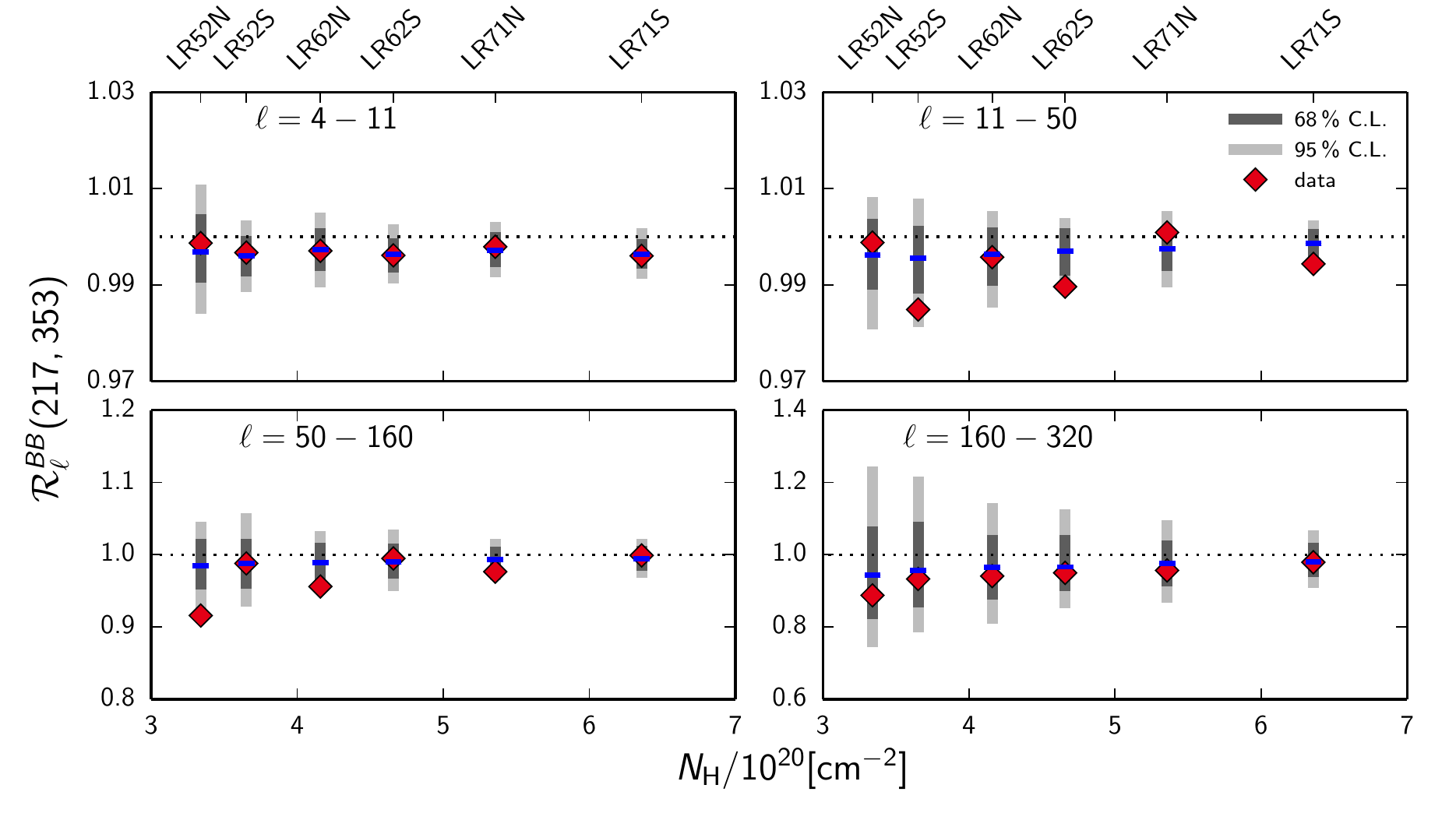}
\end{tabular}
\caption[]{Spectral correlation ratio $\mathcal{R}_{\ell}^{BB}$ versus 
the mean hydrogen column density for northern and southern splits of three large sky regions. 
The dark and light grey bars represent the 68\,\% and 95\,\% intervals computed over the 300 E2E simulations, while the blue segments mark the median values. }
\label{fig:Rell_NS_splits}
\end{figure*}

\begin{figure}[!htbp]
\includegraphics[width=0.49\textwidth]{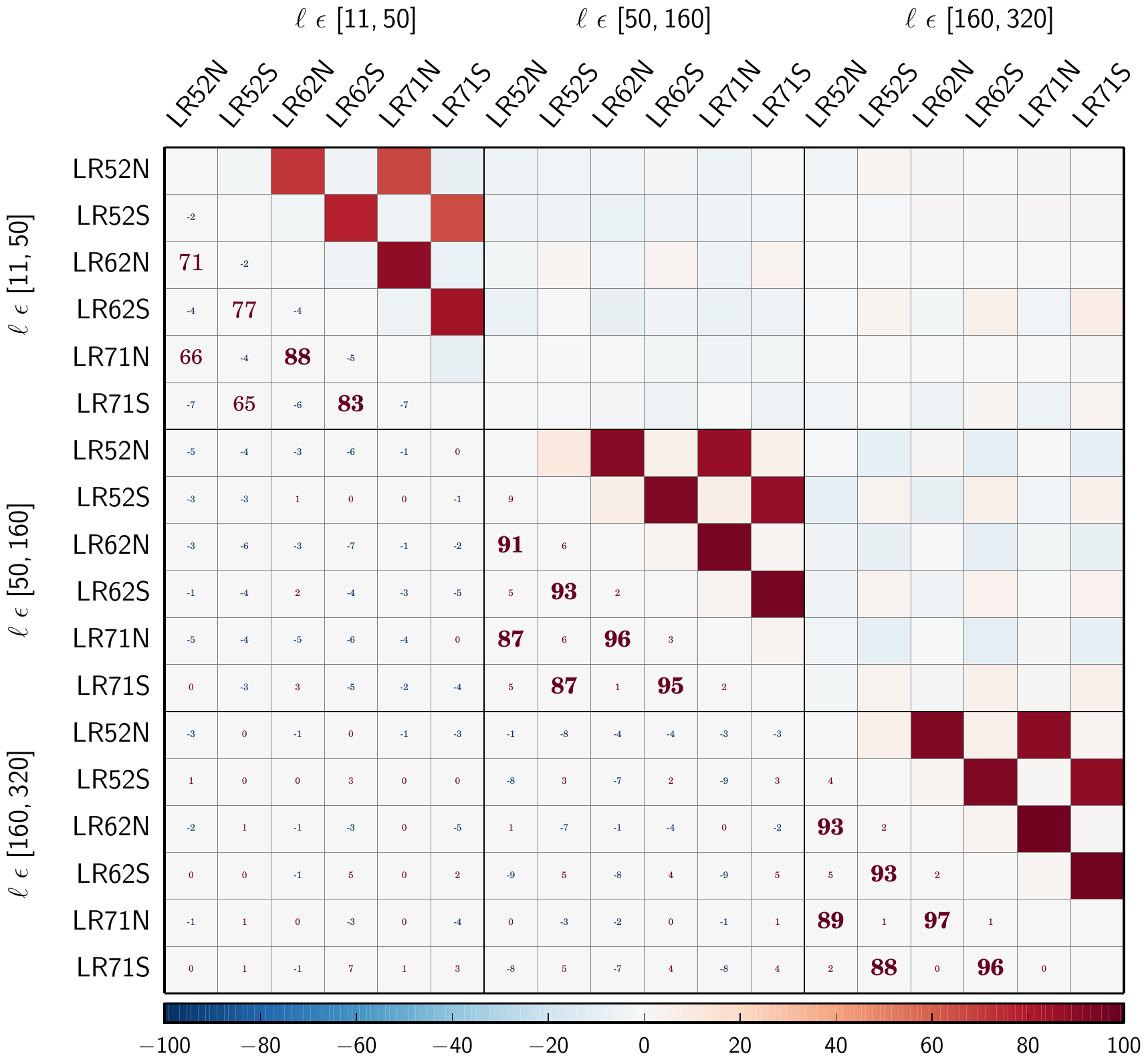}
\caption[]{$\mathcal{R}_{\ell}^{BB}$ cross-correlation measured for the north-south splits of the L52, L62, and L71 sky regions for three $\ell$ bins, expressed as percentages.}
\label{fig:correlation_NS_splits}
\end{figure}

To illustrate the correlation between nested regions, we have measured $\mathcal{R}_{\ell}^{BB}$ 
on two independent sets of sky regions, corresponding to the northern and southern Galactic parts of the LR52, LR62, and LR71 regions. 
The results of this analysis are presented in Fig.~\ref{fig:Rell_NS_splits}. The cross-correlation factors between these north-south splits,
computed from the E2E simulations, are plotted in Fig.~\ref{fig:correlation_NS_splits}. The data points within a given set (north or south) are correlated, but the two sets are independent.
For example, for the $\ell=50$--160 bin, the values of $\mathcal{R}_{\ell}^{BB}$ for the northern regions are consistently lower than those for the corresponding southern regions. 

In Table~\ref{tab:Rell_LR}, for each $\ell$ bin and sky region we list the probability, labeled PTS, 
to obtain correlation ratios smaller than the \Planck\ measurements, based on the 300 E2E simulations. The lower the value of the PTS, the more significant is the measurement.
Table~\ref{tab:Rell_splits} lists the same probability for the north-south splits. 

As in \citet{Sheehy17}, we conclude that our \Planck\ spectral correlation analysis of the 217- and 353-GHz polarization maps 
does not provide statistically significant evidence for a departure of the spectral correlation ratios from a perfect correlation (i.e. a ratio of 1).
Lower limits are listed for our set of sky regions and $\ell$ bins in Table~\ref{tab:Rell(BB)_limits}. These limits 
are from the 2.5th percentile of the 300 E2E simulations
(relating to the 
%correspond to
95\,\% confidence interval), i.e. the values of $\mathcal{R}_{\ell}^{BB}$ are smaller than these limits for only 7.5 realizations out of the 300. 
The values of $\mathcal{R}_{\ell}^{BB}$ for the $\ell$ bin 50--160 presented in figure~3 of PL are within these limits. 

\begin{table}[tbp!]
\newdimen\tblskip \tblskip=5pt
\caption{Lower limits from the two-frequency analysis of the spectral correlation ratio
$\mathcal{R}_{\ell}^{BB} (217, 353)$, from the 2.5th percentile of the E2E simulations.}
\label{tab:Rell(BB)_limits}
\vskip -5mm
%\footnotesize
\setbox\tablebox=\vbox{
 \newdimen\digitwidth
 \setbox0=\hbox{\rm 0}
 \digitwidth=\wd0
 \catcode`*=\active
 \def*{\kern\digitwidth}
 \newdimen\signwidth
 \setbox0=\hbox{+}
 \signwidth=\wd0
 \catcode`!=\active
 \def!{\kern\signwidth}
  \newdimen\dpwidth
  \setbox0=\hbox{.}
  \dpwidth=\wd0
  \catcode`?=\active
  \def?{\kern\dpwidth}
\halign{\tabskip 0pt\hbox to 2.5cm{#\leaderfil}\tabskip 0.75em&
\hfil#\hfil\tabskip 1.0em& \hfil#\hfil& \hfil#\hfil&
\hfil#\hfil& \hfil#\hfil& \hfil#\hfil\tabskip 0em\cr
\noalign{\doubleline}
\omit\hfil$\ell$ range\hfil& LR24& LR33& LR42& LR52& LR62& LR71\cr 
\noalign{\vskip 3pt\hrule\vskip 5pt}
**4--*11& \dots& 0.902& 0.971& 0.988& 0.989& 0.988\cr
*11--*50& 0.953& 0.962& 0.981& 0.987& 0.991& 0.995\cr
*50--160& 0.756& 0.854& 0.913& 0.949& 0.965& 0.980\cr
160--320& 0.361& 0.550& 0.710& 0.815& 0.878& 0.926\cr
320--500& \dots& \dots& 0.432& 0.585& 0.699& 0.786\cr
\noalign{\vskip 3pt\hrule\vskip 5pt}}}
\endPlancktable
\end{table}

\section{Electronic tables}
\label{appendix:tables}

In this Appendix we provide large tables to be made available as data files by the journal.
%%%%%%%%%%%%%%%%%%%%%%%%%%%%%%%%%
%%%%%%%%%%%%%%%%%%%%%%%%%%%%%%%%%
\begin{table*}[tb]
\begingroup
\newdimen\tblskip \tblskip=5pt
\caption{Dust polarization amplitudes for six different Galactic regions at 353\,GHz. The CMB amplitude has been subtracted from the power spectrum level. The $1\,\sigma$ error bars are derived from the E2E simulations. They do not include the 1.5\,\% uncertainty on the 353\,GHz polarization efficiency. }
\label{tab:dust_spectra} 
\nointerlineskip
\vskip -3mm
%\footnotesize
\setbox\tablebox=\vbox{
   \newdimen\digitwidth 
   \setbox0=\hbox{\rm 0} 
   \digitwidth=\wd0 
   \catcode`*=\active 
   \def*{\kern\digitwidth}
 \newdimen\signwidth
 \setbox0=\hbox{+}
 \signwidth=\wd0
 \catcode`!=\active
 \def!{\kern\signwidth}
  \newdimen\dpwidth
  \setbox0=\hbox{.}
  \dpwidth=\wd0
  \catcode`?=\active
  \def?{\kern\dpwidth}
\halign{
\hbox to 2.5cm{#\leaderfil} \tabskip 0.1em&
\hfil #\hfil\tabskip 0.5em&
\hfil #\hfil\tabskip 1.0em&
\hfil #\hfil&
\hfil #\hfil&
\hfil #\hfil\tabskip 0.5em&
\hfil #\hfil\tabskip=0pt\cr
\noalign{\doubleline}
\omit& LR24& LR33& LR42& LR52& LR62& LR71\cr
\omit\hfil$\ell_{\rm bin}$\hfil& \multispan{6} \hfil$\mathcal{D}_{\ell}^{EE}$ \hfil\cr
\noalign{\vskip -3pt}
\omit& \multispan6\hrulefill\cr
%\noalign{\doubleline}
\noalign{\vskip 3pt\hrule\vskip 3pt}
 **2--**3& *!\dots& !\dots& *\dots& *\dots& *\dots& $2527\pm69$\cr
 **4--*11& $!*43\pm*6$& $!99\pm*6$& $199\pm*7$& $371\pm*9$& $697\pm10$& $*986\pm10$\cr
 *12--*19& $!101\pm*5$& $!91\pm*5$& $140\pm*5$& $260\pm*6$& $371\pm*6$& $*739\pm*8$\cr
 *20--*39& $!*55\pm*3$& $!78\pm*3$& $133\pm*3$& $207\pm*4$& $327\pm*4$& $*543\pm*5$\cr
 *40--*59& $!*39\pm*3$& $!53\pm*3$& $*80\pm*3$& $143\pm*4$& $227\pm*5$& $*353\pm*5$\cr
 *60--*79& $!*34\pm*3$& $!47\pm*3$& $*70\pm*4$& $110\pm*4$& $181\pm*4$& $*307\pm*5$\cr
 *80--*99& $!*33\pm*4$& $!42\pm*4$& $*73\pm*4$& $116\pm*4$& $191\pm*4$& $*310\pm*5$\cr
 100--119& $!*34\pm*4$& $!45\pm*4$& $*73\pm*4$& $114\pm*4$& $174\pm*5$& $*285\pm*5$\cr
 120--139& $!*25\pm*5$& $!37\pm*4$& $*58\pm*4$& $*91\pm*5$& $156\pm*6$& $*280\pm*6$\cr
 140--159& $!*32\pm*6$& $!41\pm*5$& $*62\pm*5$& $*93\pm*5$& $145\pm*5$& $*240\pm*6$\cr
 160--179& $!*24\pm*6$& $!35\pm*6$& $*59\pm*6$& $*88\pm*6$& $136\pm*6$& $*218\pm*6$\cr
 180--199& $!*30\pm*7$& $!40\pm*6$& $*51\pm*6$& $*79\pm*7$& $119\pm*7$& $*206\pm*7$\cr
 200--219& $!*34\pm*8$& $!39\pm*7$& $*61\pm*6$& $*89\pm*6$& $138\pm*6$& $*231\pm*6$\cr
 220--239& $!*28\pm*8$& $!44\pm*8$& $*62\pm*8$& $*93\pm*8$& $130\pm*8$& $*211\pm*7$\cr
 240--259& $!*37\pm*9$& $!39\pm*9$& $*60\pm*8$& $*90\pm*8$& $126\pm*8$& $*201\pm*8$\cr
 260--279& $!*15\pm10$& $!17\pm*9$& $*40\pm*9$& $*74\pm*9$& $119\pm*8$& $*197\pm*9$\cr
 280--299& $!*15\pm11$& $!19\pm*9$& $*47\pm*9$& $*71\pm*9$& $106\pm*9$& $*175\pm*9$\cr
 300--319& $**-6\pm12$& $!*8\pm11$& $*20\pm10$& $*45\pm11$& $*86\pm10$& $*150\pm10$\cr
 320--339& $!*23\pm13$& $!36\pm13$& $*47\pm12$& $*66\pm12$& $102\pm11$& $*166\pm10$\cr
 340--359& $!*19\pm14$& $!27\pm13$& $*38\pm12$& $*63\pm12$& $*98\pm12$& $*165\pm11$\cr
 360--379& $!*23\pm16$& $!28\pm14$& $*41\pm13$& $*52\pm13$& $*84\pm12$& $*143\pm12$\cr
 380--399& $!**8\pm17$& $!*6\pm15$& $*33\pm14$& $*54\pm13$& $*85\pm12$& $*143\pm11$\cr
 400--419& $!*15\pm17$& $!33\pm15$& $*47\pm15$& $*73\pm14$& $108\pm14$& $*157\pm13$\cr
 420--439& $!*26\pm19$& $!33\pm18$& $*56\pm16$& $*71\pm15$& $*98\pm14$& $*151\pm14$\cr
 440--459& $!*34\pm21$& $!41\pm20$& $*47\pm19$& $*66\pm18$& $*88\pm17$& $*136\pm16$\cr
 460--479& $!**7\pm21$& $!23\pm19$& $*41\pm17$& $*56\pm16$& $*82\pm16$& $*128\pm15$\cr
 480--499& $!*41\pm24$& $!64\pm22$& $*72\pm21$& $*84\pm20$& $101\pm19$& $*142\pm18$\cr
 500--549& $!*29\pm18$& $!28\pm16$& $*43\pm14$& $*65\pm14$& $*93\pm13$& $*142\pm12$\cr
 550--599& $!*20\pm19$& $!41\pm17$& $*52\pm16$& $*67\pm16$& $*88\pm14$& $*131\pm13$\cr
\noalign{\vskip 3pt\hrule\vskip 3pt}
\omit\hfil$\ell_{\rm bin}$\hfil& \multispan{6} \hfil$\mathcal{D}_{\ell}^{BB}$\hfil\cr
\noalign{\vskip -3pt}
\omit& \multispan6\hrulefill\cr
\noalign{\vskip 3pt\hrule\vskip 3pt}
 **2--**3& !\dots& !\dots& *\dots& *\dots& *\dots& $*247\pm17$\cr
 **4--*11& $!36\pm*4$& $!78\pm*5$& $187\pm*8$& $365\pm*8$& $357\pm*7$& $*502\pm*5$\cr
 *12--*19& $!30\pm*3$& $!44\pm*3$& $117\pm*4$& $159\pm*4$& $282\pm*5$& $*422\pm*5$\cr
 *20--*39& $!33\pm*2$& $!44\pm*2$& $*91\pm*2$& $137\pm*2$& $185\pm*3$& $*399\pm*3$\cr
 *40--*59& $!19\pm*2$& $!25\pm*2$& $*54\pm*2$& $*86\pm*3$& $138\pm*3$& $*241\pm*3$\cr
 *60--*79& $!15\pm*2$& $!20\pm*2$& $*36\pm*2$& $*69\pm*2$& $108\pm*3$& $*170\pm*3$\cr
 *80--*99& $!16\pm*2$& $!20\pm*2$& $*36\pm*2$& $*58\pm*3$& $*97\pm*3$& $*173\pm*3$\cr
 100--119& $!15\pm*3$& $!20\pm*3$& $*33\pm*3$& $*55\pm*3$& $*92\pm*3$& $*166\pm*3$\cr
 120--139& $!16\pm*3$& $!18\pm*3$& $*31\pm*3$& $*54\pm*3$& $*84\pm*3$& $*132\pm*4$\cr
 140--159& $!12\pm*4$& $!15\pm*4$& $*24\pm*3$& $*47\pm*4$& $*76\pm*4$& $*122\pm*4$\cr
 160--179& $!14\pm*5$& $!18\pm*4$& $*25\pm*4$& $*46\pm*4$& $*68\pm*4$& $*111\pm*4$\cr
 180--199& $!12\pm*5$& $!13\pm*5$& $*23\pm*5$& $*39\pm*5$& $*62\pm*5$& $*105\pm*5$\cr
 200--219& $!19\pm*6$& $!20\pm*5$& $*30\pm*5$& $*44\pm*5$& $*65\pm*5$& $*111\pm*5$\cr
 220--239& $!28\pm*6$& $!26\pm*6$& $*31\pm*6$& $*42\pm*5$& $*64\pm*5$& $*107\pm*5$\cr
 240--259& $!12\pm*8$& $!12\pm*7$& $*20\pm*6$& $*29\pm*6$& $*50\pm*6$& $**88\pm*6$\cr
 260--279& $!12\pm*8$& $!20\pm*7$& $*29\pm*7$& $*51\pm*7$& $*70\pm*6$& $*106\pm*6$\cr
 280--299& $!12\pm*9$& $!15\pm*8$& $*23\pm*8$& $*30\pm*8$& $*50\pm*8$& $**83\pm*7$\cr
 300--319& $!21\pm10$& $!21\pm*8$& $*25\pm*8$& $*34\pm*8$& $*55\pm*8$& $**89\pm*8$\cr
 320--339& $!11\pm11$& $!15\pm10$& $*22\pm*9$& $*32\pm*9$& $*56\pm*8$& $**88\pm*8$\cr
 340--359& $!*6\pm12$& $!*7\pm11$& $*17\pm10$& $*33\pm*9$& $*52\pm*9$& $**85\pm*8$\cr
 360--379& $!16\pm14$& $!10\pm12$& $*21\pm12$& $*38\pm11$& $*53\pm11$& $**83\pm10$\cr
 380--399& $!*4\pm13$& $!*2\pm12$& $*15\pm11$& $*18\pm11$& $*35\pm10$& $**61\pm*9$\cr
 400--419& $!*8\pm15$& $!*8\pm13$& $*17\pm13$& $*22\pm12$& $*44\pm11$& $**73\pm10$\cr
 420--439& $!23\pm16$& $!23\pm15$& $*31\pm14$& $*36\pm13$& $*55\pm12$& $**77\pm12$\cr
 440--459& $*-5\pm18$& $!11\pm15$& $*20\pm15$& $*45\pm14$& $*53\pm13$& $**74\pm12$\cr
 460--479& $-22\pm19$& $-16\pm17$& $**0\pm15$& $*18\pm15$& $*30\pm14$& $**61\pm13$\cr
 480--499& $!13\pm21$& $!*7\pm19$& $*17\pm18$& $*20\pm17$& $*38\pm17$& $**71\pm15$\cr
 500--549& $!16\pm15$& $!13\pm13$& $*12\pm11$& $*21\pm11$& $*29\pm10$& $**51\pm*9$\cr
 550--599& $!13\pm17$& $!13\pm15$& $*15\pm14$& $*21\pm14$& $*34\pm13$& $**58\pm12$\cr
\noalign{\vskip 5pt\hrule\vskip 3pt}}}
\endPlancktablewide
\par
\endgroup
\end{table*}

%%%%%%%%%%%%%%%%%%%%%%%%%%%%%%%%%
%%%% Table C2 (Spectral fit E-mode) from Nico's mail October 28 th %%%%%%%%
%%% 
\begin{table*}[tbp!]
\begingroup
\newdimen\tblskip \tblskip=5pt
\caption{Parameters of spectral model fit in Sect.~\ref{sec:dust_sed} for $E$ modes.}
\label{tab:EE_fit}
\vskip -3mm
%\footnotesize
\setbox\tablebox=\vbox{
 \newdimen\digitwidth
 \setbox0=\hbox{\rm 0}
 \digitwidth=\wd0
 \catcode`*=\active
 \def*{\kern\digitwidth}
 \newdimen\signwidth
 \setbox0=\hbox{+}
 \signwidth=\wd0
 \catcode`!=\active
 \def!{\kern\signwidth}
  \newdimen\dpwidth
  \setbox0=\hbox{.}
  \dpwidth=\wd0
  \catcode`?=\active
  \def?{\kern\dpwidth}
\halign{\tabskip 0pt\hbox to 2.5cm{#\leaderfil}\tabskip 1em&
\hfil#\hfil\tabskip 0.75em& \hfil#\hfil& \hfil#\hfil&
\hfil#\hfil& \hfil#\hfil& \hfil#\hfil\tabskip 0em\cr
\noalign{\doubleline}
 \omit& LR24& LR33& LR42& LR52& LR62& LR71\cr 
\noalign{\vskip 3pt\hrule\vskip 5pt}
%\multispan2$f_{\rm sky}^{\rm eff}$ [\%]\hfil& 24& 33& 42& 52& 62& 71\cr
%\noalign{\vskip 3pt\hrule\vskip 5pt}
\omit\hfil*$\ell$ range\hfil& \multispan6\hfil$A_{\rm d}$  [$\mu$K$^2$] \hfil\cr
\noalign{\vskip -5pt}
\omit& \multispan6\hrulefill\cr
\noalign{\vskip 3pt\hrule\vskip 5pt}
**4--*11 &     $47\pm5$ &    $102\pm5$ &    $202\pm5$ & $379^{+8}_{-7}$ &    $706\pm8$ &    $994\pm8$ \cr
*12--*19 & $95^{+4}_{-3}$ & $!86^{+4}_{-3}$ &    $137\pm3$ &    $254\pm4$ &    $367\pm5$ &    $738\pm6$ \cr
*20--*39 &     $55\pm1$ &     $!77\pm2$ &    $133\pm2$ &    $205\pm3$ &    $324\pm3$ &    $539\pm3$ \cr
*40--*59 &     $38\pm2$ &     $!53\pm2$ &     $!81\pm2$ &    $143\pm3$ &    $228\pm3$ & $354^{+3}_{-4}$ \cr
*60--*79 &     $32\pm2$ &     $!46\pm2$ &     $!68\pm3$ &    $109\pm3$ &    $181\pm3$ &    $306\pm4$ \cr
*80--*99 &     $32\pm3$ & $41^{+3}_{-2}$ &     $!72\pm3$ &    $116\pm3$ &    $191\pm3$ &    $310\pm4$ \cr
100--119 &     $32\pm3$ &     $!44\pm3$ &     $!72\pm3$ &    $110\pm3$ &    $170\pm4$ &    $282\pm4$ \cr
120--139 &     $24\pm4$ &     $!35\pm3$ &     $!56\pm3$ &     $!88\pm4$ &    $153\pm4$ &    $276\pm5$ \cr
140--159 &     $33\pm5$ & $41^{+5}_{-4}$ &     $!62\pm4$ &     $!93\pm4$ & $146^{+4}_{-5}$ &    $242\pm4$ \cr
\noalign{\vskip 3pt\hrule\vskip 5pt}
\omit& \multispan6\hfil$A_{\rm s}$ [$\mu$K$^2$]\hfil\cr
\noalign{\vskip -5pt}
\omit& \multispan6\hrulefill\cr
\noalign{\vskip 3pt\hrule\vskip 5pt}
**4--*11 &  $3.5\pm0.2$ &  $8.6\pm0.3$ & $12.7\pm0.3$ & $12.8\pm0.2$ & $15.2\pm0.2$ & $18.8\pm0.2$ \cr 
*12--*19 &  $7.3\pm0.2$ &  $6.8\pm0.2$ &  $!6.1\pm0.2$ &  $!8.2\pm0.2$ &  $!9.6\pm0.2$ & $11.0\pm0.2$ \cr
*20--*39 &  $2.1\pm0.1$ &  $2.8\pm0.1$ &  $!4.2\pm0.1$ &  $!4.8\pm0.1$ &  $!5.7\pm0.1$ &  $!7.9\pm0.1$ \cr
*40--*59 &  $1.1\pm0.2$ &  $1.2\pm0.1$ & $!2.0^{+0.2}_{-0.1}$ &  $12.9\pm0.1$ &  $!3.5\pm0.1$ &  $!5.4\pm0.2$ \cr
*60--*79 &  $0.4\pm0.2$ &  $0.8\pm0.2$ &  $!1.2\pm0.2$ &  $!1.6\pm0.2$ &  $!2.0\pm0.2$ &  $!3.2\pm0.2$ \cr
*80--*99 &  $0.3\pm0.2$ &  $0.4\pm0.2$ &  $!0.5\pm0.2$ &  $!1.0\pm0.2$ &  $!1.8\pm0.2$ &  $!2.9\pm0.2$ \cr
100 --119 &  $0.7\pm0.4$ &  $0.7\pm0.4$ &  $!1.3\pm0.3$ &  $!1.6\pm0.3$ &  $!2.2\pm0.3$ &  $!3.2\pm0.3$ \cr
120--139 &  $0.0\pm0.3$ &  $0.0\pm0.4$ &  $!0.0\pm0.3$ &  $!0.0\pm0.4$ &  $!1.1\pm0.4$ &  $!2.1\pm0.4$ \cr
140--159 &  $0.0\pm0.4$ &  $0.0\pm0.2$ &  $!0.0\pm0.2$ &  $!0.0\pm0.2$ &  $!0.0\pm0.3$ &  $!0.0\pm0.8$ \cr
\noalign{\vskip 3pt\hrule\vskip 5pt}
\omit& \multispan6\hfil$\beta_{\rm d}$\hfil\cr
\noalign{\vskip -5pt}
\omit& \multispan6\hrulefill\cr
\noalign{\vskip 3pt\hrule\vskip 5pt}
**4--*11 & $!2.01^{+0.22}_{-0.20}$ & $!1.89^{+0.10}_{-0.09}$ & $!1.72\pm0.05$ & $!1.64\pm0.04$ & $!1.58\pm0.03$ & $!1.55\pm0.02$ \cr
*12--*19 & $!1.58\pm0.06$ & $!1.51\pm0.06$ & $!1.52\pm0.04$ & $!1.49\pm0.03$ & $!1.50\pm0.03$ & $!1.53\pm0.02$ \cr
*20--*39 & $!1.34\pm0.05$ & $!1.46\pm0.04$ & $!1.49\pm0.03$ & $!1.52\pm0.03$ & $!1.50\pm0.02$ & $!1.50\pm0.02$ \cr
*40--*59 & $!1.53\pm0.10$ & $!1.47\pm0.07$ & $!1.54\pm0.05$ & $!1.53\pm0.04$ & $!1.56\pm0.03$ & $!1.55^{+0.02}_{-0.03}$ \cr
*60--*79 & $!1.27\pm0.12$ & $!1.38\pm0.09$ & $!1.43\pm0.07$ & $!1.42\pm0.05$ & $!1.47\pm0.03$ & $!1.47\pm0.03$ \cr
*80--*99 & $!1.42^{+0.16}_{-0.15}$ & $!1.41\pm0.11$ & $!1.47\pm0.07$ & $!1.50\pm0.05$ & $!1.53\pm0.04$ & $!1.52\pm0.03$ \cr
100--119 & $!1.65^{+0.23}_{-0.22}$ & $!1.49^{+0.14}_{-0.13}$ & $!1.48^{+0.09}_{-0.08}$ & $!1.45\pm0.06$ & $!1.46\pm0.04$ & $!1.49\pm0.03 $ \cr
120--139 & $!1.54^{+0.35}_{-0.31}$ & $!1.45^{+0.20}_{-0.19}$ & $!1.42^{+0.12}_{-0.11}$ & $!1.42\pm0.08$ & $!1.48\pm0.05$ & $!1.50^{+0.04}_{-0.03}$ \cr
140--159 & $!2.13^{+0.40}_{-0.35}$ & $!1.98^{+0.26}_{-0.24}$ & $!1.62^{+0.14}_{-0.13}$ & $!1.64\pm0.09$ & $!1.56\pm0.06 $ & $!1.56\pm0.04$ \cr
\noalign{\vskip 3pt\hrule\vskip 5pt}
\omit& \multispan6\hfil$\beta_{\rm s}$\hfil\cr
\noalign{\vskip -5pt}
\omit& \multispan6\hrulefill\cr
\noalign{\vskip 3pt\hrule\vskip 5pt}
**4--*11  & $-3.05^{+0.13}_{-0.14}$ & $-3.10^{+0.08}_{-0.09}$ & $-3.16\pm0.06 $ & $-3.13\pm0.05$ & $-3.09\pm0.04$ & $-3.10\pm0.03$ \cr
*12--*19 & $-3.00\pm0.09$ & $-2.95\pm0.09$ & $-2.96\pm0.09$ & $-2.97\pm0.07$ & $-3.11\pm0.06$ & $-3.15\pm0.05 $ \cr
*20--*39 & $-3.17\pm0.15 $ & $-3.17\pm0.12$ & $-3.31\pm0.09$ & $-3.31\pm0.08$ & $-3.24\pm0.07$ & $-3.23\pm0.05$ \cr
*40--*59 & $-3.08\pm0.18$ & $-3.13\pm0.18$ & $-3.20\pm0.16$ & $-3.19\pm0.14$ & $-3.13^{+0.12}_{-0.13}$ & $-3.12\pm0.09$ \cr
*60--*79 & $-3.11\pm0.18$ & $-3.11\pm0.18$ & $-3.17\pm0.17$ & $-3.15^{+0.18}_{-0.17}$ & $-3.18^{+0.17}_{-0.16}$ & $-3.14^{+0.14}_{-0.15}$ \cr
*80--*99 & $-3.13\pm0.18$ & $-3.13\pm0.18$ & $-3.12\pm0.18$ & $-3.11\pm0.18$ & $-3.10\pm0.17$ & $-3.11^{+0.15}_{-0.16}$ \cr
100--119 & $-3.13\pm0.18$ & $-3.13^{+0.18}_{-0.19}$ & $-3.16\pm0.18$ & $-3.16^{+0.17}_{-0.18}$ & $-3.13\pm0.17$ & $-3.10^{+0.16}_{-0.17}$ \cr
120--139 & $-3.11\pm0.18$ & $-3.10^{+0.18}_{-0.19}$ & $-3.11^{+0.19}_{-0.18}$ & $-3.11\pm0.18$ & $-3.10^{+0.19}_{-0.18}$ & $-3.08\pm0.18$ \cr
140--159 & $-3.12\pm0.18$ & $-3.12^{+0.18}_{-0.19}$ & $-3.11\pm0.18$ & $-3.12\pm0.18$ & $-3.12\pm0.18$ & $-3.13^{+0.18}_{-0.19}$ \cr
\noalign{\vskip 3pt\hrule\vskip 5pt}
\omit& \multispan6\hfil$\rho$\hfil\cr
\noalign{\vskip -6pt}
\omit& \multispan6\hrulefill\cr
\noalign{\vskip 3pt\hrule\vskip 5pt}
**4--*11& $!0.12\pm0.05$ & $!0.49^{+0.03}_{-0.02}$ & $!0.50\pm0.01$ & $!0.35\pm0.01$ & $!0.32\pm0.01$ & $!0.31\pm0.01$ \cr
*12--*19& $!0.25\pm0.02$ & $!0.02\pm0.02$ & $!0.05\pm0.02$ & $!0.27\pm0.01$ & $!0.35\pm0.01$ & $!0.35\pm0.01$ \cr
*20--*39& $!0.12\pm0.02$ & $!0.14\pm0.02$ & $!0.22\pm0.01$ & $!0.15\pm0.01$ & $!0.16\pm0.01$ & $!0.17\pm0.01$ \cr
*40--*59& $!0.01\pm0.05$ & $!0.04\pm0.04$ & $!0.07\pm0.03$ & $!0.11\pm0.02$ & $!0.08\pm0.02$ & $!0.14\pm0.01$ \cr
*60--*79& $!0.04^{+0.13}_{-0.11}$ & $-0.02\pm$0.07 & $!0.04^{+0.05}_{-0.04}$ & $-0.01\pm0.03$ & $-0.03\pm0.02$ & $!0.06\pm0.02$ \cr
*80--*99& $!0.41^{+0.28}_{-0.19}$ & $!0.42^{+0.22}_{-0.14}$ & $!0.31^{+0.17}_{-0.11}$ & $!0.13\pm0.05$ & $!0.13\pm0.03$ & $!0.16\pm0.02$ \cr
100--119& $!0.10^{+0.17}_{-0.15}$ & $!0.00^{+0.12}_{-0.11}$ & $!0.07^{+0.07}_{-0.06}$ & $!0.06\pm0.05$ & $!0.03\pm0.03$ & $!0.04\pm0.02$ \cr
120--139& $!0.13^{+0.37}_{-0.33}$ & $!0.27^{+0.33}_{-0.24}$ & $!0.01\pm$0.23 & $-0.12^{+0.16}_{-0.23}$ & $!0.01\pm$0.06 & $!0.13^{+0.04}_{-0.03}$ \cr
140--159& $-0.64^{+0.29}_{-0.23}$ & $-0.34^{+0.39}_{-0.38}$ & $-0.14^{+0.33}_{-0.37}$ & $-0.03^{+0.30}_{-0.31}$ & $!0.25^{+0.31}_{-0.18}$ & $!0.16^{+0.18}_{-0.10}$ \cr
\noalign{\vskip 3pt\hrule\vskip 5pt}
\omit& \multispan6\hfil$\chi^2 (N_{\rm dof}=\ndofspect)$\hfil\cr
\noalign{\vskip -5pt}
\omit& \multispan6\hrulefill\cr
\noalign{\vskip 3pt\hrule\vskip 5pt}
**4--*11 &         50.4 &         14.3 &          9.8 &         36.2 &         37.5 &         35.8 \cr
*12--*19 &         25.8 &         21.9 &         16.2 &         25.1 &         13.3 &         10.8 \cr
*20--*39 &          *5.5 &         16.8 &         11.8 &          *9.8 &         14.4 &         11.0 \cr
*40--*59 &         19.0 &         18.3 &         13.4 &         12.1 &          *8.0 &         17.2 \cr
*60--*79 &         18.7 &         17.8 &         16.8 &         17.1 &         18.2 &         23.4 \cr
*80--*99 &         15.2 &         14.5 &         20.8 &         33.3 &         29.7 &         30.1 \cr
100--119 &         16.8 &         15.7 &         14.5 &         14.3 &         13.9 &          *8.5 \cr
120--139 &         19.5 &         24.2 &         29.1 &         21.1 &         19.2 &         18.9 \cr
140--159 &         29.8 &         26.4 &         22.1 &         14.6 &         17.1 &         20.3 \cr
\noalign{\vskip 3pt\hrule\vskip 5pt}}}
\endPlancktablewide
\endgroup
\end{table*}

%%%%%%%%%%%%%%%%%%%%%%%%%%%%%%%%%
%%%% Table C3 (Spectral fit B-mode) from Nico's mail January the 3rd  %%%%%%%%
%%% 
\begin{table*}[tbp!]
\newdimen\tblskip \tblskip=5pt
%\caption{Best-fit parameters resulting from the spectral model fitting procedure described in Sect.\ref{sec:dust_sed} for $B$-mode power spectra.}
\caption{Parameters of spectral model fit in Sect.~\ref{sec:dust_sed} for $B$ modes.} 
%\DS{This table also has some values without enough digits.}
\label{tab:BB_fit}
\vskip -3mm
%\footnotesize
\setbox\tablebox=\vbox{
\newdimen\digitwidth
\setbox0=\hbox{\rm 0}
\digitwidth=\wd0
\catcode`*=\active
 \def*{\kern\digitwidth}
 \newdimen\signwidth
 \setbox0=\hbox{+}
 \signwidth=\wd0
 \catcode`!=\active
 \def!{\kern\signwidth}
  \newdimen\dpwidth
  \setbox0=\hbox{.}
  \dpwidth=\wd0
  \catcode`?=\active
  \def?{\kern\dpwidth}
\halign{\tabskip 0pt\hbox to 2.5cm{#\leaderfil}\tabskip 1em&
\hfil#\hfil\tabskip 0.75em& \hfil#\hfil& \hfil#\hfil&
\hfil#\hfil& \hfil#\hfil& \hfil#\hfil\tabskip 0em\cr
\noalign{\doubleline}
\omit& LR24& LR33& LR42& LR52& LR62& LR71\cr 
\noalign{\vskip 3pt\hrule\vskip 5pt}
%\multispan2$f_{\rm sky}^{\rm eff}$ [\%]\hfil& 24& 33& 42& 52& 62& 71\cr
%\noalign{\vskip 3pt\hrule\vskip 5pt}
\omit\hfil*$\ell$ range\hfil& \multispan6\hfil$A_{\rm d}$  [$\mu$K$^2$]\hfil\cr
\noalign{\vskip -5pt}
\omit& \multispan6\hrulefill\cr
\noalign{\vskip 3pt\hrule\vskip 5pt}
**4--*11 &     $36\pm3$ &     $78\pm4$ & $188^{+6}_{-5}$ & $366^{+6}_{-5}$ &    $359\pm5$ &    $506\pm4$ \cr
*12--*19 &     $29\pm2$ &     $43\pm2$ &    $116\pm3$ &    $158\pm3$ & $282^{+4}_{-3}$ &    $422\pm4$ \cr
*20--*39 &     $31\pm1$ &     $43\pm1$ &     $!89\pm1$ &    $134\pm2$ &    $183\pm2$ &    $398\pm2$ \cr
*40--*59 &     $19\pm1$ &     $25\pm1$ &     $!53\pm1$ &     $!86\pm2$ &    $137\pm2$ &    $241\pm2$ \cr
*60--*79 &     $14\pm1$ &     $19\pm1$ &     $!35\pm1$ &     $!68\pm2$ &    $108\pm2$ &    $170\pm2$ \cr
*80--*99 &     $17\pm1$ &     $19\pm1$ &     $!36\pm2$ &     $!58\pm2$ &     $!97\pm2$ &    $173\pm2$ \cr
100--119 &     $14\pm2$ &     $18\pm2$ &     $!31\pm2$ &     $!54\pm2$ &     $!92\pm2$ &    $167\pm2$ \cr
120--139 & $14^{+3}_{-2}$ &     $18\pm2$ &     $!29\pm2$ &     $!52\pm3$ & $!83^{+3}_{-2}$ &    $131\pm3$ \cr
140--159 &      $!9\pm3$ & $14^{+3}_{-2}$ &     $!23\pm2$ &     $!46\pm3$ &     $!74\pm3$ &    $122\pm3$ \cr
\noalign{\vskip 3pt\hrule\vskip 5pt}
\omit& \multispan6\hfil$A_{\rm s}$  [$\mu$K$^2$]\hfil\cr
\noalign{\vskip -5pt}
\omit& \multispan6\hrulefill\cr
\noalign{\vskip 3pt\hrule\vskip 5pt}
**4--*11 &  $6.7\pm0.3$ &  $8.8\pm0.2$ &  $9.7\pm0.2$ & $12.4\pm0.2$ & $10.8\pm0.2$ &  $9.9\pm0.2$ \cr
*12--*19 &  $1.5\pm0.2$ &  $0.9\pm0.1$ &  $1.4\pm0.1$ &  $!2.7\pm0.1$ & $!4.2^{+0.2}_{-0.1}$ &  $6.1\pm0.2$ \cr
*20--*39 &  $0.4\pm0.1$ &  $0.4\pm0.1$ &  $1.0\pm0.1$ &  $!0.7\pm0.1$ &  $!1.2\pm0.1$ &  $3.2\pm0.1$ \cr
*40--*59 & $0.1^{+0.1}_{-0.0}$ &  $0.0\pm0.1$ &  $0.3\pm0.1$ &  $!0.2\pm0.1$ &  $!0.6\pm0.1$ &  $1.7\pm0.1$ \cr
*60--*79 &  $0.0\pm0.1$ &  $0.0\pm0.1$ &  $0.0\pm0.1$ &  $!0.0\pm0.1$ &  $!0.0\pm0.2$ &  $0.6\pm0.2$ \cr
*80--*99 &  $0.0\pm0.2$ &  $0.0\pm0.2$ &  $0.0\pm0.2$ &  $!0.0\pm0.1$ &  $!0.0\pm0.3$ &  $0.5\pm0.2$ \cr
100 --119 &  $0.0\pm0.2$ &  $0.0\pm0.3$ &  $0.0\pm0.4$ & $!0.3^{+0.3}_{-0.2}$ &  $!0.6\pm0.3$ &  $0.9\pm0.3$ \cr
120--139 &  $0.0\pm0.3$ &  $0.0\pm0.2$ &  $0.0\pm0.2$ &  $!0.0\pm0.2$ &  $!0.0\pm0.3$ &  $0.0\pm0.4$ \cr
 140--159 &  $0.0\pm0.5$ &  $0.0\pm0.6$ &  $0.0\pm0.6$ &  $!0.0\pm0.4$ &  $!0.0\pm0.4$ &  $0.0\pm0.3$ \cr
\noalign{\vskip 3pt\hrule\vskip 5pt}
\omit& \multispan6\hfil$\beta_{\rm d}$\hfil\cr
\noalign{\vskip -5pt}
\omit& \multispan6\hrulefill\cr
\noalign{\vskip 3pt\hrule\vskip 5pt}
**4--*11 & $!1.98^{+0.19}_{-0.18}$ & $!1.68\pm0.07$ & $!1.60\pm0.04$ & $!1.54\pm0.03$ & $!1.55^{+0.03}_{-0.02}$ & $!1.55\pm0.02$ \cr
*12--*19 & $!1.34^{+0.09}_{-0.10}$ & $!1.43^{+0.08}_{-0.07}$ & $!1.53\pm0.04$ & $!1.61\pm0.04$ & $!1.57\pm0.03$ & $!1.53\pm0.02$ \cr
*20--*39 & $!1.49^{+0.08}_{-0.07}$ & $!1.53\pm0.06$ & $!1.51\pm0.03$ & $!1.53\pm0.03$ & $!1.55\pm0.03$ & $!1.56\pm0.02$ \cr
*40--*59 & $!1.77^{+0.15}_{-0.14}$ & $!1.58\pm0.1$ & $!1.51\pm0.05$ & $!1.48\pm0.04$ & $!1.50\pm0.03$ & $!1.54\pm0.02$ \cr
 *60--*79 & $!1.22\pm0.14$ & $!1.4\pm0.13$ & $!1.52\pm0.08$ & $!1.51\pm0.05$ & $!1.50^{+0.04}_{-0.03}$ & $!1.52\pm0.03$ \cr
*80--*99 & $!1.50\pm0.19$ & $!1.44^{+0.16}_{-0.14}$ & $!1.42\pm0.09$ & $!1.46\pm0.06$ & $!1.49\pm0.04$ & $!1.52\pm0.03$ \cr
100--119 & $!1.64^{+0.34}_{-0.29}$ & $!1.62^{+0.24}_{-0.22}$ & $!1.56\pm0.13$ & $!1.56\pm0.08$ & $!1.54\pm0.05$ & $!1.50\pm0.03$ \cr
120--139 & $!1.73^{+0.43}_{-0.36}$ & $!1.66^{+0.28}_{-0.25}$ & $!1.80\pm0.19$ & $!1.67^{+0.11}_{-0.10}$ & $!1.66\pm0.06$ & $!1.60^{+0.05}_{-0.04}$ \cr
140--159 & $!1.05^{+0.63}_{-0.48}$ & $!1.17^{+0.31}_{-0.29}$ & $!1.22^{+0.18}_{-0.17}$ & $!1.37\pm0.10$ & $!1.48\pm0.07$ & $!1.53\pm0.05$ \cr
\noalign{\vskip 3pt\hrule\vskip 5pt}
\omit& \multispan6\hfil$\beta_{\rm s}$\hfil\cr
\noalign{\vskip -5pt}
\omit& \multispan6\hrulefill\cr
\noalign{\vskip 3pt\hrule\vskip 5pt}
**4--*11 & $-3.13^{+0.11}_{-0.12}$ & $-3.13\pm0.08$ & $-3.10\pm0.06$ & $-3.18\pm0.04$ & $-3.18^{+0.04}_{-0.05}$ & $-3.14\pm0.04$ \cr
*12--*19 & $-3.26\pm0.17$ & $-3.17\pm0.18$ & $-3.44\pm0.16$ & $-3.26^{+0.13}_{-0.14}$ & $-3.32\pm0.10$ & $-3.13\pm0.08$ \cr
 *20--*39 & $-3.14\pm0.18$ & $-3.17^{+0.17}_{-0.18}$ & $-3.12\pm0.16$ & $-3.04\pm0.17$ & $-2.95\pm0.15$ & $-2.95^{+0.08}_{-0.09}$ \cr
*40--*59 & $-3.11\pm0.18$ & $-3.11^{+0.19}_{-0.18}$ & $-3.11\pm0.18$ & $-3.12\pm0.18$ & $-3.15^{+0.17}_{-0.18}$ & $-3.22^{+0.14}_{-0.15}$ \cr
*60--*79 & $-3.11^{+0.18}_{-0.19}$ & $-3.11^{+0.19}_{-0.18}$ & $-3.08\pm0.18$ & $-3.10^{+0.19}_{-0.18}$ & $-3.09^{+0.18}_{-0.19}$ & $-3.09\pm0.18$ \cr
*80--*99 & $-3.11\pm0.18 $& $-3.12\pm0.18$ & $-3.12\pm0.18$ & $-3.11\pm0.18$ & $-3.12\pm0.18$ & $-3.12\pm0.18$ \cr
100--119 & $-3.11\pm0.18$ & $-3.11\pm0.19$ & $-3.12^{+0.19}_{-0.18}$ & $-3.12^{+0.18}_{-0.19}$ & $-3.13\pm0.18$ & $-3.15\pm0.18$ \cr
120--139 & $-3.11\pm0.18$ & $-3.12\pm0.18 $ & $-3.11\pm0.18$ & $-3.11^{+0.18}_{-0.19}$ & $-3.11^{+0.17}_{-0.18}$ & $-3.10\pm0.18$ \cr
140--159 & $-3.11\pm0.18$ & $-3.11\pm0.18$ & $-3.11\pm0.18$ & $-3.12^{+0.18}_{-0.19}$ & $-3.12\pm0.18 $& $-3.11\pm0.18$ \cr
\noalign{\vskip 3pt\hrule\vskip 5pt}
\omit& \multispan6\hfil$\rho$\hfil\cr
\noalign{\vskip -6pt}
\omit& \multispan6\hrulefill\cr
\noalign{\vskip 3pt\hrule\vskip 5pt}
**4--*11& $!0.39\pm0.04$ & $!0.55\pm0.02$ & $!0.55\pm0.01$ & $!0.69\pm0.01$ & $!0.67\pm0.01$ & $!0.48\pm0.01$ \cr
*12--*19& $-0.13\pm0.05$ & $-0.13\pm0.06$ & $!0.42\pm0.04$ & $!0.41\pm0.02$ & $!0.49\pm0.02$ & $!0.24\pm0.01$ \cr
 *20--*39& $!0.56^{+0.12}_{-0.09}$ & $!0.39^{+0.08}_{-0.07}$ & $!0.39^{+0.04}_{-0.03}$ & $!0.30\pm0.03$ & $!0.28\pm0.02$ & $!0.33\pm0.01$ \cr
 *40--*59& $-0.71^{+0.23}_{-0.19}$ & $-0.35^{+0.25}_{-0.34}$ & $!0.27^{+0.12}_{-0.08}$ & $!0.02\pm0.07$ & $!0.17\pm0.04$ & $!0.20\pm0.02$ \cr
*60--*79& $!0.45\pm0.32$ & $!0.17^{+0.40}_{-0.36}$ & $!0.42^{+0.29}_{-0.19}$ & $!0.15^{+0.28}_{-0.18}$ & $!0.10^{+0.15}_{-0.10}$ & $!0.14^{+0.05}_{-0.04}$ \cr
*80--*99& $!0.27^{+0.32}_{-0.24}$ & $-0.01\pm0.24$ & $-0.09^{+0.22}_{-0.25}$ & $-0.01\pm0.2$ & $!0.05^{+0.14}_{-0.10}$ & $!0.05\pm0.05$ \cr
100--119& $!0.23^{+0.38}_{-0.34}$ & $!0.16^{+0.34}_{-0.26}$ & $!0.26^{+0.28}_{-0.18}$ & $!0.24^{+0.21}_{-0.12}$ & $!0.06^{+0.08}_{-0.07}$ & $!0.08^{+0.05}_{-0.04}$ \cr
120--139& $!0.59^{+0.26}_{-0.31}$ & $!0.28^{+0.38}_{-0.37}$ & $!0.29^{+0.37}_{-0.31}$ & $!0.19^{+0.35}_{-0.24}$ & $!0.25^{+0.30}_{-0.16}$ & $!0.15^{+0.21}_{-0.11}$ \cr
140--159& $-0.06^{+0.42}_{-0.41}$ & $-0.13^{+0.30}_{-0.33}$ & $-0.38^{+0.22}_{-0.30}$ & $-0.36^{+0.19}_{-0.30}$ & $-0.24^{+0.18}_{-0.31}$ & $!0.06^{+0.20}_{-0.16}$ \cr
\noalign{\vskip 3pt\hrule\vskip 5pt}
\omit& \multispan6\hfil$\chi^2 (N_{\rm dof}=\ndofspect)$\hfil\cr
\noalign{\vskip -5pt}
\omit& \multispan6\hrulefill\cr
\noalign{\vskip 3pt\hrule\vskip 5pt}
**4--*11 &         19.7 &         14.1 &          4.4 &         12.2 &         12.7 &         24.1 \cr
*12--*19 &         23.1 &         14.1 &         35.0 &         17.6 &         19.2 &         43.3 \cr
*20--*39 &         21.5 &         23.5 &         19.4 &         24.7 &         24.3 &         31.8 \cr
*40--*59 &         24.4 &         10.9 &         10.9 &         11.1 &          *9.7 &          *8.8 \cr
*60--*79 &         18.8 &         25.5 &         19.5 &         20.5 &         16.8 &         14.4 \cr
*80--*99 &         22.4 &         13.4 &         16.3 &         18.3 &         16.1 &         20.1 \cr
100--119 &         29.8 &         34.7 &         40.3 &         45.5 &         53.7 &         46.5 \cr
120--139 &         21.5 &         18.2 &         23.0 &         27.7 &         27.7 &         30.0 \cr
140--159 &         27.2 &         31.4 &         29.8 &         34.8 &         34.0 &         32.4 \cr
\noalign{\vskip 3pt\hrule\vskip 5pt}}}
\endPlancktablewide
\end{table*}

%%%%%%%%%%%%%%%%%%%%%%%%%%%%%
%% Beta_dust from colour ratio: Table for Planck data
%%%%%%%%%%%%%%%%%%%%%%%%%%%%
\begin{table*}[tbp!]
\newdimen\tblskip \tblskip=5pt
\caption{Dust $TT$, $EE$, and $BB$ spectral indices from the colour ratio $\alpha_\ell^{XX} \, (217,353)$
(Eq.~(\ref{eq:color_ratio}) in Sect.~\ref{sec:betaT}) for the \Planck\ data. The error bars on $\beta_{\mathrm d}^{EE}$ and $\beta_{\mathrm d}^{BB}$ do not include the 1.5\,\% uncertainty on the 353\,GHz polarization efficiency. }
\label{tab:betadvaldata}
\vskip -3mm
%\footnotesize
\setbox\tablebox=\vbox{
 \newdimen\digitwidth
 \setbox0=\hbox{\rm 0}
 \digitwidth=\wd0
 \catcode`*=\active
 \def*{\kern\digitwidth}
 \newdimen\signwidth
 \setbox0=\hbox{+}
 \signwidth=\wd0
 \catcode`!=\active
 \def!{\kern\signwidth}
  \newdimen\dpwidth
  \setbox0=\hbox{.}
  \dpwidth=\wd0
  \catcode`?=\active
  \def?{\kern\dpwidth}
\halign{\tabskip 0pt\hbox to 3.0cm{#\leaderfil}\tabskip 0.5em&
\hfil#\hfil\tabskip 0.75em& \hfil#\hfil& \hfil#\hfil&
\hfil#\hfil& \hfil#\hfil& \hfil#\hfil\tabskip 0em\cr
\noalign{\doubleline}
\omit& LR24& LR33& LR42& LR52& LR62& LR71\cr
\noalign{\vskip 3pt\hrule\vskip 5pt}
\omit\hfil*$\ell$ range\hfil& \multispan6\hfil$\beta_{\rm d}^{TT}$\hfil\cr
\noalign{\vskip -4pt}
\omit& \multispan6\hrulefill\cr
\noalign{\vskip 3pt\hrule\vskip 5pt}
**4--*11& $1.39\pm0.01$& $1.45\pm0.01$& $1.50\pm0.00$& $1.50\pm0.00$& $1.49\pm0.00$& $1.47\pm0.00$\cr
*12--*19& $1.53\pm0.01$& $1.53\pm0.01$& $1.48\pm0.00$& $1.47\pm0.00$& $1.49\pm0.00$& $1.47\pm0.00$\cr
*20--*39& $1.42\pm0.01$& $1.47\pm0.01$& $1.47\pm0.00$& $1.48\pm0.00$& $1.49\pm0.00$& $1.48\pm0.00$\cr
*40--*59& $1.41\pm0.01$& $1.46\pm0.01$& $1.48\pm0.01$& $1.48\pm0.00$& $1.48\pm0.00$& $1.47\pm0.00$\cr
*60--*79& $1.43\pm0.01$& $1.47\pm0.01$& $1.50\pm0.01$& $1.49\pm0.00$& $1.48\pm0.00$& $1.48\pm0.00$\cr
*80--*99& $1.42\pm0.01$& $1.45\pm0.01$& $1.48\pm0.01$& $1.47\pm0.01$& $1.48\pm0.00$& $1.48\pm0.00$\cr
100--119& $1.48\pm0.01$& $1.51\pm0.01$& $1.51\pm0.01$& $1.49\pm0.01$& $1.49\pm0.00$& $1.48\pm0.00$\cr
120--139& $1.47\pm0.02$& $1.50\pm0.01$& $1.52\pm0.01$& $1.50\pm0.01$& $1.50\pm0.00$& $1.49\pm0.00$\cr
140--169& $1.52\pm0.02$& $1.53\pm0.01$& $1.53\pm0.01$& $1.50\pm0.01$& $1.50\pm0.00$& $1.50\pm0.00$\cr
\noalign{\vskip 3pt}
\noalign{\vskip 3pt\hrule\vskip 5pt}
\omit& \multispan6\hfil$\beta_{\mathrm d}^{EE}$\hfil\cr
\noalign{\vskip -4pt}
\omit& \multispan6\hrulefill\cr
\noalign{\vskip 3pt\hrule\vskip 5pt}
**4--*11& $1.67\pm0.41$& $1.72\pm0.22$& $1.65\pm0.12$& $1.55\pm0.08$& $1.53\pm0.04$& $1.52\pm0.03$\cr
*12--*19& $1.78\pm0.14$& $1.70\pm0.16$& $1.61\pm0.10$& $1.56\pm0.06$& $1.52\pm0.05$& $1.53\pm0.03$\cr
*20--*39& $1.32\pm0.12$& $1.49\pm0.11$& $1.50\pm0.06$& $1.55\pm0.05$& $1.53\pm0.03$& $1.52\pm0.02$\cr
*40--*59& $1.53\pm0.21$& $1.46\pm0.16$& $1.48\pm0.11$& $1.52\pm0.07$& $1.53\pm0.06$& $1.53\pm0.04$\cr
*60--*79& $1.47\pm0.27$& $1.44\pm0.19$& $1.49\pm0.15$& $1.41\pm0.10$& $1.45\pm0.07$& $1.46\pm0.05$\cr
*80--*99& $1.56\pm0.32$& $1.57\pm0.24$& $1.54\pm0.15$& $1.51\pm0.11$& $1.53\pm0.07$& $1.52\pm0.05$\cr
100--119& $1.73\pm0.37$& $1.51\pm0.24$& $1.54\pm0.16$& $1.53\pm0.11$& $1.53\pm0.08$& $1.52\pm0.05$\cr
120--139& $1.64\pm0.53$& $1.69\pm0.34$& $1.58\pm0.22$& $1.55\pm0.16$& $1.59\pm0.10$& $1.57\pm0.06$\cr
140--169& $1.97\pm0.53$& $1.92\pm0.38$& $1.60\pm0.24$& $1.60\pm0.17$& $1.54\pm0.11$& $1.53\pm0.07$\cr
\noalign{\vskip 3pt}
\noalign{\vskip 3pt\hrule\vskip 5pt}
\omit& \multispan6\hfil$\beta_{\mathrm d}^{BB}$\hfil\cr
\noalign{\vskip -4pt}
\omit& \multispan6\hrulefill\cr
\noalign{\vskip 3pt\hrule\vskip 5pt}
**4--*11& $2.10\pm0.37$& $1.71\pm0.20$& $1.57\pm0.12$& $1.51\pm0.06$& $1.48\pm0.05$& $1.50\pm0.03$\cr
*12--*19& $1.33\pm0.28$& $1.50\pm0.19$& $1.55\pm0.10$& $1.60\pm0.08$& $1.56\pm0.05$& $1.52\pm0.04$\cr
*20--*39& $1.66\pm0.16$& $1.54\pm0.12$& $1.55\pm0.07$& $1.57\pm0.05$& $1.58\pm0.04$& $1.56\pm0.02$\cr
*40--*59& $1.67\pm0.26$& $1.58\pm0.21$& $1.55\pm0.11$& $1.46\pm0.08$& $1.50\pm0.05$& $1.53\pm0.03$\cr
*60--*79& $1.37\pm0.33$& $1.49\pm0.27$& $1.54\pm0.17$& $1.53\pm0.10$& $1.48\pm0.07$& $1.49\pm0.05$\cr
*80--*99& $1.25\pm0.36$& $1.35\pm0.31$& $1.37\pm0.19$& $1.44\pm0.14$& $1.47\pm0.09$& $1.49\pm0.06$\cr
100--119& $2.07\pm0.57$& $1.89\pm0.41$& $1.77\pm0.24$& $1.64\pm0.15$& $1.49\pm0.09$& $1.46\pm0.05$\cr
120--139& $2.05\pm0.63$& $1.54\pm0.48$& $1.89\pm0.31$& $1.80\pm0.17$& $1.66\pm0.11$& $1.62\pm0.08$\cr
140--169& $2.44\pm1.10$& $1.41\pm0.67$& $1.31\pm0.39$& $1.36\pm0.21$& $1.52\pm0.14$& $1.52\pm0.09$\cr
\noalign{\vskip 3pt}
 \noalign{\vskip 3pt\hrule\vskip 5pt}}}
\endPlancktablewide
\end{table*}

%??? would be nice to force a new page here
\clearpage
\newpage

\begin{table*}[tbp!]
\newdimen\tblskip \tblskip=5pt
\caption{Values of the spectral correlation ratio $\mathcal{R}_{\ell}^{BB} (217, 353)$ for six LR regions and five $\ell$ bins.}
\label{tab:Rell_LR}
\vskip -3mm
%\footnotesize
\setbox\tablebox=\vbox{
 \newdimen\digitwidth
 \setbox0=\hbox{\rm 0}
 \digitwidth=\wd0
 \catcode`*=\active
 \def*{\kern\digitwidth}
 \newdimen\signwidth
 \setbox0=\hbox{+}
 \signwidth=\wd0
 \catcode`!=\active
 \def!{\kern\signwidth}
  \newdimen\dpwidth
  \setbox0=\hbox{.}
  \dpwidth=\wd0
  \catcode`?=\active
  \def?{\kern\dpwidth}
\halign{\tabskip 0pt\hbox to 2.5cm{#\leaderfil}\tabskip 1em&
\hfil#\hfil\tabskip 0.75em& \hfil#\hfil& \hfil#\hfil&
\hfil#\hfil& \hfil#\hfil& \hfil#\hfil\tabskip 0em\cr
\noalign{\doubleline}
\omit& LR24& LR33& LR42& LR52& LR62& LR71\cr 
\noalign{\vskip 3pt\hrule\vskip 5pt}
\omit\hfil*$\ell$ range\hfil& \multispan6\hfil$R_{\ell}^{BB}$\hfil\cr
\noalign{\vskip -5pt}
\omit& \multispan6\hrulefill\cr
\noalign{\vskip 3pt\hrule\vskip 5pt}
**4--*11& 1.014& $0.996$& $0.996$& $0.998$& $0.997$& $0.997$\cr
*11--*50& 0.977& $0.990$& $0.990$& $0.992$& $0.993$& $0.998$\cr
*50--160& 0.822& $0.886$& $0.932$& $0.954$& $0.976$& $0.989$\cr
160--320& 0.479& $0.607$& $0.842$& $0.911$& $0.945$& $0.970$\cr
320--500& 0.745& $0.788$& $0.724$& $0.941$& $0.973$& $0.980$\cr
\noalign{\vskip 3pt\hrule\vskip 5pt}
\omit& \multispan6\hfil$\rm{PTS}_{\rm{HM}}$ [\%]\hfil\cr
\noalign{\vskip -6pt}
\omit& \multispan6\hrulefill\cr
\noalign{\vskip 3pt\hrule\vskip 5pt}
**4--*11& 84.4& $70.0$& $75.0$& $79.7$& $69.7$& $88.0$\cr
*11--*50& 29.7& $60.3$& $28.7$& $17.7$& $*9.0$& $44.0$\cr
*50--160& 11.3& $*9.0$& $*6.3$& $*4.3$& $11.7$& $27.7$\cr
160--320& *7.6& $*3.0$& $23.3$& $29.7$& $30.7$& $34.7$\cr
320--500& 58.7& $57.3$& $37.2$& $59.1$& $61.7$& $63.0$\cr
\noalign{\vskip 3pt\hrule\vskip 5pt}}}
\endPlancktablewide
\end{table*}

\begin{table*}[tbp!]
\newdimen\tblskip \tblskip=5pt
\caption{
Values of the spectral correlation ratio $\mathcal{R}_{\ell}^{BB} (217, 353)$ for northern and southern portions of LR regions and five $\ell$ bins.}
\label{tab:Rell_splits}
\vskip -3mm
%\footnotesize
\setbox\tablebox=\vbox{
 \newdimen\digitwidth
 \setbox0=\hbox{\rm 0}
 \digitwidth=\wd0
 \catcode`*=\active
 \def*{\kern\digitwidth}
 \newdimen\signwidth
 \setbox0=\hbox{+}
 \signwidth=\wd0
 \catcode`!=\active
 \def!{\kern\signwidth}
  \newdimen\dpwidth
  \setbox0=\hbox{.}
  \dpwidth=\wd0
  \catcode`?=\active
  \def?{\kern\dpwidth}
\halign{\tabskip 0pt\hbox to 2.5cm{#\leaderfil}\tabskip 1em&
\hfil#\hfil\tabskip 0.75em& \hfil#\hfil& \hfil#\hfil&
\hfil#\hfil& \hfil#\hfil& %  \hfil#\hfil& \hfil#\hfil& \hfil#\hfil& & \hfil#\hfil&
\hfil#\hfil& \hfil#\hfil& \hfil#\hfil\tabskip 0em\cr
\noalign{\doubleline}
\omit& LR42N& LR42S& LR52N& LR52S& LR62N& LR62S& LR71N& LR71S\cr 
\noalign{\vskip 3pt\hrule\vskip 5pt}
$f_{\rm sky}^{\rm eff}$& 23& 17& 27& 24& 32& 29& 36& 34\cr 
$\ainten$& 0.109& 0.096& 0.128& 0.133& 0.159& 0.169& 0.204& 0.232\cr 
$N_{\rm H}$& 2.83& 2.62& 3.34& 3.65& 4.16& 4.66& 5.36& 6.36\cr 
\noalign{\vskip 3pt\hrule\vskip 5pt}
\omit\hfil*$\ell$ range\hfil& \multispan8\hfil$R_{\ell}^{BB}$\hfil\cr
\noalign{\vskip -5pt}
\omit& \multispan8\hrulefill\cr
\noalign{\vskip 3pt\hrule\vskip 5pt}
**4--*11& 1.005& $0.991$& $0.999$& $0.997$& $0.997$& $0.996$& $0.998$& $0.996$\cr
*11--*50& 0.987& $0.990$& $0.999$& $0.985$& $0.996$& $0.990$& $1.001$& $0.994$\cr
*50--160& 0.902& $0.999$& $0.916$& $0.988$& $0.956$& $0.995$& $0.976$& $0.999$\cr
160--320& 0.815& $0.889$& $0.888$& $0.933$& $0.941$& $0.950$& $0.956$& $0.979$\cr
320--500& 0.750& $0.729$& $0.885$& $1.095$& $0.858$& $1.123$& $0.948$& $1.007$\cr
\noalign{\vskip 3pt\hrule\vskip 5pt}
\omit& \multispan8\hfil$\rm{PTS}_{\rm{HM}}$ [\%]\hfil\cr
\noalign{\vskip -6pt}
\omit& \multispan8\hrulefill\cr
\noalign{\vskip 3pt\hrule\vskip 5pt}
**4--*11& 56.7& $19.3$& $59.7$& $56.0$& $48.7$& $44.7$& $58.7$& $44.7$\cr
*11--*50& 24.0& $40.7$& $65.0$& $*4.0$& $47.3$& $*2.3$& $86.7$& $*1.3$\cr
*50--160& *1.7& $64.3$& $*0.7$& $50.7$& $*2.3$& $58.3$& $*6.0$& $70.0$\cr
160--320& 21.3& $41.7$& $29.3$& $42.3$& $38.3$& $39.7$& $34.0$& $48.0$\cr
320--500& 44.4& $44.7$& $51.4$& $69.5$& $41.1$& $76.3$& $52.7$& $66.3$\cr
\noalign{\vskip 3pt\hrule\vskip 5pt}}}
\endPlancktablewide
\end{table*}

\end{document}